\documentclass[journal, 12pt,onecolumn, draftclsnofoot]{IEEEtran}
\usepackage{ragged2e}
\usepackage{floatrow}
\usepackage{amsthm}
\usepackage[utf8]{inputenc}
\usepackage[english]{babel}
\theoremstyle{remark}
\newtheorem{theorem}{Theorem}
\newtheorem{lemma}{Lemma}

\newtheorem{remark}{Remark}

\newtheorem{Proposition}{Proposition}

\usepackage{multicol}
\usepackage{epsfig}
\usepackage{epstopdf}
\usepackage{float}
\usepackage{perpage}
\MakeSorted{figure}
\MakeSorted{table}
\usepackage{array}
\usepackage{graphicx}
\usepackage{amsfonts}
\usepackage{amssymb}
\usepackage{soul}
\let\labelindent\relax
\usepackage{enumitem}
\usepackage[english]{babel}
\usepackage[utf8]{inputenc}
\usepackage[linesnumbered,ruled,vlined]{algorithm2e}

\usepackage{hyperref}
\hypersetup{linktocpage} % links to only the page number
  \hypersetup{
    colorlinks,
    citecolor=black,
    filecolor=black,
    linkcolor=black,
    urlcolor=black
}

\usepackage{cite}
\DeclareMathAlphabet\mathbfcal{OMS}{cmsy}{b}{n}
\ifCLASSINFOpdf
\else
\fi
\usepackage[cmex10]{amsmath}
\usepackage{array}
\usepackage{fixltx2e}
\newcounter{mytempeqncnt}
\usepackage[linesnumbered,ruled,vlined]{algorithm2e}
\usepackage[font=footnotesize,labelfont=bf, figurename=Fig.]{caption} %scriptsize\\
\usepackage{subcaption}
\usepackage{fix-cm}

\usepackage{etoolbox} % for \iftoggle
\newtoggle{SINGLE_COL}
\toggletrue{SINGLE_COL}     % for single column format
\newtoggle{BIG_EQUATION}
\toggletrue{BIG_EQUATION}  % for two column format
\togglefalse{BIG_EQUATION}  % for single column format

\iftoggle{SINGLE_COL}{
\setlength{\abovedisplayskip}{3pt}
\setlength{\belowdisplayskip}{3pt}
}{}

\begin{document}

%\title{QoS-Constrained Energy-Efficient Optimization for Multi-Pair Two-Way AF Full-Duplex Massive MIMO Relaying}
\title{Full-Duplex Massive MIMO Multi-Pair Two-Way AF Relaying: Energy Efficiency Optimization}
\author{% <-this % stops a space
\iftoggle{SINGLE_COL}{\vspace*{-0.2in}}{}
Ekant Sharma, Rohit Budhiraja,  K Vasudevan and Lajos Hanzo, \textit{Fellow IEEE}
\thanks{Ekant Sharma, Rohit Budhiraja  and K Vasudevan are with the Department of Electrical Engineering, Indian Institute of Technology,
        Kanpur, 208016, India.   email: \{ekant, rohitbr, vasu@\}iitk.ac.in. \newline Lajos Hanzo is with the School of
Electronics and Computer Science, University of Southampton, Southampton,
SO17 1BJ, U.K. (e-mail: lh@ecs.soton.ac.uk).
        %{\small email: \{ekant, rohitbr, vasu@\}iitk.ac.in }
        \newline Part of this work will be presented in IEEE International Symposium on Personal, Indoor and Mobile Radio Communications (PIMRC) 2017.
        }%
       
       }

\maketitle
%---------------------------------------------------------------
\iftoggle{SINGLE_COL}{\vspace*{-0.9in}}{}
\begin{abstract}\iftoggle{SINGLE_COL}{\vspace*{-0.15in}}{}
We consider two-way amplify and forward  relaying, where multiple full-duplex user pairs  exchange information via a shared full-duplex massive multiple-input multiple-output (MIMO) relay. Most of the previous massive MIMO relaying works maximize the spectral efficiency (SE). By contrast, we maximize the non-convex energy efficiency (EE) metric by approximating it as a pseudo-concave problem, which is then solved using the classic Dinkelbach approach. We also maximize the EE of the least energy-efficient user {relying} on the max-min approach. For solving these optimization problems, we derive closed-form lower bounds for the ergodic achievable rate both for maximal-ratio combining and zero-forcing processing at the relay, by using minimum mean squared error channel estimation. We numerically characterize the accuracy of the lower bounds derived. We also
compare the SE and EE of the proposed design to those of the existing full-duplex systems and quantify the significant improvement achieved by the proposed algorithm. 
We also compare the EE of the proposed full-duplex system to that of its half-duplex counterparts, and characterize the self-loop and inter-user interference regimes, for which the proposed full-duplex system succeeds in outperforming the half-duplex ones.
\end{abstract}
%---------------------------------------------------------------
%---------------------------------------------------------------
\iftoggle{SINGLE_COL}{\vspace*{-0.25in}}{}
\begin{IEEEkeywords}\iftoggle{SINGLE_COL}{\vspace*{-0.15in}}{}
Energy efficiency, full-duplex, relay.
\end{IEEEkeywords}
%---------------------------------------------------------------
\IEEEpeerreviewmaketitle
\iftoggle{SINGLE_COL}{\vspace*{-0.35in}}{}
%---------------------------------------------------------------
\section{Introduction}\iftoggle{SINGLE_COL}{\vspace*{-0.1in}}{}
%---------------------------------------------------------------
Relay-based communication has been extensively investigated to expand the coverage, improve the diversity, increase the data rate, and reduce the power consumption of wireless communication systems \cite{Lee_Hanzo,ZhangCLVH15}. The current generation of relays is predominantly half-duplex, as a benefit of their implementational simplicity. A half-duplex relay requires two channel uses to send a data packet from the transmitter to the receiver, since the relay cannot transmit and receive with in the same time slot. Full-duplex technology is becoming  popular as a benefit of its increased throughput\cite{ZhangLVH16,ZhangCLVH15,fd_relay_si_can_wichman,fd_tut_ref_ashu,DBLP:journals/corr/NadhSSAG16}.
%,  demonstrated a significant reduction in the loop interference, caused due to transmission and reception on the same channel. 
A full-duplex one-way relay~\cite{fd_relay_si_can_young,fd_relay_si_can_wichman} transmits and receives at the same time, hence theoretically it doubles the spectral efficiency (SE) of a half-duplex one-way relay~\cite{ZhangMC11},~\cite{HammerstromW07}. 

%The full-duplex technology is being actively investigated to improve the SE of one-way and two-way communication.
%In full-duplex relaying \cite{DBLP:conf/spawc/AlvesCSL13,DBLP:conf/iswcs/ZhangTH13a}, a user sends one data unit to the destination user in one channel use via a relay. 

Full-duplex two-way relaying \cite{Zheng15,ChoiL14,DBLP:journals/twc/ZhangMDXK16}, wherein two users exchange their data over a single channel use via a relay, further improves the SE. Two-way full-duplex relaying has recently been extended to multi-pair two-way full-duplex relaying \cite{DBLP:journals/jsac/NgoSML14,DBLP:journals/jsac/ZhangCSX16,Zhang2016Chen} wherein multiple user pairs exchange their data via a shared relay in a single channel use. A multi-pair two-way full-duplex relay system has the following interference sources: i) co-channel (inter-pair) interference due to multiple users simultaneously accessing the channel; ii) self-loop interference at the relay and at the users; and iii) inter-user interference caused due to simultaneous transmission and~reception by full-duplex~nodes. 

%In literature, different signal processing  techniques exist to suppress inter-pair/multi-user interference \cite{DBLP:journals/tsp/JoungS10} and loop interference \cite{riihonen2009spatial,DBLP:journals/corr/NadhSSAG16}.

%Relay based transmission enables the wireless network to work in a cooperative manner \cite{DBLP:journals/tcom/Bhatnagar15a}.  Generally, amplify-and-forward (AF) or the decode-and-forward (DF) protocol is considered at the relay, where AF based relay amplifies and re-transmit the received signal and DF based relay re-encodes and re-transmit the received signal. 
 
Massive multiple-input multiple-output (MIMO) systems have become popular, since they cancel co-channel interference by using simple linear transmit processing schemes, such as, zero-forcing transmission (ZFT) and  maximal-ratio transmission (MRT)  \cite{DBLP:journals/twc/Marzetta10,DBLP:journals/tcom/NgoLM13,LuLSAZ14,DBLP:journals/spm/RusekPLLMET13}, which significantly improve their SE.
%\colb{why have you added the following text?}
%\colr{The power radiated by users could be made inversely proportional to the number of antennas or to the square-root of the number of antennas depending on whether channel state information is known or unknown, respectively \cite{DBLP:journals/tcom/NgoLM13}. Massive MIMO however relies on favourable propagation, where it is assumed that vector valued channels corresponding to different users grow asymptotically orthogonal as the number of antennas increases. On the limit of antennas, only slowly varying large-scale fading remains whereas the effect of small-scale fading, uncorrelated noise, and intra-cellular interference vanish. The only impairment which does not vanish is the inter-cell interference which is caused by the reuse of pilot symbols in other cells, commonly known as pilot contamination \cite{DBLP:journals/twc/Marzetta10}}.
%Reference~\cite{CuiSJ14} investigates thespectral efficieny and energy efficiency of one-way full-duplex , it reveals that the effect of small-scale fading and inter-pair interference averages out.
Massive MIMO technology is also being incorporated into multi-pair full-duplex relays for mitigating the self-loop interference at the relay, and the inter-pair co-channel interference \cite{DBLP:journals/jsac/NgoSML14,DBLP:journals/jsac/ZhangCSX16,Zhang2016Chen,DBLP:journals/twc/DaiD16,mm_relay_hong}. Ngo~\textit{et~al.}~\cite{DBLP:journals/jsac/NgoSML14} derived the achievable rate and a power allocation scheme for maximizing the ergodic sum-rate for one-way decode and forward full-duplex massive MIMO-aided relaying.  Zhang~\textit{et~al.} \cite{DBLP:journals/jsac/ZhangCSX16} proposed four power scaling schemes for two-way full-duplex massive MIMO relaying to improve both its SE and its energy efficiency (EE). Zhang~\textit{et~al.}~\cite{Zhang2016Chen} developed a power allocation scheme for maximizing the sum-rate of multi-pair two-way full-duplex massive MIMO amplify-and-forward (AF) relaying by using maximal-ratio combining (MRC)/MRT processing at the relay, and by using least squares (LS) channel estimation.
Dai~\textit{et~al.}\cite{DBLP:journals/twc/DaiD16} considered a half-duplex multi-pair two-way massive MIMO  AF relay and derived closed-form achievable rate expressions and a power allocation scheme for maximizing the sum-rate under realistic imperfect channel state information (CSI). Cui~\textit{et~al.}~\cite{mm_relay_hong} developed power scaling schemes for half-duplex massive MIMO-aided one-way relay systems.
%, shows that by decreasing the transmit power under massive relay antennas the loop interference can be reduced, in addition the inter-pair interference and inter-user interference can be eliminated with large number of antennas.

%Reference~\cite{DBLP:journals/jsac/NgoSML14}, \cite{DBLP:journals/twc/DaiD16} optimally allocates power at the relay to maximize ergodic sum-rate for FD decode and forward one-way relaying and half-duplex amplify and forward two-way relaying, respectively. 

%As the number of antenna increases, the ergodic rate and instantaneous rate tends to be same due to law of large numbers. In \cite{DBLP:journals/jsac/NgoSML14},  \cite{DBLP:journals/twc/DaiD16} the sum-rate is optimized subject to the power constraints at user and the relay. The quality-of-service (QoS) constraints which affect the user experience are not taken into account. The works in \cite{DBLP:journals/jsac/NgoSML14}, \cite{DBLP:journals/twc/DaiD16} either consider full-duplex decode and forward one-way relay or half-duplex amplify and forward two-way relay.

%In future generation mobile networks there is an increase in demand of network capacity by thousand fold giving rise to energy related pollution.

The energy cost of wireless network operation has increased to almost $50\%$ of the total operational cost \cite{DBLP:journals/cm/CorreiaZBFJGAP10}. The EE metric, which relies on the Pareto-optimality between throughput and energy consumption, has recently drawn attention as a {potent} performance measure. The EE has recently been optimized for both conventional MIMO~\cite{ee_tut_bhargava,ee_relay_owr_csi_zappone} and for single-hop massive MIMO systems~\cite{NguyenDNT17,LiFLL17}. Nguyen~\textit{et~al.}~\cite{NguyenDNT17} optimized the downlink EE of a cell-free single-hop massive MIMO system using ZF precoding. Li~\textit{et~al.}~\cite{LiFLL17} optimized both the achievable rate and the EE of a small-cell based massive MIMO single-hop full-duplex system. It is anticipated that a paradigm shift towards multi-component Pareto-optimization is about to take place, leading to an entire optimal Pareto-front of solutions \cite{FeiLYXCH17,AlanisBBNH15}.
%This work also proposes power allocation algorithms to maximize EE andspectral efficieny.

%Liu~\textit{et~al.} in \cite{liu2016energy} investigated the trade-off between the EE andspectral efficieny for single-hop massive MIMO systems with linear precoding and transmit antenna selection. Reference \cite{LiuTGXYX17} derives the EE of the single-hop uplink massive MIMO system when non-linear successive-interference cancellation receivers are employed at the base station. 

The existing literature of massive MIMO relays, on the other hand, has either optimized the achievable rate \cite{DBLP:journals/jsac/NgoSML14,Zhang2016Chen,DBLP:journals/twc/DaiD16} or analyzed the EE \cite{DBLP:journals/jsac/ZhangCSX16,mm_relay_hong}. The EE optimization for massive MIMO relaying has not been investigated at the time of writing, except for a recent study in \cite{Tan2017Lv} which optimized the \textit{asymptotic} EE for a multi-pair \textit{one-way decode and forward} massive MIMO \textit{half-duplex} relay. To the best of our knowledge, the EE of multi-pair \textit{two-way AF} massive MIMO \textit{full-duplex} relay system has not been considered in the literature for a realistic finite-cardinality antenna-set. Hence we fill this gap. Due to the self-loop interference and the coupled channels {encountered} in \textit{two-way full-duplex AF relaying}, the power allocation scheme of \cite{Tan2017Lv} cannot be applied to our system. Against this {backdrop}, we list the \textbf{main contributions} of~this~paper. %\st{Further the work in} \cite{Tan2017Lv} \st{does not consider quality-of-service (QoS) constraints, which are important in practical systems.} 

%We consider practical scenario, in which channel estimates are not perfect and has to be estimated at the relay using pilots and also there exist inter-user interference between different users which are nearby each other.  
%Energy efficiency is  defined by fractional functions, and a key-role in the  optimization of energy efficiency is played by fractional programming. The resultant problem is generally a non-convex FP problem.

1) We derive closed-form lower bounds for the achievable rate of the multi-pair two-way AF full-duplex massive MIMO relay for an arbitrary number of relay antennas. We consider both MRC/MRT and zero-forcing reception (ZFR)/ZFT processing at the relay, whilst relying on the minimum mean square error (MMSE) relay channel estimation. In contrast to~\cite{Zhang2016Chen}, which derives a closed-form lower bound for MRC/MRT processing alone, which is based on the LS channel estimation, we derive new bounds both for MRC/MRT and for ZFR/ZFT processing based on the MMSE channel estimation. These closed-form achievable-rate expressions have not been derived for an arbitrary number of relay antennas in the massive MIMO~relaying~literature. \newline 
%\colr{\st{Through simulation results we have shown that MMSE estimator considered in this work performs superior than LS estimator discussed in} \cite{Zhang2016Chen}.}
2) We optimally allocate power to maximize the EE  by using the closed-form achievable rate expressions derived. The EE maximization, which has a  non-convex objective, is solved by proposing an algorithm wherein we first approximate the objective as a pseudo-convex function, and later {choose} Dinkelbach's approach. This contribution is significantly different from \cite{Tan2017Lv}, which considers asymptotic EE optimization for decode and forward one-way half-duplex relay. The achievable rate expressions, and consequently the EE optimization, developed herein are applicable to any antenna configuration. {Furthermore}, the expressions and the analysis developed herein are significantly more complex due to the coupling of channels in AF relaying, and both the self-loop as well as the inter-user interference imposed by the full-duplex nodes. We numerically compare the performance of the proposed EE algorithm to the equal-power approach of \cite{DBLP:journals/jsac/ZhangCSX16}. Furthermore, we show that the EE optimization algorithm can also be used for optimizing the EE under specific Quality-of-Service (QoS) constraints. We investigate the effect of QoS constraints on the EE performance.
\newline%Finally, an algorithm is proposed to optimally allocate the power in order to maximize EE, based on  \cite{dinkelbach1967nonlinear}.\newline
3) We also maximize the EE by using the max-min fairness criterion; the problem has a non-differentiable objective. We solve this problem by first using the sequential convex programming approach to approximate the objective function by a quasi-concave function, and later by using the generalized Dinkelbach's method of \cite{crouzeix1991algorithms}. \newline %An optimal power allocation algorithm is proposed based on generalized Dinkelbach's procedure \cite{crouzeix1991algorithms}.\newline
%4) The current work \cite{Tan2017Lv} 
%\newline
%4) We also maximize the SE for MRC/MRT and ZFR/ZFT processing with MMSE channel estimation. We note that the SE of the system considered herein with  MMSE channel estimation and MRC and ZF processing has not been studied before. We show that for small number of relay antennas, the MRC, with optimal power allocation, performs better that equal-power ZF processing.
4) The proposed EE maximization framework can also maximize the SE for both MRC/MRT and ZFR/ZFT processing. In contrast to~\cite{Zhang2016Chen}, which derives the lower bound and maximizes the SE (not EE) for the MRC/MRT  processing alone relying on LS channel estimation, we {conceive} a more general approach than that of~\cite{Zhang2016Chen}. We quantify the considerably improved SE of ZFR/ZFT processing and MMSE channel estimation over the LS-based MRC/MRT processing of \cite{Zhang2016Chen}.\newline
%5) We compare the optimized SE of the full-duplex system derived in this work with the optimized SE of half-duplex system derived in \cite{DBLP:journals/twc/DaiD16}. We numerically determine the loop and inter-user interference values for which the MMSE-channel estimation based full-duplex system has better SE than a half-duplex system in \cite{DBLP:journals/twc/DaiD16}.  \newline
5) The proposed full-duplex EE optimization framework can also be used for evaluating the EE of massive MIMO half-duplex AF systems which has not been investigated in the open literature. We compare the EE of both the full-duplex and of the half-duplex systems and numerically quantify the self-loop and
inter-user interference values for which a full-duplex system has a better EE than a half-duplex system. We also show  the significantly improved EE of optimal power allocation over the equal-power EE analyzed in \cite{mm_relay_hong}.

The rest of the paper is organized as follows.  We present our system model in Section~\ref{sys_model}, and discuss the MMSE channel estimation in Section~\ref{ch_est_ref}. The relay processing is discussed in Section~\ref{relay_pre_des}, while the achievable rates are analyzed in Section~\ref{rate_ana_ref}. {Our} energy-efficient optimization problems are formulated in Section~\ref{PA}, while {our} performance improvements over \cite{DBLP:journals/jsac/ZhangCSX16,mm_relay_hong,Zhang2016Chen} are presented in Section~\ref{simu_sec_ref}. The paper is concluded in Section~\ref{conclude_ref}.

\textit{Notations}: The boldface capital and small letter represents matrix and vector, respectively, $\mathbb{C}^{r\times s}$ denotes a complex matrix of dimension $r \times s$. The superscript $(\cdot)^{T}$, $(\cdot)^{H}$, $(\cdot)^{*}$ denotes the transpose, Hermitian and conjugate operations, respectively. The $\mbox{diag}(\mathbf{x})$ denotes a (square) diagonal matrix with elements $\mathbf{x}$ on its main diagonal, $\mathbf{I}_{Q}$ denotes an $Q  \times Q$ identity matrix, and $\mathbf{1}_{k}$ denotes a $2K\times 1$ vector consisting of value one at $k$th row and zero otherwise. The expectation and trace operations are denoted by $\mathbb{E}[\cdot]$ and $\mbox{tr}\{\cdot\}$, respectively. The notation $\mathcal{CN}(\mathbf{0},\mathbf{K)}$ represents a circularly-symmetric complex Gaussian random vector with covariance matrix $\mathbf{K}$.
\iftoggle{SINGLE_COL}{\vspace*{-25pt}}{}
%---------------------------------------------------------------
\section{System Model}
\label{sys_model}
\begin{figure}[!htb]
	\centering
	\includegraphics[scale=\iftoggle{SINGLE_COL}{0.51}{0.49}]{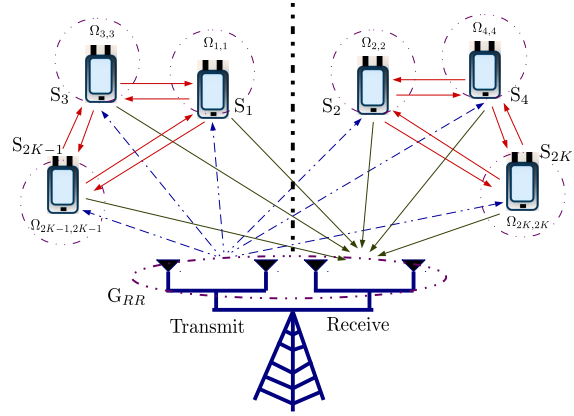}\iftoggle{SINGLE_COL}{\vspace*{-5pt}}{}
	\caption{Multi-pair two-way full-duplex AF massive MIMO relay system: All users and the relay simultaneously transmit and receive which leads to self-loop interference at the relay $\mathbf{G}_{RR}$ (shown by dotted ellipse), at the user $\Omega_{k,k}$ (shown by dotted circle) and inter-user interference (shown by solid red line). The users on either side of the relay (marked with bold dotted line at the center) are isolated.}
	\label{system_model}	
\end{figure}

\iftoggle{SINGLE_COL}{\vspace*{-5pt}}{}
%---------------------------------------------------------------
We consider the multi-pair two-way AF full-duplex relaying shown in Fig.~\ref{system_model}, where $K$ full-duplex user pairs communicate via a single full-duplex relay within the same time-frequency resource. We assume that the user $S_{2m-1}$ for $m=1,\cdots,K$ on one side of the relay, wants to send as well as receive from the user $S_{2m}$ on the other side of the relay. We also assume that there are no direct links between the user-pairs $(S_{2m-1},S_{2m})$ due to {a high} path loss and heavy shadowing, which is commonly assumed in the multi-user two-way massive MIMO relaying literature \cite{DBLP:journals/jsac/NgoSML14,DBLP:journals/jsac/ZhangCSX16}. Furthermore, the relay has $N$ transmit and $N$ receive antennas, while each user has one transmit and one receive antenna. The users on either side of the relay interfere with each other, due to full-duplex architecture; the interference caused is termed as inter-user~interference.

%---------------------------------------------------------------
\iftoggle{SINGLE_COL}{\vspace*{-12pt}}{}
%---------------------------------------------------------------
%The channels between the users and the relay are independent and identically distributed (i.i.d) with Rayleigh distribution. 

At time instant $n$, each user $S_{k}$, $k =1$ to $2K$,\footnote{To avoid repetition, we assume that $k =1$ to $2K$ throughout this paper.} transmits the signal $\sqrt{p_{k}}x_{k}(n)$ to the relay, and simultaneously the relay broadcasts a vector $\mathbf{x}_{R}(n)\in\mathbb{C}^{N\times 1}$ to all the users. Here the term $p_k$ denotes the transmit power of the $k$th user. The signal received at the relay and at the user $S_{k}$ are given by\iftoggle{SINGLE_COL}{\vspace*{-5pt}}{}
%---------------------------------------------------------------
\begin{align}
\mathbf{y}_{R}(n) &= \sum\limits_{k=1}^{2K}\sqrt{p_{k}} \mathbf{g}_{k}x_{k}(n) +\mathbf{G_{RR}}{\mathbf{x}}_{R}(n)+\mathbf{z}_{R}(n)\label{yrn}\iftoggle{SINGLE_COL}{}{\nonumber\\
&}= \tilde{\mathbf{G}}\mathbf{x}(n) +\mathbf{G_{RR}}{\mathbf{x}}_{R}(n)+\mathbf{z}_{R}(n),\iftoggle{SINGLE_COL}{\\[-6pt]}{\\}
y_{k}(n)&= \mathbf{f}_{k}^{T}\mathbf{x}_{R}(n)+\sum_{i,k\in U_{k}}\Omega_{k,i}\sqrt{p_{i}}x_{i}(n)+z_{k}(n)\label{ykn}.
\end{align}
%---------------------------------------------------------------
Here $\mathbf{g}_k \in \mathbb{C}^{N\times 1}$ and $\mathbf{f}_k \in \mathbb{C}^{N\times 1}$ denote the channels {spanning} from the transmit antenna of the $k$th user to the relay's receive antenna array, and from the relay's  transmit antenna array to the receive antenna of the $k$th user, respectively. We now introduce the matrix $\mathbf{G} = \left[\mathbf{g}_{1},\,\mathbf{g}_{2},\,\mathbf{g}_{3},\,\cdots,\, \mathbf{g}_{2K}\right]\in\mathbb{C}^{N\times 2K}$ and the matrix $\mathbf{F}=\left[\mathbf{f}_{1},\,\mathbf{f}_{2},\,\mathbf{f}_{3},\,\cdots,\, \mathbf{f}_{2K}\right]\in\mathbb{C}^{N\times 2K}$ (to be used later in the sequel). Furthermore we have, $\tilde{\mathbf{G}} = \mathbf{G}\mathbf{P}$ where $\mathbf{P}=\mbox{diag}\left\{\sqrt{p_{1}},\,\sqrt{p_{2}},\cdots\,\sqrt{p_{2K}}\right\}$ with $0\leq p_{k}\leq P^{\max}$. The signal received at the relay and at the users are interfered by their own transmit signal, which is {termed} as the self-loop interference. In \eqref{yrn} and \eqref{ykn}, $\mathbf{G}_{RR}$ and ${\Omega}_{k,k}$ denote the self-loop interference at the relay and at the user $S_{k}$, respectively. The entries of the matrix  $\mathbf{G}_{RR}$ are independent and identically distributed (i.i.d.) with distribution $\mathcal{C}\mathcal{N}(0,\sigma_{LIR}^{2})$ \cite{DBLP:journals/jsac/ZhangCSX16}. 
Furthermore ${\Omega}_{k,k}$ is distributed as $\mathcal{C}\mathcal{N}(0,\sigma_{k,k}^2)$. The terms $\Omega_{k,i}\, (k,i\in U_{k},i\neq k)$, which denotes the inter-user interference channel, are independent and are distributed as  $\mathcal{C}\mathcal{N}(0,\sigma_{k,i}^{2})$ \cite{DBLP:journals/jsac/ZhangCSX16}, where the set obeys $U_{k}=\left[1,3,5,\cdots,2K-1\right]$ for odd $k$ and $U_{k}=\left[2,4,6,\cdots,2K\right]$ for even $k$. We also define the vector $\mathbf{x}(n)= \left[x_{1}(n),\,x_{2}(n),\,x_{3}(n),\,\cdots,\,x_{2K}(n)\right]^{T}\in\mathbb{C}^{N\times 1}$ with $\mathbb{E}\left[\mathbf{x}(n)\mathbf{x}^{H}(n)\right]= \mathbf{I}_{2K}$. The vector $\mathbf{z}_{R}(n)\in\mathbb{C}^{N\times 1}$ and the scalar ${z}_{k}(n)$ are additive white Gaussian noise (AWGN) processes at the relay and at the user $S_{k}$. The elements of $\mathbf{z}_{R}$ and the scalar ${z}_{k}(n)$ are modeled  as i.i.d. $\mathcal{C}\mathcal{N}(0,\sigma_{nr}^2)$ and $\mathcal{C}\mathcal{N}(0,\sigma_{n}^2)$, respectively. %The relay has maximum transmit power $P_r$ and its transmit signal is scaled to satisfy its power constraint  $\mbox{Tr}\left\{\mathbb{E}\left[\mathbf{x}_{R}(n)\mathbf{x}_{R}^{H}(n)\right]\right\}= P_R$.
\iftoggle{SINGLE_COL}{\vspace*{-0.10in}}{}
\begin{remark}
The channel matrices account for both small-scale and large-scale fading; we therefore have $\mathbf{G} = \mathbf{H}_{u}\mathbf{D}_{u}^{1/2}$ and $\mathbf{F} = \mathbf{H}_{d}\mathbf{D}_{d}^{1/2}$. Here the small-scale fading matrices $\mathbf{H}_{u}$ and $\mathbf{H}_{d}$ have i.i.d. $\mathcal{C}\mathcal{N}(0,1)$ elements, while the $k$th element of the large-scale diagonal fading matrices $\mathbf{D}_{u}$ and $\mathbf{D}_{d}$ are denoted as $\sigma_{g,k}^{2}$ and $\sigma_{f,k}^{2}$, respectively.
%The inter-element distance in transmit/receive arrays is assumed to be smaller than the distance between the two arrays which leads to the channel between the transmit and receive antennas to be independent.
\end{remark}
\iftoggle{SINGLE_COL}{\vspace*{-10pt}}{}
In the first time slot $(n=1)$, the relay only receives the signal from the users, and does not transmit. The signals received at the relay and at the user $S_{k}$ are given respectively as 
\begin{eqnarray}
\mathbf{y}_{R}(1) &=& \tilde{\mathbf{G}}\mathbf{x}(1) +\mathbf{z}_{R}(1)\iftoggle{SINGLE_COL}{}{\nonumber\\}\iftoggle{BIG_EQUATION}{}{,\quad}
y_{k}(1)\iftoggle{SINGLE_COL}{=}{&=&} \sum_{i,k\in U_{k}}\Omega_{k,i}\sqrt{p_{i}}x_{i}(1)+z_{k}(1).
\end{eqnarray}
At the $n$th time slot, the relay linearly precodes its received signal $\mathbf{y}_{R}(n-1)$ using a matrix $\mathbf{W}$, such that
\iftoggle{SINGLE_COL}{\vspace*{-0.25in}}{}\begin{align}\label{xrn}
\mathbf{x}_{R}(n) &= \alpha \mathbf{W}\mathbf{y}_{R}(n-1),
\end{align}
where $\alpha$ is the scaling factor chosen to satisfy the relay's power constraint. The relay's transmit signal $\mathbf{x}_{R}(n)$, before applying any self-loop interference cancellation technique, can be {re-formulated} by iteratively substituting (\ref{yrn}) into \eqref{xrn} as
\begin{align}
\label{inf_memory}
\hspace{-0.1in}\mathbf{x}_{R}(n) &= s\left[\mathbf{x}(n-\nu)+\mathbf{x}(n-2\nu)+\cdots\iftoggle{SINGLE_COL}{}{\right.\nonumber\\&\left.\,\,\,}
+\,\mathbf{z}_{R}(n-\nu)+\mathbf{z}_{R}(n-2\nu)+\cdots \right].
\end{align}
Here $s[\cdot]$, as discussed in \cite{DBLP:journals/jsac/ZhangCSX16}, is a function involving both vector and matrix operations, while $\nu$ is the relay's processing  delay ($\nu = 1$ in this paper). The relay transmit signal, in the above form which assumes no interference cancellation, is difficult to analyze \cite{DBLP:journals/jsac/ZhangCSX16}. 
%The loop interference channel between two antennas that are closed to each other is modelled as Rician distribution with large loop $K$ factor. However, after using loop interference mitigation techniques the line of sight component becomes weak and residual loop interference (RLI) is modelled as Rician distribution with small $K$ factor. In case of MIMO, the RLI is modelled as Rayleigh fading.
The recent full-duplex studies e.g. in \cite{DBLP:journals/jsac/ZhangCSX16,DBLP:journals/corr/NadhSSAG16,fd_relay_si_can_wichman} have shown that the self-loop interference can be significantly suppressed, and the residual self-loop interference at the relay can be modeled as an additional Gaussian noise source. We also apply these cancellation techniques, and replace $\mathbf{x}_{R}(n)$ in the self-loop interference term $\mathbf{G}_{RR}\mathbf{x}_{R}(n)$ in \eqref{yrn} with a Gaussian noise $\tilde{\mathbf{x}}_{R}(n)$, which represents the residual self-loop interference with the power constraint $\left\{\mathbb{E}\left[\tilde{\mathbf{x}}_{R}(n)\tilde{\mathbf{x}}_{R}^{H}(n)\right]\right\} =\frac{P_{R}}{N}\mathbf{I}_N$~\cite{DBLP:journals/jsac/ZhangCSX16}. We~therefore~have
%There are various solution proposed in the relaying literature~\cite{riihonen2009spatial}, that significantly reduce the self-loop interference caused by $\mathbf{x}_{R}$, so that the residual self-loop interference can be replaced by $\tilde{\mathbf{x}}_{R}(n)$, modelled as an additional Gaussian noise source associated with $\left\{\mathbb{E}\left[\tilde{\mathbf{x}}_{R}(n)\tilde{\mathbf{x}}_{R}^{H}(n)\right]\right\} =\frac{P_{R}}{N}\mathbf{I}_N$ \cite{DBLP:journals/jsac/ZhangCSX16}.
%Previous researches in FD single-input single-output (SISO) relay scenario model the RLI as additional relay input Gaussian noise. Similarly, in FD MIMO relay scenario, after some ingenious loop interference mitigation schemes (such as time domain or spatial suppression or both of them), loop interference can be significantly degrdaded and the RLI is so weak that it can also be regarded as additional noise. 
%Hence in this paper, we replace $\mathbf{x}_{R}$ in the loop interference term $\mathbf{G}_{RR}\mathbf{x}_{R}(n)$ in \eqref{} by a Gaussian noise source $\tilde{\mathbf{x}}_{R}$ with the same power limitation to represent the RLI signal.} Therefore, the relay's received signal in \eqref{yrn} can be re-formulated~as
\iftoggle{SINGLE_COL}{\vspace*{-0.15in}}{}
\begin{align}\label{modyr}
\tilde{\mathbf{y}}_{R}(n) &= \tilde{\mathbf{G}}\mathbf{x}(n) +\mathbf{G_{RR}}\tilde{\mathbf{x}}_{R}(n)+\mathbf{z}_{R}(n).
\end{align}
After loop interference suppression, the relay's transmit signal in (\ref{xrn}) is re-expressed using (\ref{modyr})~as
\iftoggle{SINGLE_COL}{\vspace*{-0.35in}}{}
\begin{align}\label{modixrn1}
\mathbf{x}_{R} (n)&= \alpha \mathbf{W}\tilde{\mathbf{y}}_{R}(n-1).
\end{align}
We observe from \eqref{modixrn1} that $\mathbf{x}_{R} (n)$ now does not have infinite memory of $\mathbf{x}(n)$, which was the case earlier in \eqref{inf_memory}. It  is a function of $\mathbf{x}(n-1)$. The time labels from the model in \eqref{modixrn1} can now be dropped for the sake of brevity. We re-write (\ref{modixrn1}), using (\ref{modyr}), after dropping the labels as
\begin{align}\label{modixrn2}
\mathbf{x}_{R} &= \alpha \mathbf{W}\tilde{\mathbf{y}}_{R}= \alpha \mathbf{W}\tilde{\mathbf{G}}\mathbf{x} + \alpha \mathbf{W} \mathbf{G_{RR}}\tilde{\mathbf{x}}_{R} + \alpha \mathbf{W}\mathbf{z}_{R}.
\end{align}
The relay's transmit signal should satisfy its transmit power constraint, so we have
\begin{align}\label{exppr}
{P}_{R} &= \mbox{Tr}\left\{\mathbb{E}\left[\mathbf{x}_{R}\mathbf{x}^{H}_{R}\right]\right\}\iftoggle{SINGLE_COL}{}{\nonumber\\
&\hspace{-0.25in}}= \mathbb{E}\left[\|\alpha \mathbf{W}\tilde{\mathbf{G}}\mathbf{x}\|^{2}\right]+\mathbb{E}\left[\|\alpha \mathbf{W} \mathbf{G_{RR}}\tilde{\mathbf{x}}_{R}\|^{2}\right]+\mathbb{E}\left[\|\alpha \mathbf{W}\mathbf{z}_{R}\|^{2}\right],
\end{align}
which leads to \iftoggle{SINGLE_COL}{\vspace*{-0.05in}}{}
\begin{equation}\label{alpha}
\alpha = \sqrt{\frac{P_{R}}{\mathbb{E}\left[\|\mathbf{W}\tilde{\mathbf{G}}\mathbf{x}\|^{2}\right]+\mathbb{E}\left[\| \mathbf{W} \mathbf{G_{RR}}\tilde{\mathbf{x}}_{R}\|^{2}\right]+\mathbb{E}\left[\| \mathbf{W}\mathbf{z}_{R}\|^{2}\right]}}.
\end{equation}
We next re-formulate the signal received at the user $S_{k}$ in (\ref{ykn}), by using (\ref{modixrn2}), as
\begin{align}
y_{k} %&=& \mathbf{f}_{k}^{T}\mathbf{x}_{R} + \sum\limits_{i,k\in U{k}}\Omega_{k,i}\sqrt{p_{k}}x_{i} + z_{k} \nonumber\\
&= \alpha \mathbf{f}_{k}^{T}\mathbf{W}\tilde{\mathbf{G}}\mathbf{x} + \alpha \mathbf{f}_{k}^{T}\mathbf{W} \mathbf{G_{RR}}\tilde{\mathbf{x}}_{R} + \alpha \mathbf{f}_{k}^{T}\mathbf{W}\mathbf{z}_{R}\iftoggle{SINGLE_COL}{}{\nonumber\\
&\quad}+\sum\limits_{i,k\in U{k}}\Omega_{k,i}\sqrt{p_{i}}x_{i} + z_{k} \nonumber\iftoggle{SINGLE_COL}{\\[-6pt]}{\\}
&= \underbrace{\alpha \mathbf{f}_{k}^{T}\mathbf{W}\sqrt{p_{k^{'}}}{\mathbf{g}_{k^{'}}}x_{k^{'}}}_{\text{desired signal}} + \underbrace{\alpha \mathbf{f}_{k}^{T}\mathbf{W}\sqrt{p_{k}}{\mathbf{g}_{k}}x_{k}}_{\text{self-interference}}\iftoggle{SINGLE_COL}{}{\nonumber\\
&}+ \underbrace{\alpha \mathbf{f}_{k}^{T}\mathbf{W}\sum\limits_{i\neq k,k^{'}}^{2K}\sqrt{p_{i}}\mathbf{g}_{i}x_{i}}_{\text{inter-pair interference}}+\underbrace{\alpha \mathbf{f}_{k}^{T}\mathbf{W} \mathbf{G_{RR}}\tilde{\mathbf{x}}_{R}}_{\text{amplified self-loop interference}}\nonumber\iftoggle{SINGLE_COL}{\\[-8pt]}{\\}
& + \underbrace{\alpha \mathbf{f}_{k}^{T}\mathbf{W}\mathbf{z}_{R}}_{\text{amplified noise from relay}}+\underbrace{\sum\limits_{i,k\in U{k}}\Omega_{k,i}\sqrt{p_{i}}x_{i}}_{\substack{\text{self-loop interference}\\ \text{and inter-user interference}}} + \underbrace{z_{k}}_{\text{AWGN at $S_{k}$}}.\iftoggle{SINGLE_COL}{}{\nonumber\\}
\end{align}
Here we have $(k,k^{'})=(2m-1,2m)$ or $(2m,2m-1)$, where $m=1,\cdots,K$ denotes the user pair, which exchange information with one another.

In this work we assume that the relay estimates the channels $\mathbf{G}$ and $\mathbf{F}$ and uses them to design the precoder $\mathbf{W}$. The relay then transmits the self-interference cancellation (SIC) coefficient $\hat{\mathbf{f}}_{k}^T\mathbf{W}\hat{\mathbf{g}}_{k}$ for each user, where $\hat{\mathbf{f}}_{k}$ and $\hat{\mathbf{g}}_{k}$ are the estimated channel coefficients. The signal received at user $S_{k}$ after SIC is\iftoggle{SINGLE_COL}{\vspace*{-5pt}}{}
\begin{align}\label{yktilde}
\tilde{y}_{k} &= \underbrace{\alpha \mathbf{f}_{k}^{T}\mathbf{W}\sqrt{p_{k^{'}}}{\mathbf{g}_{k^{'}}}x_{k^{'}}}_{\text{desired signal}} +\underbrace{\alpha\sqrt{p_{k}}\lambda_{k}x_{k}}_{\text{residual interference}}\iftoggle{SINGLE_COL}{}{\nonumber\\
&}+ \underbrace{\alpha \mathbf{f}_{k}^{T}\mathbf{W}\sum\limits_{i\neq k,k^{'}}^{2K}\sqrt{p_{i}}\mathbf{g}_{i}x_{i}}_{\text{inter-pair interference}}+\underbrace{\alpha \mathbf{f}_{k}^{T}\mathbf{W} \mathbf{G_{RR}}\tilde{\mathbf{x}}_{R}}_{\text{amplified self-loop interference}}\nonumber\iftoggle{SINGLE_COL}{\\[-8pt]}{\\}
&+ \underbrace{\alpha \mathbf{f}_{k}^{T}\mathbf{W}\mathbf{z}_{R}}_{\text{amplified noise from relay}}+\underbrace{\sum\limits_{i,k\in U{k}}\Omega_{k,i}\sqrt{p_{i}}x_{i}}_{\substack{\text{self-loop interference}\\ \text{and inter-user interference}}} + \underbrace{z_{k}}_{\text{AWGN at $S_{k}$}}.\iftoggle{SINGLE_COL}{}{\nonumber\\}
\end{align}
Here $\lambda_{k} = {\mathbf{f}}_{k}^{T}\mathbf{W}{{\mathbf{g}_{k}}}-\hat{\mathbf{f}}_{k}^{T}\mathbf{W}{\hat{\mathbf{g}_{k}}}$ is the residual self-interference. Before designing the relay's precoder~$\mathbf{W}$, we briefly digress to discuss the MMSE channel estimation process.

\iftoggle{SINGLE_COL}{\vspace*{-0.25in}}{}
\section{Channel Estimation}\label{ch_est_ref}
\iftoggle{SINGLE_COL}{\vspace*{-0.10in}}{}
During the channel estimation phase, all $2K$ users simultaneously transmit pilot sequences of length $\tau\leq T$ symbols to the relay, where $T$ is the channel's coherence interval. Let $\sqrt{\tau P_{\rho}}\boldsymbol{\varphi}\in\mathbb{C}^{2K\times \tau}$ denote the pilot symbols transmitted from $2K$ users with $P_{\rho}$ being the transmit power of each pilot symbol. The pilots are assumed to be orthogonal so that $\boldsymbol{\varphi}\boldsymbol{\varphi}^{H}= \mathbf{I}_{2K}$, which requires that $\tau \geq 2K$\cite{Biguesh}. 
%During the channel estimation phase, using the procedure similar to \cite{DBLP:journals/jsac/NgoSML14}, the relay performs MMSE channel estimation of $\mathbf{G}$ and $\mathbf{F}$ \cite{kay1993fundamentals}. 
The MMSE channel estimates $\hat{\mathbf{G}}$ and $\hat{\mathbf{F}}$ are given as \cite{DBLP:journals/jsac/NgoSML14,kay1993fundamentals}
\begin{eqnarray}\label{est_ch_ref}
\mathbf{G}=\hat{\mathbf{G}}+\mathbf{E}_{g}, \text{   and   } \mathbf{F} = \hat{\mathbf{F}}+\mathbf{E}_{f}.
\end{eqnarray}
Here $\mathbf{E}_{g}$ and $\mathbf{E}_{f}$ represents the estimation error matrices of $\mathbf{G}$ and $\mathbf{F}$, respectively. The estimated channel matrices $\hat{\mathbf{G}}$ and $\hat{\mathbf{F}}$ are independent of the error matrices ${\mathbf{E}_{g}}$ and ${\mathbf{E}_{f}}$, respectively \cite{kay1993fundamentals}. The rows of the matrices $\hat{\mathbf{G}}$ and $\hat{\mathbf{F}}$ are distributed as $\mathcal{C}\mathcal{N}(0,\hat{\mathbf{D}}_{u})$ and $\mathcal{C}\mathcal{N}(0,\hat{\mathbf{D}}_{d})$ respectively, where $\hat{\mathbf{D}}_{u}=\mbox{diag}\left\{\hat\sigma_{g,1}^{2},\,\hat\sigma_{g,2}^{2},\,\cdots,\,\hat\sigma_{g,2K}^{2}\right\}$ and $\hat{\mathbf{D}}_{d}=\mbox{diag}\left\{\hat\sigma_{f,1}^{2},\,\hat\sigma_{f,2}^{2},\,\cdots,\,\hat\sigma_{f,2K}^{2}\right\}$, with $\hat\sigma_{g,k}^{2}=\frac{\tau P_{\rho}\sigma_{g,k}^{4}}{\tau P_{\rho}\sigma_{g,k}^{2}+1}$ and $\hat\sigma_{f,k}^{2}=\frac{\tau P_{\rho}\sigma_{f,k}^{4}}{\tau P_{\rho}\sigma_{f,k}^{2}+1}$ {\cite{kay1993fundamentals}}. Hence ${\mathbf{E}_{g}}\sim \mathcal{C}\mathcal{N}(0,\mathbf{D}_{u}-\hat{\mathbf{D}}_{u})$ and ${\mathbf{E}_{f}}\sim \mathcal{C}\mathcal{N}(0,\mathbf{D}_{d}-\hat{\mathbf{D}}_{d})$, with $\mathbf{D}_{u}-\hat{\mathbf{D}}_{u} = \mbox{diag}\left\{\sigma_{\xi,g,1}^{2},\,\sigma_{\xi,g,2}^{2},\,\sigma_{\xi,g,3}^{2},\,\cdots,\,\sigma_{\xi,g,2K}^{2}\right\}$ and $\mathbf{D}_{d}-\hat{\mathbf{D}}_{d} = \mbox{diag}\left\{\sigma_{\xi,f,1}^{2},\,\sigma_{\xi,f,2}^{2},\,\sigma_{\xi,f,3}^{2},\,\cdots,\,\sigma_{\xi,f,2K}^{2}\right\}$ with  $\sigma_{\xi,g,k}^{2}=\frac{\sigma_{g,k}^{2}}{\tau P_{\rho}\sigma_{g,k}^{2}+1}$ and $\sigma_{\xi,f,k}^{2}=\frac{\sigma_{f,k}^{2}}{\tau P_{\rho}\sigma_{f,k}^{2}+1}$.

\iftoggle{SINGLE_COL}{\vspace*{-0.20in}}{}
\section{Relay precoder design}
\label{relay_pre_des}\iftoggle{SINGLE_COL}{\vspace*{-0.10in}}{}
We design our relay precoder based on: i) MRC/MRT; and ii) ZFR/ZFT processing. \iftoggle{SINGLE_COL}{\vspace*{-0.3in}}{}
\subsection{MRC/MRT processing}\iftoggle{SINGLE_COL}{\vspace*{-0.10in}}{}
The MRC/MRT precoder using the estimated CSI is formulated as
\begin{eqnarray}\label{wmr}
\mathbf{W} &=& \hat{\mathbf{F}}^{*}\mathbf{T}\hat{\mathbf{G}}^{H},
\end{eqnarray}
where $\mathbf{T} = \mbox{blkdiag}\left\{\mathbf{T}_{1},\,\mathbf{T}_{2},\,\mathbf{T}_{3},\cdots,\, \mathbf{T}_{K}\right\}$. The permutation matrix $\mathbf{T}_{m} = \left[0\, 1;\, 1\, 0\right]$ permutes the transmit data of each user pair to ensure that the transmit data reaches the intended receiver. We next state the following proposition, whose proof is relegated to Appendix \ref{alpha_mrc_append}, to simplify the scaling factor $\alpha$ in \eqref{alpha}.\iftoggle{SINGLE_COL}{\vspace{-0.10in}}{}
\begin{Proposition}\label{pre1}
For the MRC/MRT precoder, we have
\begin{eqnarray}\label{alphamrproof}
\alpha &=& \sqrt{\frac{P_{R}}{N^{2}\left({\Psi}+\sigma_{nr}^{2}+P_{R}\sigma_{LIR}^{2}\right)\hat{\Phi}+ N^{3}\hat{\Upsilon}
}},
\end{eqnarray}
where
$\Psi= \sum\limits_{i=1}^{K}\left(p_{2i-1}\sigma_{g,2i-1}^{2}+p_{2i}\sigma_{g,2i}^{2}\right),\hat{\Phi} = \sum\limits_{i=1}^{K}\left(\hat\sigma_{g,2i-1}^{2}\hat\sigma_{f,2i}^{2}+\hat\sigma_{g,2i}^{2}\hat\sigma_{f,2i-1}^{2}\right) \text{ and }\iftoggle{SINGLE_COL}{$
\\$}{}\hat{\Upsilon} = \sum\limits_{i=1}^{K}\left(p_{2i-1}\hat\sigma_{g,2i-1}^{4}\hat\sigma_{f,2i}^{2}+p_{2i}\hat\sigma_{g,2i}^{4}\hat\sigma_{f,2i-1}^{2}\right).$
\end{Proposition}\iftoggle{SINGLE_COL}{\vspace{-0.05in}}{}
%\begin{proof}
%Refer to Appendix \ref{alpha_mrc_append}.
%\end{proof}
\iftoggle{SINGLE_COL}{}{
\begin{figure*}
\normalsize
% Store the current equation number.
\setcounter{mytempeqncnt}{\value{equation}}
% Set the equation number to one less than the one
% desired for the first equation here.
% The value here will have to changed if equations
% are added or removed prior to the place these
% equations are referenced in the main text.
\setcounter{equation}{37}
\begin{align}
\mbox{SNR}_{k} &= \frac{p_{k^{'}}|{\mathbf{f}}_{k}^{T}\mathbf{W}{{\mathbf{g}_{k^{'}}}}|}{p_{k}|\lambda_{k}|^{2}+\sum\limits_{i\neq k,k^{'}}^{2K}|{\mathbf{f}}_{k}^{T}\mathbf{W}{{\mathbf{g}_{i}}}|^{2}+\|{\mathbf{f}}_{k}^{T}\mathbf{W}\mathbf{G}_{RR}\|^{2}\frac{P_{R}}{N}+\|{\mathbf{f}}_{k}^{T}\mathbf{W}\|^{2}\sigma_{n_{R}}^{2}+\frac{1}{\alpha^{2}}\sum\limits_{i,k\in U{k}}\sigma_{k,i}^{2}{p_{i}} + \frac{1}{\alpha^{2}}\sigma_{n}^{2}}\label{snrk}\setcounter{equation}{17}\\
\mbox{SNR}_{k,\mbox{lower}} &= \frac{\alpha^{2}p_{k^{'}}\left|\mathbb{E}\left[{\mathbf{f}}_{k}^{T}\mathbf{W}{{\mathbf{g}_{k^{'}}}}\right]\right|^{2}}{\alpha^{2}p_{k^{'}}\mbox{var}\left[{\mathbf{f}}_{k}^{T}\mathbf{W}{{\mathbf{g}_{k^{'}}}}\right]+\alpha^{2}p_{k}\mbox{SI}_{k}+\alpha^{2}\mbox{IP}_{k}+\alpha^{2}\mbox{NR}_{k}+\alpha^{2}\mbox{LIR}_{k}+\mbox{UI}_{k}+\mbox{NU}_{k}}\setcounter{equation}{22}\label{gammalower}
\end{align}
\setcounter{equation}{38}
\hrule
\end{figure*}
}
\iftoggle{SINGLE_COL}{\vspace*{-0.15in}}{}
\subsection{ZFR/ZFT processing}\iftoggle{SINGLE_COL}{\vspace*{-0.10in}}{}
%The inter-pair interference which exists due to the sharing of same time-frequency resources is nulled out by using ZF processing, where each stream is projected onto the orthogonal complement of the inter-pair interference. However, due to the imperfect channel information inter-pair interference still exist. 
The ZFR/ZFT precoder using the estimated CSI is formulated as
\begin{eqnarray}\label{wzf}\setcounter{equation}{16}
\mathbf{W} &=& \hat{\bar{\mathbf{F}}}^{*}\mathbf{T}\hat{\bar{\mathbf{G}}}^{H},\label{wzf}
\end{eqnarray}
where $\hat{\bar{\mathbf{F}}}=\hat{\mathbf{F}}\left(\hat{\mathbf{F}}^{H}\hat{\mathbf{F}}\right)^{-1}$ and $\hat{\bar{\mathbf{G}}}=\hat{\mathbf{G}}\left(\hat{\mathbf{G}}^{H}\hat{\mathbf{G}}\right)^{-1}$.
In the next proposition, which is proved in Appendix \ref{alpha_zf_append}, we simplify the scaling factor $\alpha$ in \eqref{alpha}.
\iftoggle{SINGLE_COL}{\vspace*{-0.05in}}{}
\begin{Proposition}\label{pre2}
For the ZFR/ZFT precoder, we have
\begin{eqnarray}\label{alphazfproof}
\alpha &=& \displaystyle{\sqrt{\frac{P_{R}}{\hat\lambda +\hat{\eta}\left(\sum\limits_{i=1}^{2K} p_{i}\sigma_{\xi,g,i}^{2}+\sigma_{nr}^{2}+P_{R}\sigma_{LIR}^{2}\right)}}},
\end{eqnarray}
where
$\hat\lambda = \sum\limits_{i=1}^{2K}\frac{p_{i^{'}}}{\left(N-2K-1\right)\hat\sigma_{f,i}^{2}},\, \hat{\eta} = \sum_{j=1}^{2K}\frac{1}{\left(N-2K-1\right)^{2}\hat\sigma_{f,j}^{2}\hat\sigma_{g,	j^{'}}^{2}}.$
%\begin{eqnarray}
%\hat\lambda &=& \sum\limits_{i=1}^{2K}\frac{p_{i^{'}}}{\left(N-2K-1\right)\hat\sigma_{f,i}^{2}}\iftoggle{SINGLE_COL}{}{\nonumber\\\hat{\eta} &=&}\iftoggle{BIG_EQUATION}{}{,\,\,\hat{\eta} =}  \sum_{j=1}^{2K}\frac{1}{\left(N-2K-1\right)^{2}\hat\sigma_{f,j}^{2}\hat\sigma_{g,	j^{'}}^{2}}\nonumber.
%\end{eqnarray}
\end{Proposition}\iftoggle{SINGLE_COL}{\vspace*{-0.05in}}{}
%\begin{proof}
%Refer to Appendix \ref{alpha_zf_append}
%\end{proof}
\iftoggle{SINGLE_COL}{\vspace*{-0.00in}}{}
\section{Achievable sum-rate of MRC/MRT and ZFR/ZFT precoders}
\label{rate_ana_ref}\iftoggle{SINGLE_COL}{\vspace*{-0.10in}}{}
In this section, we calculate lower bounds on the instantaneous sum-rate for both MRC/MRT and ZFR/ZFT precoders. The instantaneous $\mbox{SNR}_{k}$ at the user $S_{k}$ can be expressed using (\ref{yktilde})~as~in \iftoggle{SINGLE_COL}{}{(\ref{snrk}) (shown at the top of this page).}
\iftoggle{BIG_EQUATION}{}{\vspace{-0.25in}
\begin{align}\label{snrk}
\mbox{SNR}_{k} &= \frac{p_{k^{'}}|{\mathbf{f}}_{k}^{T}\mathbf{W}{{\mathbf{g}_{k^{'}}}}|}{p_{k}|\lambda_{k}|^{2}+\iftoggle{SINGLE_COL}{\hspace{-0.1in}}{}\sum\limits_{i\neq k,k^{'}}^{2K}\iftoggle{SINGLE_COL}{\hspace{-0.05in}}{}|{\mathbf{f}}_{k}^{T}\mathbf{W}{{\mathbf{g}_{i}}}|^{2}+\|{\mathbf{f}}_{k}^{T}\mathbf{W}\mathbf{G}_{RR}\|^{2}\frac{P_{R}}{N}+\|{\mathbf{f}}_{k}^{T}\mathbf{W}\|^{2}\sigma_{n_{R}}^{2}+\frac{1}{\alpha^{2}}\iftoggle{SINGLE_COL}{\hspace{-0.10in}}{}\sum\limits_{i,k\in U{k}}\iftoggle{SINGLE_COL}{\hspace{-0.10in}}{}\sigma_{k,i}^{2}{p_{i}} \iftoggle{SINGLE_COL}{\hspace{-0.05in}}{}+ \frac{1}{\alpha^{2}}\sigma_{n}^{2}}.\iftoggle{SINGLE_COL}{}{\nonumber\\}
\end{align}
}
\iftoggle{SINGLE_COL}{\vspace{-0.99cm}}{}
The sum-rate of the system is
\begin{eqnarray}\setcounter{equation}{19}
\iftoggle{SINGLE_COL}{\hspace{0.45in}}{}R = \mathbb{E}\left\{\sum_{k=1}^{2K}\log_{2}\left(1+\mbox{SNR}_{k}\right)\right\}.\label{sumexact}
\end{eqnarray}
\iftoggle{SINGLE_COL}{}{
\begin{figure*}
\normalsize
% Store the current equation number.
\setcounter{mytempeqncnt}{\value{equation}}
\setcounter{equation}{24}
\begin{align}
\hspace{-0.0in}\mbox{SNR}_{k}^{\mbox{mrc}}
(p_{k},P_{R}) &= \frac{a_{k}p_{k^{'}}}{\displaystyle{\sum\limits_{i=1}^{2K}\left(b_{k,i}^{(1)}+b_{k,i}^{(2)}P_{R}^{-1}+\sum\limits_{i,k\in U_{k}}p_{i}P_{R}^{-1}b_{k,i}^{(3)}\right)p_{i}+c_{k}p_{k}+\left(d_{k}^{(1)}+d_{k}^{(2)}P_{R}+d_{k}^{(3)}P_{R}^{-1}\right)+\sum\limits_{i,k \in U_{k}}p_{i}f_{k,i}(P_{R})}}\label{gammamrct1}\\
\hspace{-0.4in}\mbox{SNR}_{k}^{\mbox{zf}}
(p_{k},P_{R}) &= \frac{u_{k}p_{k^{'}}}{\displaystyle{\sum_{i=1}^{2K}\left(d^{(1)}_{k,i}+d^{(2)}_{k,i}P_{R}^{-1}+\sum_{i,k\in U_{k}}p_{i}P_{R}^{-1}d^{(3)}_{k,i}\right)p_{i}+\left(v^{(1)}_{k}+v^{(2)}_{k}P_{R}+v^{(3)}_{k}P_{R}^{-1}\right)+\sum_{i,k\in U_{k}}p_{i}u_{k,i}(P_{R})}}\label{gammazft2}
\end{align}
\setcounter{equation}{19}
\hrule
\end{figure*}
}

Next we derive a lower bound on the achievable rate using the method of~\cite{DBLP:journals/twc/JoseAMV11}. For the $(k,k^{'})$ pair, the signal received by the $k$th user can be written as (see \eqref{yktilde}) 
\begin{eqnarray}
\tilde{y}_{k} = \underbrace{\alpha\sqrt{p_{k^{'}}}\mathbb{E}\left[{\mathbf{f}}_{k}^{T}\mathbf{W}{{\mathbf{g}_{k^{'}}}}\right]x_{k^{'}}}_{\text{desired signal}} + \underbrace{\tilde{n}_{k}}_{\text{effective noise}},
\end{eqnarray}
where \iftoggle{SINGLE_COL}{\vspace*{-0.34in}}{}
\iftoggle{SINGLE_COL}{}{
\begin{eqnarray}
\tilde{n}_{k} &=& \alpha \sqrt{p_{k^{'}}}\left(\mathbf{f}_{k}^{T}\mathbf{W}\mathbf{g}_{k^{'}}-\mathbb{E}\left[{\mathbf{f}}_{k}^{T}\mathbf{W}{{\mathbf{g}_{k^{'}}}}\right]\right)x_{k^{'}} + \alpha\sqrt{p_{k}}\lambda_{k}x_{k}\nonumber\\
&&+\alpha \mathbf{f}_{k}^{T}\mathbf{W}\sum\limits_{i\neq k,k^{'}}^{2K}\sqrt{p_{i}}\mathbf{g}_{i}x_{i} +\alpha \mathbf{f}_{k}^{T}\mathbf{W} \mathbf{G_{RR}}\tilde{\mathbf{x}}_{R}\nonumber\\
&& + \alpha \mathbf{f}_{k}^{T}\mathbf{W}\mathbf{z}_{R}+\sum\limits_{i,k\in U{k}}\Omega_{k,i}\sqrt{p(k)}x_{i} + z_{k}.
\end{eqnarray}
}
\iftoggle{BIG_EQUATION}{}{
\begin{eqnarray}
\tilde{n}_{k} &=& \alpha \sqrt{p_{k^{'}}}\left(\mathbf{f}_{k}^{T}\mathbf{W}\mathbf{g}_{k^{'}}-\mathbb{E}\left[{\mathbf{f}}_{k}^{T}\mathbf{W}{{\mathbf{g}_{k^{'}}}}\right]\right)x_{k^{'}} + \alpha\sqrt{p_{k}}\lambda_{k}x_{k}+\alpha \mathbf{f}_{k}^{T}\mathbf{W}\sum\limits_{i\neq k,k^{'}}^{2K}\sqrt{p_{i}}\mathbf{g}_{i}x_{i} \nonumber\iftoggle{SINGLE_COL}{\\[-10pt]}{\\}
&& +\alpha \mathbf{f}_{k}^{T}\mathbf{W} \mathbf{G_{RR}}\tilde{\mathbf{x}}_{R} + \alpha \mathbf{f}_{k}^{T}\mathbf{W}\mathbf{z}_{R}+\sum\limits_{i,k\in U{k}}\Omega_{k,i}\sqrt{p(k)}x_{i} + z_{k}.
\end{eqnarray}
}
The value of $\mathbb{E}\left[{\mathbf{f}}_{k}^{T}\mathbf{W}{{\mathbf{g}_{k^{'}}}}\right]$ can be calculated by exploiting the knowledge of channel distribution. We observe that the desired signal and the effective noise are uncorrelated. Similar to \cite{DBLP:journals/tit/HassibiH03,HoydisBD13}, we only exploit the knowledge of $\mathbb{E}\left[{\mathbf{f}}_{k}^{T}\mathbf{W}{{\mathbf{g}_{k^{'}}}}\right]$ in the detection. We use the central limit theorem to treat the uncorrelated additive noise $\tilde{n}(k)$ as the worst-case Gaussian noise, when computing the sum-rate.
For massive MIMO systems, the central limit theorem provides a tight statistical lower bound on the achievable rate. This fact is extensively exploited for deriving the SE and EE expressions in massive MIMO systems\cite{HoydisBD13,DBLP:journals/jsac/NgoSML14,DBLP:journals/twc/DaiD16}.
The lower bound on the achievable sum-rate, consequently, becomes
\begin{eqnarray}
R_{\mbox{\,lower}} = \left[\sum\limits_{k=1}^{2K}\log_{2}\left(1+\mbox{SNR}_{k,\mbox{lower}}\right)\right],
\end{eqnarray}\iftoggle{SINGLE_COL}{\vspace{-0.10in}}{}
where \iftoggle{SINGLE_COL}{}{$\mbox{SNR}_{k,\mbox{lower}}$ is given by (\ref{gammalower}) (shown at the top of last page).}
\iftoggle{BIG_EQUATION}{}{
\begin{align}\label{gammalower}
\mbox{SNR}_{k,\mbox{lower}} = \frac{\alpha^{2}p_{k^{'}}\left|\mathbb{E}\left[{\mathbf{f}}_{k}^{T}\mathbf{W}{{\mathbf{g}_{k^{'}}}}\right]\right|^{2}}{\alpha^{2}p_{k^{'}}\mbox{var}\left[{\mathbf{f}}_{k}^{T}\mathbf{W}{{\mathbf{g}_{k^{'}}}}\right]+\alpha^{2}p_{k}\mbox{SI}_{k}+\alpha^{2}\mbox{IP}_{k}+\alpha^{2}\mbox{NR}_{k}+\alpha^{2}\mbox{LIR}_{k}+\mbox{UI}_{k}+\mbox{NU}_{k}}.
\end{align}
}
%\begin{figure*}
%\begin{eqnarray}\label{gammalower}
%\mbox{SNR}_{k,\mbox{lower}} = \frac{\alpha^{2}p_{k^{'}}\left|\mathbb{E}\left[{\mathbf{f}}_{k}^{T}\mathbf{W}{{\mathbf{g}_{k^{'}}}}\right]\right|^{2}}{\alpha^{2}p_{k^{'}}\mbox{var}\left[{\mathbf{f}}_{k}^{T}\mathbf{W}{{\mathbf{g}_{k^{'}}}}\right]+\alpha^{2}p_{k}\mbox{SI}_{k}+\alpha^{2}\mbox{IP}_{k}+\alpha^{2}\mbox{NR}_{k}+\alpha^{2}\mbox{LIR}_{k}+\mbox{UI}_{k}+\mbox{NU}_{k}}
%\end{eqnarray}
%\hrule
%\end{figure*}
In (\ref{gammalower}), the residual self-interference after SIC $(\mbox{SI})$, the inter-pair interference $(\mbox{IP})$, the amplified noise from the relay $(\mbox{NR})$, the amplified self-loop interference at relay $(\mbox{LIR})$, self-loop interference and inter-user interference $(\mbox{UI})$, and the noise at the user $(\mbox{NU})$, are given as follows:
\iftoggle{SINGLE_COL}{}{
\setcounter{equation}{23}}
\begin{eqnarray}
\mbox{SI}_{k}&=& \mathbb{E}\left[|{\mathbf{f}}_{k}^{T}\mathbf{W}{{\mathbf{g}_{k}}}-\hat{\mathbf{f}}_{k}^{T}\mathbf{W}{\hat{\mathbf{g}}_{k}}|^{2}\right],\iftoggle{SINGLE_COL}{}{\nonumber\\\mbox{IP}_{k} &=&}\iftoggle{SINGLE_COL}{\,\,\mbox{IP}_{k} =}{} \sum\limits_{i\neq k,k^{'}}^{2K}p_{i}\mathbb{E}\left[|{\mathbf{f}}_{k}^{T}\mathbf{W}{{\mathbf{g}_{i}}}|^{2}\right],\nonumber\iftoggle{SINGLE_COL}{\\[-6pt]}{\\}
\mbox{NR}_{k}&=& \mathbb{E}\left[|{\mathbf{f}}_{k}^{T}\mathbf{W}\mathbf{z}_{R}|^{2}\right],\,\,\mbox{LIR}_{k}= \mathbb{E}\left[|{\mathbf{f}}_{k}^{T}\mathbf{W}\mathbf{G}_{RR}\tilde{\mathbf{x}}|^{2}\right],\nonumber\iftoggle{SINGLE_COL}{\\[-8pt]}{\\}
\mbox{UI}_{k}&=& \sum\limits_{i,k\in U{k}}p_{i}\mathbb{E}\left[|\Omega_{k,i}{x_{i}}|^{2}\right],\,\,\mbox{NU}_{k}= \mathbb{E}\left[|z_{k}|^{2}\right].
\end{eqnarray}
We further simplify the $\mbox{SNR}_{k,\mbox{lower}}$ expressions for both the MRC/MRT and ZFR/ZFT precoders in the following theorems. \iftoggle{SINGLE_COL}{\vspace{-0.10in}}{}
\begin{theorem}\label{theorem1}
The achievable rate of user $S_k$ for a finite number of receive antennas at the relay relying on MMSE channel estimate based MRC/MRT processing is lower bounded as follows: \iftoggle{SINGLE_COL}{\\}{}$\log_{2}\left\{1+ \mbox{SNR}_{k}^{\mbox{mrc}}
(p_{k},P_{R})\right\}$, where we have:\iftoggle{SINGLE_COL}{}{$\mbox{SNR}_{k}^{\mbox{mrc}}(p_{k},P_{R})$ is given by} \iftoggle{SINGLE_COL}{}{(\ref{gammamrct1}), shown at the top of this page, with
\begin{align}
a_{k} &= N^{2}\hat\sigma_{f,k}^{4}\hat\sigma_{g,k^{'}}^{4},\nonumber\\
b_{k,i}^{(1)} &= \eta_{k,i} = \hat{\Phi}\sigma_{f,k}^{2}\sigma_{g,i}^{2} +  N\left(\sigma_{f,k}^{2}\hat\sigma_{g,i}^{4}\hat\sigma_{f,i^{'}}^{2}+\sigma_{g,i}^{2}\hat\sigma_{f,k}^{4}\hat\sigma_{g,k^{'}}^{2}\right),\nonumber\\
b_{k,i}^{(2)} &= \sigma_{n}^{2}\left(\hat{\Phi}\sigma_{g,i}^{2}+N\hat\sigma_{g,i}^{4}\hat\sigma_{f,i^{'}}^{2}\right),\nonumber\\\
b_{k,i}^{(3)} &= \sigma_{k,i}^{2}\left(\hat{\Phi}\sigma_{g,i}^{2}+N\hat\sigma_{g,i}^{4}\hat\sigma_{f,i^{'}}^{2}\right),\nonumber\\
c_{k}&= -\left(\hat{\Phi}\hat\sigma_{f,k}^{2}\hat\sigma_{g,k}^{2}+N\left(\hat\sigma_{f,k}^{4}\hat\sigma_{g,k}^{2}\hat\sigma_{g,k^{'}}^{2}+\hat\sigma_{f,k}^{2}\hat\sigma_{g,k}^{4}\hat\sigma_{f,k^{'}}^{2}\right)\right),\nonumber\\
d_{k}^{(1)} &= \left(\sigma_{LIR}^{2}\sigma_{n}^{2}+\sigma_{nr}^{2}\sigma_{f,k}^{2}\right)\hat{\Phi}+N\sigma_{nr}^{2}\hat\sigma^{4}_{f,k}\hat\sigma_{g,k^{'}}^{2},\nonumber\\
d_{k}^{(2)} &= \sigma_{LIR}^{2}\left(\sigma_{f,k}^{2}\hat{\Phi}+N\hat\sigma_{f,k}^{4}\hat\sigma_{g,k^{'}}^{2}\right),\,\,d_{k}^{(3)} = \sigma_{nr}^{2}\sigma_{n}^{2}\hat{\Phi},\nonumber\\
&\hspace{-0.2in}f_{k,i}(P_{R})=\left(P_{R}^{-1}e_{k,i}^{(1)}+e_{k,i}^{(2)}\right),\nonumber\\
e_{k,i}^{(1)} &= \sigma_{nr}^{2}\sigma_{k,i}^{2}\hat{\Phi},\,\,
e_{k,i}^{(2)} = \sigma_{LIR}^{2}\sigma_{k,i}^{2}\hat{\Phi}\nonumber.
\end{align}
\end{theorem}
\begin{proof}
Refer to Appendix \ref{gammrc}.
\end{proof}
}
\iftoggle{BIG_EQUATION}{}{
\begin{align}\label{gammamrct1}
\iftoggle{BIG_EQUATION}{}{&\hspace{-0.0in}}\iftoggle{SINGLE_COL}{}{\hspace{-0.8in}}\mbox{SNR}_{k}^{\mbox{mrc}}
(p_{k},P_{R})= \iftoggle{BIG_EQUATION}{}{\nonumber\\
&} \frac{a_{k}p_{k^{'}}}{\displaystyle{\sum\limits_{i=1}^{2K}\hspace{-0.05in}\left(b_{k,i}^{(1)}+b_{k,i}^{(2)}P_{R}^{-1}+\iftoggle{SINGLE_COL}{\hspace{-0.1in}}{}\sum\limits_{i,k\in U_{k}}\iftoggle{SINGLE_COL}{\hspace{-0.05in}}{}p_{i}P_{R}^{-1}b_{k,i}^{(3)}\right)\iftoggle{SINGLE_COL}{\hspace{-0.05in}}{}p_{i}\iftoggle{SINGLE_COL}{\hspace{-0.05in}}{}+\iftoggle{SINGLE_COL}{\hspace{-0.05in}}{}c_{k}p_{k}\iftoggle{SINGLE_COL}{\hspace{-0.05in}}{}+\iftoggle{SINGLE_COL}{\hspace{-0.05in}}{}\left(d_{k}^{(1)}+d_{k}^{(2)}P_{R}+d_{k}^{(3)}P_{R}^{-1}\right)\hspace{-0.05in}+\hspace{-0.1in}\sum\limits_{i,k \in U_{k}}\iftoggle{SINGLE_COL}{\hspace{-0.05in}}{}p_{i}f_{k,i}(P_{R})}}\iftoggle{SINGLE_COL}{}{\nonumber\\},
\end{align}
where $a_{k} = N^{2}\hat\sigma_{f,k}^{4}\hat\sigma_{g,k^{'}}^{4}$, $b_{k,i}^{(1)} = \eta_{k,i} = \hat{\Phi}\sigma_{f,k}^{2}\sigma_{g,i}^{2} +  N\left(\sigma_{f,k}^{2}\hat\sigma_{g,i}^{4}\hat\sigma_{f,i^{'}}^{2}+\sigma_{g,i}^{2}\hat\sigma_{f,k}^{4}\hat\sigma_{g,k^{'}}^{2}\right)$,
\\$b_{k,i}^{(2)} = \sigma_{n}^{2}\left(\hat{\Phi}\sigma_{g,i}^{2}+N\hat\sigma_{g,i}^{4}\hat\sigma_{f,i^{'}}^{2}\right)$, $b_{k,i}^{(3)} = \sigma_{k,i}^{2}\left(\hat{\Phi}\sigma_{g,i}^{2}+N\hat\sigma_{g,i}^{4}\hat\sigma_{f,i^{'}}^{2}\right)$, \\$c_{k}= -\left(\hat{\Phi}\hat\sigma_{f,k}^{2}\hat\sigma_{g,k}^{2}+N\left(\hat\sigma_{f,k}^{4}\hat\sigma_{g,k}^{2}\hat\sigma_{g,k^{'}}^{2}+\hat\sigma_{f,k}^{2}\hat\sigma_{g,k}^{4}\hat\sigma_{f,k^{'}}^{2}\right)\right)$, \\$d_{k}^{(1)} = \left(\sigma_{LIR}^{2}\sigma_{n}^{2}+\sigma_{nr}^{2}\sigma_{f,k}^{2}\right)\hat{\Phi}+N\sigma_{nr}^{2}\hat\sigma^{4}_{f,k}\hat\sigma_{g,k^{'}}^{2}$, $d_{k}^{(2)} = \sigma_{LIR}^{2}\left(\sigma_{f,k}^{2}\hat{\Phi}+N\hat\sigma_{f,k}^{4}\hat\sigma_{g,k^{'}}^{2}\right)$, 
$d_{k}^{(3)} = \sigma_{nr}^{2}\sigma_{n}^{2}\hat{\Phi}$, 
$f_{k,i}(P_{R})=\left(P_{R}^{-1}e_{k,i}^{(1)}+e_{k,i}^{(2)}\right)$, $e_{k,i}^{(1)} = \sigma_{nr}^{2}\sigma_{k,i}^{2}\hat{\Phi}$, $e_{k,i}^{(2)} = \sigma_{LIR}^{2}\sigma_{k,i}^{2}\hat{\Phi}$.
\end{theorem}\iftoggle{SINGLE_COL}{\vspace*{-0.1in}}{}
\begin{proof}
Refer to Appendix \ref{gammrc}.
\end{proof}
}
\iftoggle{SINGLE_COL}{\vspace*{-0.15in}}{}
\begin{theorem}\label{theorem2}
The achievable rate of user $S_k$  for a finite number of receive antennas at the relay with MMSE channel estimate based ZFR/ZFT processing is lower bounded as \iftoggle{SINGLE_COL}{\\}{}$\log_{2}\left\{1+ \mbox{SNR}_{k}^{\mbox{zf}}
(p_{k},P_{R})\right\}$, where \iftoggle{SINGLE_COL}{}{$\mbox{SNR}_{k}^{\mbox{zf}}
(p_{k},P_{R})$ is given by} \iftoggle{SINGLE_COL}{}{(\ref{gammazft2}), shown at the top of this page, with
\begin{align}
u_{k} &= 1,\nonumber\\
d^{(1)}_{k,i} &= \frac{1}{(N-2K-1)}\left(\frac{\sigma_{\xi,f,k}^{2}}{\hat\sigma_{f,i^{'}}^{2}}+ \frac{\sigma_{\xi,g,i}^{2}}{\hat\sigma_{g,k^{'}}^{2}}\right)+ \sigma_{\xi,f,k}^{2}\sigma_{\xi,g,i}^{2}\hat{\eta},\nonumber\\
d^{(2)}_{k,i} &= \sigma_{n}^{2}\left(\frac{1}{\left(N-2K-1\right)\hat\sigma_{f,i^{'}}^{2}} +\hat{\eta}\sigma_{\xi,g,i}^{2}\right),\nonumber\\
d^{(3)}_{k,i} &= \sigma_{k,i}^{2}\left(\frac{1}{\left(N-2K-1\right)\hat\sigma_{f,i^{'}}^{2}} +\hat{\eta}\sigma_{\xi,g,i}^{2}\right),\nonumber\\
v^{(1)}_{k} &= \sigma_{nr}^{2}\left(\frac{1}{(N-2K-1)\hat\sigma_{g,k^{'}}^{2}}+ \sigma_{\xi,f,k}^{2}\hat{\eta}\right)+\hat\eta\sigma_{LIR}^{2}\sigma_{n}^{2},\nonumber\\
v^{(2)}_{k} &= \sigma_{LIR}^{2}\left(\frac{1}{(N-2K-1)\hat\sigma_{g,k^{'}}^{2}}+\sigma_{\xi,f,k}^{2}\hat{\eta}\right),\nonumber\\
&\hspace{-0.2in}u_{k,i}(P_{R})=\left(w^{(1)}_{k,i}+P_{R}^{-1}w^{(2)}_{k,i}\right)\nonumber\\
v^{(3)}_{k}&= \hat\eta\sigma_{nr}^{2}\sigma_{n}^{2},\,\,\,w^{(1)}_{k,i} = \hat\eta\sigma_{k,i}^{2}\sigma_{LIR}^{2},\,\,\,w^{(2)}_{k,i} = \hat\eta\sigma_{k,i}^{2}\sigma_{nr}^{2}\nonumber.
\end{align}
\end{theorem}
\begin{proof}
Refer to Appendix \ref{gamzf}.
\end{proof}
}
\iftoggle{BIG_EQUATION}{}{
\begin{align}\label{gammazft2}
\iftoggle{BIG_EQUATION}{}{&\hspace{-0.0in}}\iftoggle{SINGLE_COL}{}{\hspace{-0.8in}}\mbox{SNR}_{k}^{\mbox{zf}}
(p_{k},P_{R})= \iftoggle{BIG_EQUATION}{}{\nonumber\\
&\hspace{-0.0in}}  \frac{u_{k}p_{k^{'}}}{\displaystyle{\sum_{i=1}^{2K}\left(d^{(1)}_{k,i}+d^{(2)}_{k,i}P_{R}^{-1}+\iftoggle{SINGLE_COL}{\hspace{-0.1in}}{}\sum_{i,k\in U_{k}}\iftoggle{SINGLE_COL}{\hspace{-0.05in}}{}p_{i}P_{R}^{-1}d^{(3)}_{k,i}\right)p_{i}+\left(v^{(1)}_{k}+v^{(2)}_{k}P_{R}+v^{(3)}_{k}P_{R}^{-1}\right)+\iftoggle{SINGLE_COL}{\hspace{-0.1in}}{}\sum_{i,k\in U_{k}}p_{i}u_{k,i}(P_{R})}},\iftoggle{SINGLE_COL}{}{\nonumber\\}
\end{align}
where \iftoggle{SINGLE_COL}{
$u_{k}= 1$, $d^{(1)}_{k,i} = \frac{1}{(N-2K-1)}\left(\frac{\sigma_{\xi,f,k}^{2}}{\hat\sigma_{f,i^{'}}^{2}}+ \frac{\sigma_{\xi,g,i}^{2}}{\hat\sigma_{g,k^{'}}^{2}}\right)+ \sigma_{\xi,f,k}^{2}\sigma_{\xi,g,i}^{2}\hat{\eta}$,
\\$d^{(2)}_{k,i} = \sigma_{n}^{2}\left(\frac{1}{\left(N-2K-1\right)\hat\sigma_{f,i^{'}}^{2}} +\hat{\eta}\sigma_{\xi,g,i}^{2}\right)$, $d^{(3)}_{k,i} = \sigma_{k,i}^{2}\left(\frac{1}{\left(N-2K-1\right)\hat\sigma_{f,i^{'}}^{2}} +\hat{\eta}\sigma_{\xi,g,i}^{2}\right)$,\\ $v^{(1)}_{k} = \sigma_{nr}^{2}\left(\frac{1}{(N-2K-1)\hat\sigma_{g,k^{'}}^{2}}+ \sigma_{\xi,f,k}^{2}\hat{\eta}\right)+\hat\eta\sigma_{LIR}^{2}\sigma_{n}^{2}$, $v^{(2)}_{k} = \sigma_{LIR}^{2}\left(\frac{1}{(N-2K-1)\hat\sigma_{g,k^{'}}^{2}}+\sigma_{\xi,f,k}^{2}\hat{\eta}\right)$,\\ $v^{(3)}_{k} = \hat\eta\sigma_{nr}^{2}\sigma_{n}^{2}$, $u_{k,i}(P_{R})=\left(w^{(1)}_{k,i}+P_{R}^{-1}w^{(2)}_{k,i}\right)$, $w^{(1)}_{k,i} = \hat\eta\sigma_{k,i}^{2}\sigma_{LIR}^{2}$, $w^{(2)}_{k,i} = \hat\eta\sigma_{k,i}^{2}\sigma_{nr}^{2}$}{}
%----------------------------------------------------------
\iftoggle{SINGLE_COL}{}{where 
\begin{eqnarray}
\hspace{0.5in}u_{k} &=& 1,\,\,d^{(1)}_{k,i} = \frac{1}{(N-2K-1)}\left(\frac{\sigma_{\xi,f,k}^{2}}{\hat\sigma_{f,i^{'}}^{2}}+ \frac{\sigma_{\xi,g,i}^{2}}{\hat\sigma_{g,k^{'}}^{2}}\right)+ \sigma_{\xi,f,k}^{2}\sigma_{\xi,g,i}^{2}\hat{\eta},\nonumber\\
d^{(2)}_{k,i} &=& \sigma_{n}^{2}\left(\frac{1}{\left(N-2K-1\right)\hat\sigma_{f,i^{'}}^{2}} +\hat{\eta}\sigma_{\xi,g,i}^{2}\right),\,\,
d^{(3)}_{k,i} = \sigma_{k,i}^{2}\left(\frac{1}{\left(N-2K-1\right)\hat\sigma_{f,i^{'}}^{2}} +\hat{\eta}\sigma_{\xi,g,i}^{2}\right),\nonumber\\
v^{(1)}_{k} &=& \sigma_{nr}^{2}\left(\frac{1}{(N-2K-1)\hat\sigma_{g,k^{'}}^{2}}+ \sigma_{\xi,f,k}^{2}\hat{\eta}\right)+\hat\eta\sigma_{LIR}^{2}\sigma_{n}^{2},\nonumber\\
v^{(2)}_{k} &=& \sigma_{LIR}^{2}\left(\frac{1}{(N-2K-1)\hat\sigma_{g,k^{'}}^{2}}+\sigma_{\xi,f,k}^{2}\hat{\eta}\right),\,
v^{(3)}_{k} &=& \hat\eta\sigma_{nr}^{2}\sigma_{n}^{2},\,
u_{k,i}(P_{R})&=&\left(w^{(1)}_{k,i}+P_{R}^{-1}w^{(2)}_{k,i}\right)\,w^{(1)}_{k,i} = \hat\eta\sigma_{k,i}^{2}\sigma_{LIR}^{2},\,\,\,w^{(2)}_{k,i} = \hat\eta\sigma_{k,i}^{2}\sigma_{nr}^{2}\nonumber.
\end{eqnarray}}
\end{theorem}\iftoggle{SINGLE_COL}{\vspace*{-0.1in}}{}
\begin{proof}
Refer to Appendix \ref{gamzf}.
\end{proof}
}
\iftoggle{SINGLE_COL}{\vspace*{-0.20in}}{}
\section{Energy-efficient optimization}\label{PA}\iftoggle{SINGLE_COL}{\vspace*{-0.15in}}{}
We now optimally allocate the power for maximizing the EE subject to the rate required by each user. The EE (in bits/Joule/Hz) is defined~as~{\cite{DBLP:journals/jsac/ZhangCSX16,ee_relay_owr_csi_zappone}}\iftoggle{SINGLE_COL}{\vspace*{-0.05in}}{}
\begin{eqnarray}\setcounter{equation}{27}
\mbox{EE} = \frac{\tilde{R}(p_{k},P_{R})}{P_{T}(p_{k},P_{R})}.
\label{EE_def}
\end{eqnarray}
The numerator in the EE is the SE, which also includes the channel estimation overhead, and is given~by
\iftoggle{SINGLE_COL}{\vspace*{-0.2in}}{}\begin{align}\label{R}
\tilde{R}(p_{k},P_{R}) &= \left(1-\frac{\tau}{T}\right)\sum\limits_{k=1}^{2K}\log_{2}\left(1+\mbox{SNR}^{\zeta}_{k}(p_{k},P_{R})\right).
\end{align}
The expression of $\mbox{SNR}^{\zeta}_{k}$, where $\zeta \in(\mbox{mrc},\mbox{zf})$ are given in (\ref{gammamrct1}) and (\ref{gammazft2}), respectively. The denominator in the EE denotes the overall power consumed  by the system \cite{LiFLL17}
\begin{eqnarray}
P_{T}(p_{k},P_{R}) = \sum\limits_{k=1}^{2K}p_{k}+P_{R}+P_{c}.
\end{eqnarray}
The terms $p_{k}$ and $P_{R}$ denote the transmit power of user $S_k$ and of the relay, respectively.  The term $P_{c}$ represents the fixed circuit power used by the system.\iftoggle{SINGLE_COL}{\vspace{-0.10in}}{}
\begin{remark}
We assume the power amplifiers efficiency to be unity both at the user and at the relay as in \cite{ZapponeCJB13,LiFLL17,tspZapponeJB14} for mathematical simplicity.  Also, $P_{c}$ takes into account the power required by the different components, namely by the transceiver's radio-frequency chain, the oscillator and the power consumption of channel estimation \cite{LiFLL17,ZapponeCJB13,tspZapponeJB14}. This assumption is commonly exploited in the massive MIMO literature for mathematical simplicity \cite{LiFLL17}, and it does not affect the overall behavior of the system considered.
\end{remark}\iftoggle{SINGLE_COL}{\vspace{-0.10in}}{}
%is also considered to compute the power consumed in the system. where $P_{c}$ is fixed quantity which does not depend on powers $p_{k}$ and $P_{R}$ or SE $\tilde{R}(p_{k},P_{R})$. 
Before formulating the related optimization problem, we briefly discuss the terminologies used in geometric and fractional programming from \cite{boyd2007tutorial} and \cite{ZapponeJ15} respectively, which will be used in the sequel. Geometric and fractional programming have earlier been used for power allocation in \cite{ZapponeJ15,my_paper1,my_paper_twc_jt,my_paper_twr_qos}.

A real valued function $f(x)$ of the form $f(x)=c\,x_{1}^{a_{1}}x_{2}^{a_{2}}\cdots x_{n}^{a_{n}}$, where $c>0$, $a_{i}\in \mathbb{R}$ and $\mathbf{x}\in\mathbb{R}^{n}_{++}$, is referred to as a monomial function. The sum of one or more monomials, i.e. $f(x)=\sum_{m=1}^{K}c_m\,x_{1}^{a_{1m}}x_{2}^{a_{2m}}\cdots x_{n}^{a_{nm}}$, where $c_{m}>0$ is termed as a posynomial. Monomials are closed both under multiplication and division, whereas posynomials are closed under addition and multiplication, but not under division. The ratio of a posynomial and monomial is a posynomial. A geometric program (GP) has a posynomial objective and upper bounded posynomial inequality~constraints.  

%The objective represents the energy-efficient metric to optimize, the set $\mathcal{X}$ models the constraints.

A fractional program is of the form {${u(x)/v(x)}$}, so that the optimization variable obeys $x\in\mathcal{X}$, where $u:\mathcal{C}\subset \mathbb{R}^{n}\rightarrow\mathbb{R}$, $v:\mathcal{C}\subset \mathbb{R}^{n}\rightarrow\mathbb{R}_{x+}$ and $\mathcal{X}\subset\mathcal{C}\subset\mathbb{R}^{n}.$ 
Since the objective is a fraction, the problem is not guaranteed to be always convex, even if both $u$ and $v$ are affine functions. For maximizing fractions, mostly two classes of generalized concave functions, namely quasi-concave (QC) functions and pseudo-concave (PC) functions are used.
If $\mathcal{C}\subset \mathbb{R}^{n}$ is a convex set, then $r:\mathcal{C}\rightarrow \mathbb{R}$ is QC if $r(\lambda x_{1}+(1-\lambda)x_{2})\geq \min\{r(x_{1},x_{2})\}$ for all $x_{1},x_{2}\in\mathcal{C}$ and $\lambda\in[0;1]$. Similarly, $r:\mathcal{C}\rightarrow \mathbb{R}$ is PC if it is differentiable and, $r(x_{2})<r(x_{1})\Rightarrow \triangledown (r(x_{2}))^{T}(x_{1}-x_{2})>0$  for all $x_{1},x_{2}\in\mathcal{C}$. A local maximum of optimization associated with the PC objective {constitutes} a global maximum, whereas under a QC objective it is not necessarily a global maximum. For example, when the objective has a concave numerator and a convex denominator, the fractional program is a PC and its stationary point is its global maximizer. Such a problem belongs to the class of concave-convex fractional programs (CCFP)~\cite{ZapponeJ15}.

With this information {in mind}, we optimize the EE in the next section.
\iftoggle{SINGLE_COL}{\vspace*{-0.25in}}{}
\subsection{EE maximization}\label{EE_opt}\iftoggle{SINGLE_COL}{\vspace*{-0.10in}}{}
The EE maximization problem is formulated as\iftoggle{SINGLE_COL}{\vspace{-0.05in}}{}
%------------------------------------------------------------ 
\begin{subequations}
\begin{alignat}{2}
\mathbf{P1}:& \underset{p_{k},P_{R}}{\mbox{Max}} &&\frac{\left(1-\frac{\tau}{T}\right)\sum\limits_{k=1}^{2K}\log_{2}\left(1+\mbox{SNR}^{\zeta}_{k}(p_{k},P_{R})\right)}{\sum\limits_{k=1}^{2K}p_{k}+P_{R}+P_{c}}\iftoggle{SINGLE_COL}{\\[-8pt]}{\\}
&\text{ s.t. } 
%&&\log_{2} \left(1+{\mbox{SNR}^{\zeta}_{k}(p_{k},P_{R})}\right)\geq r_{k}\\
&& 0\leq p_{k}\leq P^{\max},0\leq P_{R}\leq P_{R}^{\max}\iftoggle{SINGLE_COL}{\\[-8pt]}{\\}
&&&\sum_{k=1}^{2K}p_{k}+P_{R}\leq P_t^{\max}.
\label{problem0_1}
\end{alignat}
\end{subequations}
 The first two constraint specify the peak transmit power of the user and the relay i.e. $P^{\max}$  and $P_R^{\max}$ respectively. The last constraint impose a constraint of $P_t^{\max}$ on the total system transmit power. The problem \textbf{P1} can be re-cast~as\iftoggle{SINGLE_COL}{\vspace{-0.05in}}{}
% \begin{equation}
% \begin{aligned}
% \textbf{P2}: &  \underset{p_{k},P_{R},\Gamma_{k}}{\mbox{Max}} &&\frac{\left(1-\frac{\tau}{T}\right)\log_{2}\prod\limits_{k=1}^{2K}\left(1+\Gamma_{k}\right)}{\sum\limits_{k=1}^{2K}p_{k}+P_{R}+P_{c}}\\
% &\text{s.t.} 
% &&\Gamma_{k}=\mbox{SNR}^{\zeta}_{k}(p_{k},P_{R})\\
% &&&\mbox{ISNR}^{\zeta}_{k}(p_{k},P_{R})\leq 1/(2^{r_{k}}-1)\\
% &&&0\leq p_{k}\leq P_{0},0\leq P_{R}\leq P_{R,0}\\
% &&&\sum_{k=1}^{2K}p_{k}+P_{R}\leq P
% \end{aligned}
% \end{equation}
%-----------------------------------------------------------
\begin{subequations}
\begin{alignat}{2}
\textbf{P2}: &  \underset{p_{k},P_{R},\Gamma_{k}}{\mbox{Max}} &&\frac{\left(1-\frac{\tau}{T}\right)\log_{2}\prod\limits_{k=1}^{2K}\left(1+\Gamma_{k}\right)}{\sum\limits_{k=1}^{2K}p_{k}+P_{R}+P_{c}}\iftoggle{SINGLE_COL}{\\[-8pt]}{\\}
&\text{s.t.} 
%&&\mbox{ISNR}^{\zeta}_{k}(p_{k},P_{R})\leq 1/(2^{r_{k}}-1)\label{const2}\\
&&0\leq p_{k}\leq P^{\max},0\leq P_{R}\leq P_{R}^{\max}\label{const3}\iftoggle{SINGLE_COL}{\\[-8pt]}{\\}
&&&\sum_{k=1}^{2K}p_{k}+P_{R}\leq P_t^{\max}\label{const4}\iftoggle{SINGLE_COL}{\\[-8pt]}{\\}
&&&\mbox{ISNR}^{\zeta}_{k}(p_{k},P_{R}) \leq \Gamma_{k}^{-1}.\label{const1}
\end{alignat}
\end{subequations}
%-----------------------------------------------------------
%\colr{The objective function in Problem \textbf{P3} increases with $\Gamma_{k}$, hence the constraint \eqref{const1} is satisfied for any optimal solution of Problem \textbf{P3}.} 
%Posynomial are not closed under division and numerator of objective is consequently neither a posynomial nor a monomial. 
%The problem P2 cannot be solved using fractional programming tools as it is a non-concave fractional programming problem. 
%Further, we consider the related problem in which the equality constraint of Problem \textbf{P2} is replaced with inequality constraints.
The symbol $\Gamma_{k}=\mbox{SNR}^{\zeta}_{k}$ denotes an auxiliary variable  and the term $\mbox{ISNR}^{\zeta}_{k}$ denotes the inverse of $\mbox{SNR}^{\zeta}_{k}$, i.e. $\mbox{ISNR}^{\zeta}_{k}=1/\mbox{SNR}^{\zeta}_{k}$. We observe from \eqref{gammamrct1} and \eqref{gammazft2}  that the $\mbox{SNR}^{\zeta}_{k}$ is a ratio of a monomial and of a posynomial, the numerator of the objective in \textbf{P2} becomes a ratio of two posynomials, which is not a posynomial and hence non-convex. We also note that the constraints in \textbf{P2} are upper-bounded posynomials and are therefore convex. We now approximate the numerator as a monomial such that \textbf{P2} becomes a CCFP. To this end, we use the following lemma from \cite{DBLP:journals/tvt/WeeraddanaCLE11}.
%Then the numerator of problem P2 can be cast as a GP, with the monomial objective and upper bounded posynomial constraints. The following lemma  is used to obtain the monomial.
\iftoggle{SINGLE_COL}{\vspace{-0.10in}}{}
\begin{lemma}\label{lemma11}
Consider a monomial function $q(\nu_{k})=\delta_{k}\nu_{k}^{\alpha_{k}}$ ($\nu_{k}> 0$), which is used for approximating $s(\nu_{k})=1+\nu_{k}$ near an arbitrary point $\tilde{\nu}_{k}>0$. For the above approximation, the following two conditions hold.\iftoggle{SINGLE_COL}{\vspace{-0.05in}}{}
\begin{enumerate}
\iftoggle{SINGLE_COL}{\vspace{-0.05in}}{}\item For the best monomial local approximation, the parameters $\alpha_{k}$ and $\delta_{k}$ are given by
\begin{eqnarray}\label{bound}
\alpha_{k} = \tilde{\nu}_{k}(1+\tilde{\nu}_{k})^{-1},\, \delta_{k} = \tilde{\nu}_{k}^{-\alpha_{k}}(1+\tilde{\nu}_{k}).
\end{eqnarray}\iftoggle{SINGLE_COL}{\vspace{-0.05in}}{}
\iftoggle{SINGLE_COL}{\vspace{-0.25in}}{}\item For all $\nu_{k}>0$, $s(\nu_{k})\geq q(\nu_{k})$.
\end{enumerate}\iftoggle{SINGLE_COL}{\vspace{-0.05in}}{}
\end{lemma}
%\begin{proof}
%See proof in \cite{DBLP:journals/tvt/WeeraddanaCLE11}, Lemma 1.
%\end{proof}
Using Lemma \ref{lemma11}, the numerator of the objective function can be approximated as \iftoggle{SINGLE_COL}{\\}{}$\log_{2}\prod\limits_{k=1}^{2K}\left[\delta_{k}(p_{k},P_{R})\Gamma_{k}^{\alpha_{k}(p_{k},P_{R})}\right]$, where $\alpha_{k}(p_{k},P_{R}) = \tilde{\Gamma}_{k}\left(1+\tilde{\Gamma}_{k}\right)^{-1}$ and \iftoggle{SINGLE_COL}{\\}{} $\delta_{k}(p_{k},P_{R}) \hspace{-0.05in}=\hspace{-0.05in} \left(\tilde{\Gamma}_{k}\right)^{-\alpha_{k}(p_{k},P_{R})}\hspace{-0.05in}\left(1+\tilde{\Gamma}_{k}\right)$, where $\tilde{\Gamma}_{k}$ is an initial value approximation for $\Gamma_{k}$.
Given the approximated objective, the optimization \textbf{P2} can be formulated as follows.
\begin{equation}\label{P3}
\begin{aligned}
\mathbf{P3}:& \underset{p_{k},P_{R},\Gamma_{k}}{\mbox{Max}} &&\frac{\log_{2}\prod\limits_{k=1}^{2K}\left[\delta_{k}(p_{k},P_{R})\Gamma_{k}^{\alpha_{k}(p_{k},P_{R})}\right]}{\sum\limits_{k=1}^{2K}p_{k}+P_{R}+P_{c}}\iftoggle{SINGLE_COL}{\\[-8pt]}{\\}
&\text{ s.t. } 
&& (\ref{const3}), (\ref{const4}), (\ref{const1}).
% &&\log_{2} \left(1+{\mbox{SNR}^{\zeta}_{k}(p_{k},P_{R})}\right)\geq r_{k}\\
% &&&0\leq p_{k}\leq P_{0},0\leq P_{R}\leq P_{R,0}\\
% &&&\sum_{k=1}^{2K}p_{k}+P_{R}\leq P
% \label{problem0_1}
\end{aligned}
\end{equation}
Here we have dropped the constant $\left(1-\frac{\tau}{T}\right)$ from the objective. The optimization \textbf{P3} is now a CCFP. We use the following result, proved in \cite{dinkelbach1967nonlinear,ZapponeJ15}, to solve~it.\iftoggle{SINGLE_COL}{\vspace{-0.05in}}{}
\begin{Proposition}\label{prepo3}
%Consider the CCFP $g(x)=u(x)/v(x)$, with differentiable and concave $u$. Then the function $g(x)$ is PC and a stationary point $x^{*}$ is a stationary point of $g(x)$, then it is a global maximizer of $g(x)$. %The function of real variable
Consider the CCFP $g(x)=u(x)/v(x)$, with $u$ being non-negative, differentiable and concave, while $v$ being positive, differentiable and convex. Then the function $g(x)$ is a PC and a stationary point $x^{*}$ of $g(x)$ is its global maximizer. %$g(x)$.
The problem of maximizing $g(x)$ is equivalent to finding the positive zero of $D(\lambda)$, which is defined as 
\begin{eqnarray}
D(\lambda)\triangleq \underset{x}{\mbox{max}}\left\{u(x)-\lambda v(x)\right\}.
\end{eqnarray}
The function $D(\lambda)$ is convex, continuous and strictly monotonically decreasing and its zero is found using Dinkelbach's algorithm~\cite{dinkelbach1967nonlinear}.
%whose basic idea is to solve a sequence of auxiliary problem  which converges to a global solution.
\end{Proposition}
Let us now exploit the monomial approximation and Dinkelbach's algorithm to solve the EE problem \textbf{P3}, as illustrated in Algorithm~\ref{algorithm_EE}\footnote{{The Algorithm \ref{algorithm_EE} uses the monomial, obtained using Lemma~\ref{lemma11}, to cast the problem \textbf{P3} as a GP. Note that this algorithm is not optimal as it approximates the posynomial with a monomial. It is a heuristic that often computes the globally optimal power allocation \cite{ZapponeBSJ17_global}.
The monotonic optimization, which obtains global optimal solution in finite time \cite{ZapponeBSJ17_global}, can also be used to solve problem P2 without any approximation. But the worst-case complexity of monotonic optimization increases at least exponentially in $2K$ \cite{ZapponeBSJ17_global}. The proposed sequential fractional solution can solve the problem in polynomial time with affordable complexity.
Further the proposed algorithm  -- as concluded by exhaustive numerical studies in \cite{ZapponeBSJ17_global} -- is an excellent heuristic that achieves a global optimum solution.}}.

\begin{algorithm}
\DontPrintSemicolon % Some LaTeX compilers require you to use \dontprintsemicolon instead
\KwIn{Given a tolerance $\epsilon > 0$ and the maximum number of iterations $L$, set $m = 1$ and $\lambda^{[1]}=0$. Calculate the initial values $p_{k},P_{R}$ and $\tilde{\Gamma}_{k}^{[1]}$ by allocating equal power to all users i.e. $p_{k}=P_{R}/2K$, and $P_{R}=P/2$.}
\KwOut{$p_{k}^{*}$ and $P_{R}^{*}$ as the solutions.}
\For{$m \gets 1$ \textbf{to} $L$}{
Given a feasible $p_{k},\,P_{R}$, compute  $\alpha_{k}^{[m]}(p_{k},P_{R})$ and $\delta_{k}^{[m]}(p_{k},P_{R})$.\;
    Solve the GP to calculate $p_{k}$, $P_{R}$ and $\Gamma_{k}$.
\begin{equation}
\begin{aligned}
& \underset{p_{k},P_{R},\Gamma_{k}}{\mbox{Max}} &&\left\{\log_{2}\prod\limits_{k=1}^{2K}\delta^{[m]}_{k}(p_{k},P_{R}){\Gamma}_{k}^{\alpha^{[m]}_{k}(p_{k},P_{R})}\iftoggle{SINGLE_COL}{}{\right.\nonumber\\
&&&\hspace{0.5in}\left.}-\lambda^{[m]}\left(\sum\limits_{k=1}^{2K}p_{k}+P_{R}+P_{c}\right) \right\}\iftoggle{SINGLE_COL}{\\[-6pt]}{\\}
&\text{ s.t. } &&\beta^{-1}\tilde{\Gamma}_{k}^{[m]} \leq \Gamma_{k}\leq\beta\tilde{\Gamma}_{k}^{[m]}\iftoggle{SINGLE_COL}{\\[-8pt]}{\\}
&&& \eqref{const3}, \eqref{const4}, \eqref{const1}.
\label{problem0_1}
\end{aligned}
\end{equation}\;\iftoggle{SINGLE_COL}{\vspace*{-0.45in}}{}
{Based on the solutions $P_{R}$ and $p_{k}$ of GP, compute $\tilde{\Gamma}_{k}$, and \iftoggle{SINGLE_COL}{\vspace{-0.05in}}{}
\begin{align}
D(\lambda^{[m]})&=\log_{2}\prod\limits_{k=1}^{2K}\delta^{[m]}_{k}(p_{k},P_{R})\tilde{\Gamma}_{k}^{\alpha_{k}^{[m]}(p_{k},P_{R})}\iftoggle{SINGLE_COL}{}{\nonumber\\
&\hspace{.8in}}-\lambda^{[m]}\left(\sum\limits_{k=1}^{2K}p_{k}+P_{R}+P_{c}\right).\nonumber
\end{align}}\;\iftoggle{SINGLE_COL}{\vspace*{-0.45in}}{}
Do until convergence \iftoggle{SINGLE_COL}{\hspace{6in}}{\hspace{4in}}
\lIf{$D(\lambda_{k}^{[m]}) > \epsilon$}{$\tilde{\Gamma}_{k}^{[m+1]}=\tilde{\Gamma}_{k}$, and 
\begin{align}
\lambda^{[m+1]}&= \frac{\log_{2}\prod\limits_{k=1}^{2K}\delta^{[m]}_{k}(p_{k},P_{R})\tilde{\Gamma}_{k}^{\alpha_{k}^{[m]}(p_{k},P_{R})}}
{\sum\limits_{k=1}^{2K}p_{k}+P_{R}+P_{c}}\nonumber.
\end{align}\iftoggle{SINGLE_COL}{\vspace*{-0.54in}}{}
    }
    \lElse{
      $\mbox{break}.$
    }
}\iftoggle{SINGLE_COL}{\vspace*{-0.1in}}{}
\Return{$p_{k}^{*}$ and $P_{R}^{*}.$}\;
\caption{EE maximization algorithm}\label{algorithm_EE}
\end{algorithm}\iftoggle{SINGLE_COL}{\vspace{-0.05in}}{}
\begin{remark}
The first inequality constraint in Algorithm~\ref{algorithm_EE}, also termed as the trust region constraint \cite{boyd2007tutorial}, is added to confine the domain of variable $\Gamma$ to a region around the current guess $\tilde{\Gamma}^{[m]}$. In most practical cases, $\alpha=1.1$ provides a good accuracy/speed trade-off for the monomial approximations \cite{boyd2007tutorial,DBLP:journals/tvt/WeeraddanaCLE11}.
\end{remark}\iftoggle{SINGLE_COL}{\vspace{-0.05in}}{}
\iftoggle{SINGLE_COL}{\vspace*{-0.25in}}{}
\subsection{Energy efficient optimization relying on the max-min approach}
We next maximize the EE of the least energy-efficient user using the classic fractional programming approach. The max-min optimization can be cast as
\begin{equation}
\begin{aligned}
\mathbf{P4}:& \underset{p_{k},P_{R},\Gamma_{k}}{\mbox{Max}} &&\underset{1\leq k\leq 2K}{\mbox{min}}\frac{\log_{2}\left[\delta_{k}(p_{k},P_{R})\Gamma_{k}^{\alpha_{k}(p_{k},P_{R})}\right]}{\displaystyle{p_{k}+\frac{P_{R}+P_{c}}{2K}}}\iftoggle{SINGLE_COL}{\\[-8pt]}{\\}
&\text{ s.t. } && (\ref{const3}), (\ref{const4}),(\ref{const1}).
\label{problem0_1}
\end{aligned}
\end{equation}
Here $\delta_{k}(p_{k},P_{R})$ and $\alpha_{k}(p_{k},P_{R})$ are calculated similar to (\ref{bound}). The problem \textbf{P4} now belongs to a class of a max-min fractional program (MMFP). We next use the following proposition from \cite{crouzeix1991algorithms,ZapponeJ15} to solve our MMFP.\iftoggle{SINGLE_COL}{\vspace{-0.25in}}{}
\begin{Proposition}
The MMFP optimization problem is stated~as
\begin{eqnarray}\label{probprep4}
&& \underset{x}{\mbox{Max}} \,\, \underset{1\leq k\leq 2K}{\mbox{min}}\frac{u_{k}(x)}{v_{k}(x)}\nonumber\\
&&\hspace{0in}\mbox{s.t.}\,\,\,c_{i}(x),\,\, i = 1,2,3,\cdots, I.
\end{eqnarray}
For a non-negative and concave $u_{k}$,  positive and convex $v_{k}$, $\forall k = 1,2,3,\cdots,2K$, as well as convex $c_{i}$,  $\forall i = 1,2,3,\cdots,I$,  each ratio of the objective is a QC function. Furthermore, if $z(x)= \underset{1\leq k\leq 2K}{\mbox{min}} y_{k}(x)$, so that $y_{k}(x)$ is a QC, $\forall k=1,2,3,\cdots,2K$, then $z(x)$ is also a QC function. The auxiliary function of our real variable is defined as
\begin{eqnarray}
D(\lambda)\triangleq \underset{x}{\mbox{Max}}\left\{\underset{1\leq k \leq 2K}{\mbox{min}}\left\{u(x)-\lambda v(x)\right\}\right\}.
\end{eqnarray}
\end{Proposition}
%Even if $u(x)/v(x)$ would be pseudo-concave function, the problem in (\ref{probprep4}) is quasi-concave, since $\mbox{min}(\cdot)$ is non-differentiable. It is possible to have stationary points which are not global maximizers, as the objective function is not PC. 
The generalized Dinkelbach algorithm \cite{crouzeix1991algorithms} solves a sequence of convex problems to obtain the global solution of MMFP, as illustrated in Algorithm~\ref{algorithm2}.

\begin{algorithm}
\DontPrintSemicolon % Some LaTeX compilers require you to use \dontprintsemicolon instead
\KwIn{Given a tolerance $\epsilon > 0$ and the maximum number of iterations $L$, set $m = 1$ and $\lambda^{[1]}=0$, calculate the initial values $p_{k},P_{R}$ and $\tilde{\Gamma}_{k}^{[1]}$ by allocating equal power to all users i.e. $p_{k}=P_{R}/2K$, and $P_{R}=P/2$.}
\KwOut{$p_{k}^{*}$ and $P_{R}^{*}$ as the solutions.}
\For{$m \gets 1$ \textbf{to} $L$}{
Given a feasible $p_{k},\,P_{R}$, compute  $\alpha_{k}^{[m]}(p_{k},P_{R})$ and $\delta_{k}^{[m]}(p_{k},P_{R})$.\;
    Solve the GP to calculate $p_{k}$, $P_{R}$ and $\Gamma_{k}$. 
\begin{equation}
\begin{aligned}
& \underset{p_{k},P_{R},\Gamma_{k}}{\mbox{Max}} &&{\mbox{min}}\left\{\log_{2}\left[\delta_{k}^{[m]}(p_{k},P_{R}){\Gamma}_{k}^{\alpha^{[m]}_{k}(p_{k},P_{R})}\right]\iftoggle{SINGLE_COL}{}{\right.\\
&&&\left.\hspace{0.5in}}-\lambda^{[m]}\left(p_{k}+\frac{P_{R}+P_{c}}{2K}\right) \right\}\iftoggle{SINGLE_COL}{\\[-8pt]}{\\}
&\text{ s.t. } && \eqref{const3}, \eqref{const4}, \eqref{const1}\iftoggle{SINGLE_COL}{\\[-8pt]}{\\}
&&&\beta^{-1}\tilde{\Gamma}_{k}^{[m]} \leq \Gamma_{k}\leq\beta\tilde{\Gamma}_{k}^{[m]}.\nonumber
\label{problem0_1}
\end{aligned}
\end{equation}\;\iftoggle{SINGLE_COL}{\vspace*{-0.32in}}{}
{Based on the solutions $P_{R}$ and $p_{k}$ of GP, compute $\tilde{\Gamma}_{k}$, and 
\begin{align}
D(\lambda^{[m]})&=\underset{1\leq k\leq 2K}{\mbox{min}} \left\{\log_{2}\left[\delta_{k}^{[m]}(p_{k},P_{R})\tilde{\Gamma}_{k}^{\alpha^{[m]}_{k}(p_{k},P_{R})}\right]\iftoggle{SINGLE_COL}{}{\right.\nonumber\\
&\hspace{1.0in}\left.}-\lambda^{[m]}\left(p_{k}+\frac{P_{R}+P_{c}}{2K}\right) \right\}\nonumber. 
\end{align}}\;\iftoggle{SINGLE_COL}{\vspace*{-0.25in}}{}
Update $\tilde{\Gamma}_{k}^{[m+1]}=\tilde{\Gamma}_{k}$, and\iftoggle{SINGLE_COL}{\vspace*{-0.3in}}{} \begin{align}
\lambda^{[m+1]} &= \underset{1\leq k\leq 2K}{\mbox{min}} \frac{\log_{2}\left[\delta_{k}^{[m]}(p_{k},P_{R})\tilde{\Gamma}_{k}^{\alpha^{[m]}_{k}(p_{k},P_{R})}\right]}{p_{k}+\frac{P_{R}+P_{c}}{2K}}\nonumber.
\end{align}\iftoggle{SINGLE_COL}{\vspace*{-0.25in}}{}
}
\Return{$p_{k}^{*}$ and $P_{R}^{*}$.}\;
\caption{Max-min EE optimization algorithm}\label{algorithm2}
\end{algorithm}
\iftoggle{SINGLE_COL}{\vspace*{-0.4501in}}{}
\section{Simulation Results}
\label{simu_sec_ref}\iftoggle{SINGLE_COL}{\vspace*{-0.10in}}{}
Let us now investigate the EE and the max-min algorithms using Monte-Carlo simulations to highlight their advantage. For this study, we set the length of the coherence interval to $T=200$ symbols and the pilot length to $\tau=2K$. We also set the large scale fading matrices $\mathbf{D}_{u}$ and $\mathbf{D}_{d}$, similar to \cite{DBLP:journals/tcom/NgoLM13,DBLP:journals/twc/DaiD16}  as
%We have considered two different scenarios, channel $1$ in which both $\mathbf{D}_{u}$ and $\mathbf{D}_{u}$ are taken to be same, i.e.
%\begin{eqnarray}
%\mathbf{D}_{u}=\mathbf{D}_{d}&=& \mbox{diag}\left[0.749,\,0.246 ,\,0.125 ,\,0.635,\, 4.468,\right.\nonumber\\
%&& \left. 0.031,\,0.064,\, 0.257, \,0.195,\, 0.315\right]
%\end{eqnarray}
%and channel $2$ given by
\begin{eqnarray}
\mathbf{D}_{u} &=& \mbox{diag}\left[0.749,\,0.045 ,\,0.246 ,\,0.121,\, 0.125,\iftoggle{SINGLE_COL}{}{\right.\nonumber\\
&& \left.} 0.142,\,0.635,\, 0.256, \,0.021,\, 0.123\right],\, \mbox{and}\nonumber\iftoggle{SINGLE_COL}{\\[-8pt]}{\\}
\mathbf{D}_{d} &=& \mbox{diag}\left[0.257,\, 0.856,\, 1.000,\, 0.899,\, 0.014,\iftoggle{SINGLE_COL}{}{\right.\nonumber\\
&& \left.} 0.759,\, 0.315,\, 0.432,\, 0.195,\, 0.562\right].
\end{eqnarray}
%We first study the performance of EE maximization algorithm.
Furthermore, the maximum transmit power of each user is $P^{\max}\hspace{-0.05in}=\hspace{-0.05in}10\mbox{ dBm}$, the maximum relay transmit power is $P_R^{\max}\hspace{-0.05in}=\hspace{-0.05in}23\mbox{ dBm}$, and the circuit power is $P_{c}\hspace{-0.05in}=\hspace{-0.05in}30$~dBm {\cite{DBLP:journals/twc/DaiD16}}. {For this analysis, we consider $K\hspace{-0.05in}=\hspace{-0.05in}5$ user pairs and $N=500$ relay antennas.} The noise variances are set to $\sigma_{n}^{2}\hspace{-0.05in}=\hspace{-0.05in}\sigma_{nr}^{2}\hspace{-0.05in}=\hspace{-0.05in}\sigma^2$ and {we define $\eta\hspace{-0.05in}=\hspace{-0.05in}P_t^{\max}/\sigma^2$}, where $P_t^{\max}$ is the maximum total~transmit~power~of~the~system.
\iftoggle{SINGLE_COL}{\vspace*{-0.2in}}{} 
\subsection{EE maximization}\iftoggle{SINGLE_COL}{\vspace*{-0.1in}}{}
We investigate the performance of the EE maximization algorithm for both MRC/MRT and ZFR/ZFT processing.  {The values of the self-loop interference $\sigma_{LIR}^{2}$ and {inter-user} interference  $\sigma_{UI}^{2}\triangleq\sigma_{k,j}^{2}$ for $k,j=1,\cdots,2K$ are assumed to be $0$~dB with respect to $\sigma^{2}$.} Before investigating the performance, we have to decide the pilot transmission power $P_\rho$, which affects the channel estimation and consequently the overall performance. To determine $P_\rho$, {we maximize the EE  by varying $P_\rho$ from $-10$ dBm to $40$ dBm with two different values of $\eta$, i.e. $0$~dB and $20$~dB}. We see from Fig.~\ref{eevsNQoSzf_Algo2} that $P_\rho=20$~dBm is capable of achieving the maximum EE for both MRC/MRT and ZFR/ZFT designs. We therefore set $P_{\rho}=20$~dBm for the rest of the EE analysis. We also see that a lower pilot power degrades the ZFR/ZFT EE performance more than that of the MRC/MRT design. This is because the ZFR/ZFT design depends on the composite channels~of~the~users.

%-------------------------------------------------------------------------
\iftoggle{SINGLE_COL}{\begin{figure}[htbp]}{\begin{figure}[tbp]}
    \centering\iftoggle{SINGLE_COL}{\vspace*{-5pt}}{}
  \iftoggle{SINGLE_COL}{\begin{subfigure}[b]{.40\linewidth}}{\begin{subfigure}[b]{1\linewidth}}
    \includegraphics[width=\linewidth]{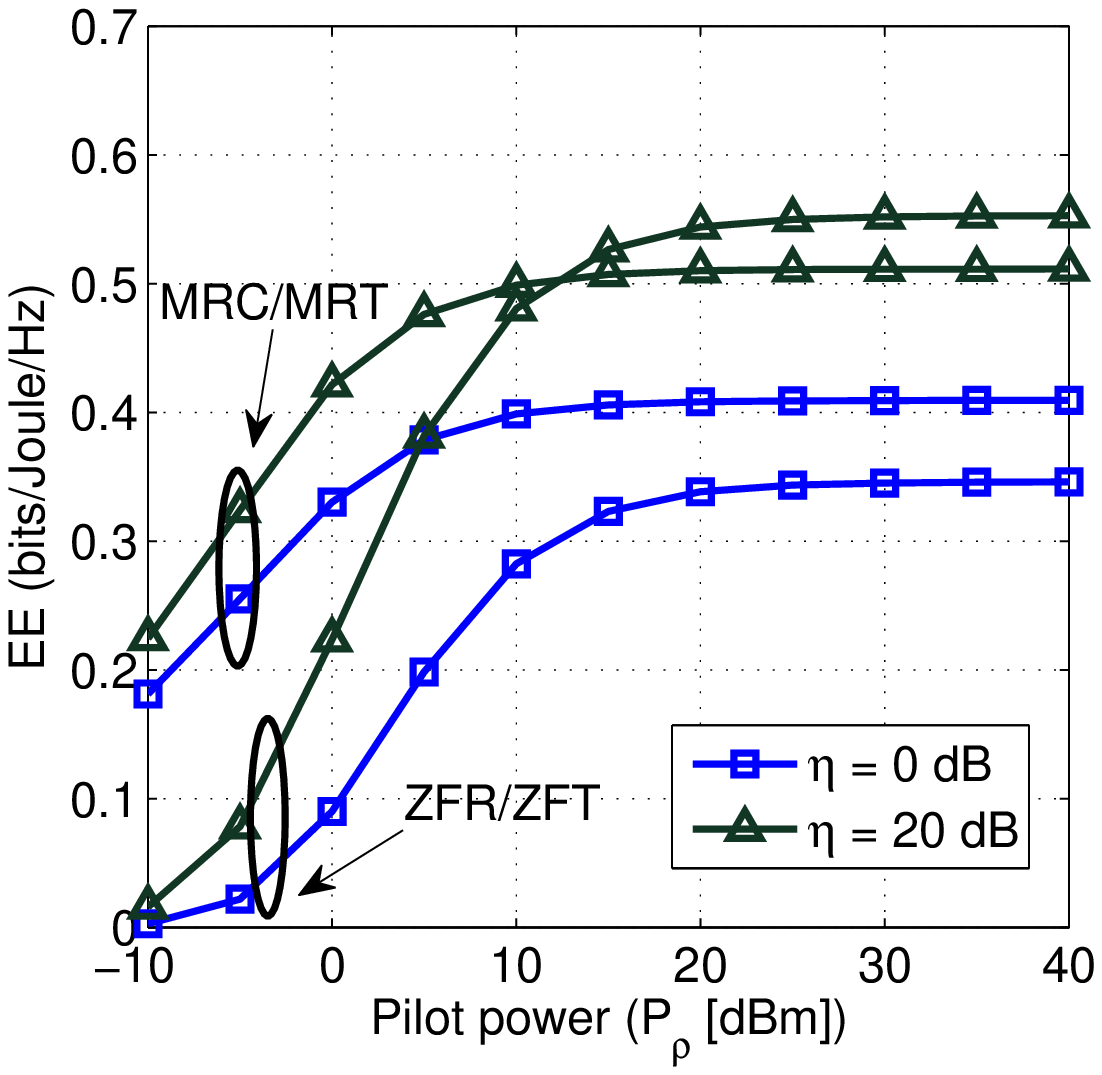}\iftoggle{SINGLE_COL}{\vspace*{-10pt}}{}
    \caption{\small }
    \label{eevsNQoSzf_Algo2}
  \end{subfigure}
  \iftoggle{SINGLE_COL}{\begin{subfigure}[b]{.40\linewidth}}{\begin{subfigure}[b]{1\linewidth}}
    \includegraphics[width=\linewidth]{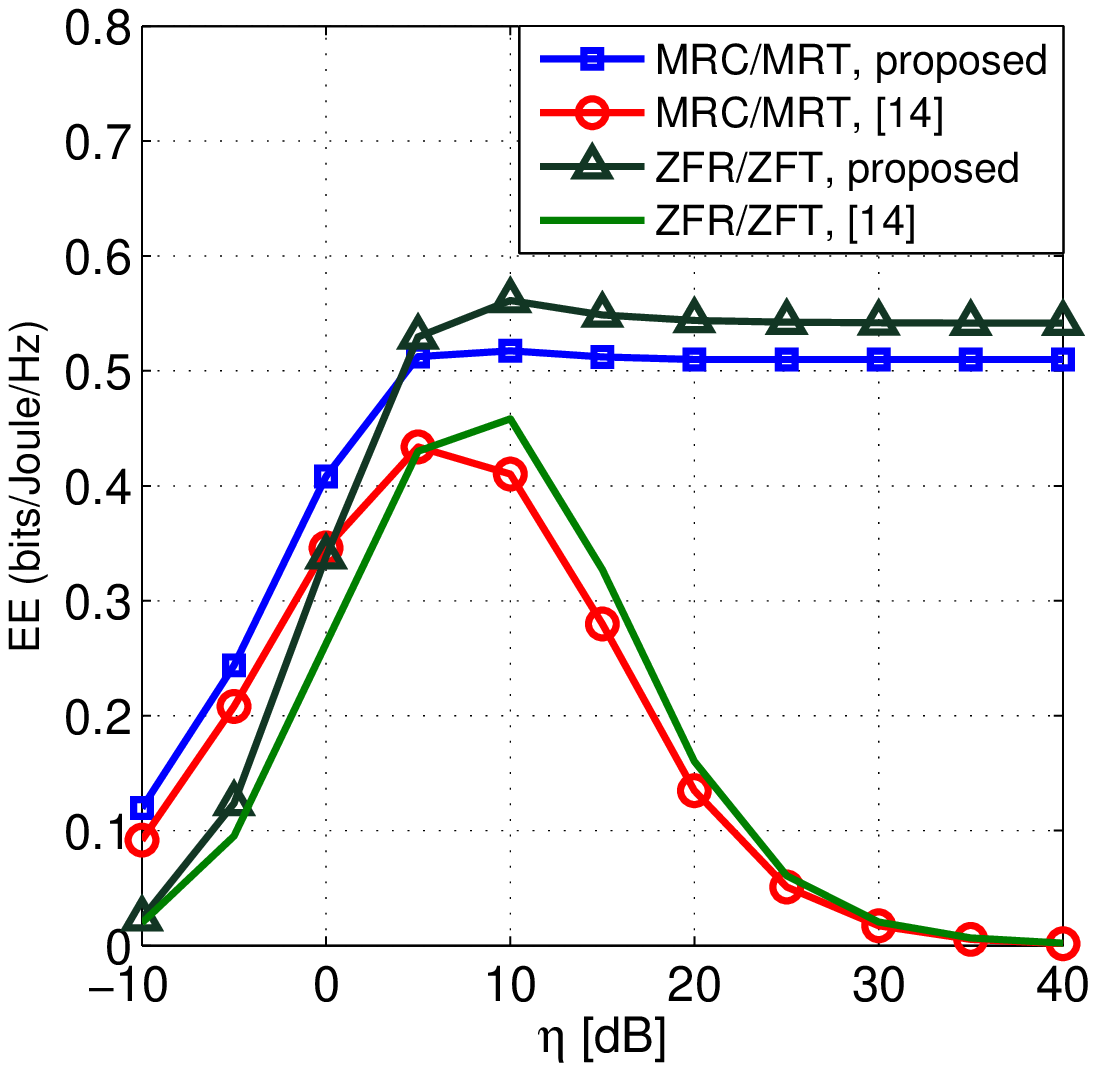}\iftoggle{SINGLE_COL}{\vspace*{-10pt}}{}
    \caption{\small }
    \label{eevsPow_MR_ZF}
  \end{subfigure}\iftoggle{SINGLE_COL}{\vspace*{-15pt}}{}
  \caption{\small EE comparison of MRC/MRT and ZFR/ZFT processing versus  a) $P_{\rho}$; b) $\eta\hspace{-0.05in}=\hspace{-0.05in}P_t^{\max}/\sigma^2$ for $P_{\rho} = 20$~dBm pilot transmit power. For both these figures we have $K=5$ user pairs and $N=500$ relay antennas.}
\label{}
  \end{figure}\iftoggle{SINGLE_COL}{\vspace*{-10pt}}{}
%------------------------------------------------------------ 

%-------------------------------------------------------------------------
%\iftoggle{SINGLE_COL}{\begin{figure}[htbp]}{\begin{figure}[tbp]}
%    \centering\iftoggle{SINGLE_COL}{\vspace*{-5pt}}{}
%    \includegraphics[width=\iftoggle{SINGLE_COL}{.45\linewidth}{1\linewidth}]{Algo1_MaxEE_EEvsPP_MR_ZF_LIR_0_UI_0_Pow_0_20.eps}\iftoggle{SINGLE_COL}{\vspace*{-12pt}}{}
%  \caption{\small {EE versus $P_{\rho}$ comparison of MRC/MRT  and ZFR/ZFT processing, for $K=5$ user pairs and $N=500$ relay antennas.}}
%    \label{eevsNQoSzf_Algo2}
%  \end{figure}
%%------------------------------------------------------------ 

{We plot in Fig.~\ref{eevsPow_MR_ZF} the EE versus $\eta$ for both MRC/MRT and ZFR/ZFT processing. We compare the performance of the proposed EE algorithm to \cite{DBLP:journals/jsac/ZhangCSX16}, where Zhang~\textit{et~al.} have derived the asymptotic SE and EE for two-way amplify-and-forward full-duplex relaying systems assuming that all the $2K$ users are allocated equal power. We see that the proposed algorithm yields a higher EE for both MRC/MRT and ZFR/ZFT than that in\cite{DBLP:journals/jsac/ZhangCSX16}}. We also see that the achieved EE remains constant for $\eta\ge 7$~dB for the MRC/MRT scheme, since with $\eta=7$~dB, the system achieves the maximum EE. After attaining the maximum EE, the system does not use any additional power, since any additional power usage would reduce the EE. We observed this undesired behavior in \cite{DBLP:journals/jsac/ZhangCSX16}, where the system keeps using the available power beyond $\eta=7$~dB and hence the EE reduces. We also see that for $\eta\leq 0$~dB the MRC/MRT has better EE than~ZFR/ZFT.

The  proposed EE optimization algorithm (Algorithm~\ref{algorithm_EE}) can also optimize the EE under QoS constraints that are specified as the rate $r_{k}$ required by each user
\begin{equation}
\log_{2} \left(1+{\mbox{SNR}^{\zeta}_{k}(p_{k},P_{R})}\right)\geq r_{k}.
\end{equation}
It is readily seen that these constraints are posynomial and the GP in Step-3 of the algorithm is still a GP. We now maximize in Fig.~\ref{eevsppQoSzf} the EE for ZFR/ZFT processing under per-user QoS requirements of $0.2$~bps/Hz, $0.5$~bps/Hz and $0.7$~bps/Hz. We see that the QoS constraints of $0.2$~bps/Hz, $0.5$~bps/Hz and $0.7$~bps/Hz cannot be met for $\eta\leq -5$~dB, $\eta\leq 0$~dB and $\eta\leq 5$~dB, respectively. We observe that for $\eta\ge 5$~dB, the EE is the same, both with and without QoS. This implies that the system is capable of satisfying the QoS constraints, while maximizing the~EE.
\iftoggle{SINGLE_COL}{\begin{figure}[htbp]}{\begin{figure}[tbp]}
    \centering\iftoggle{SINGLE_COL}{\vspace*{-5pt}}{}
  \iftoggle{SINGLE_COL}{\begin{subfigure}[b]{.40\linewidth}}{\begin{subfigure}[b]{1\linewidth}}
    \includegraphics[width=\linewidth]{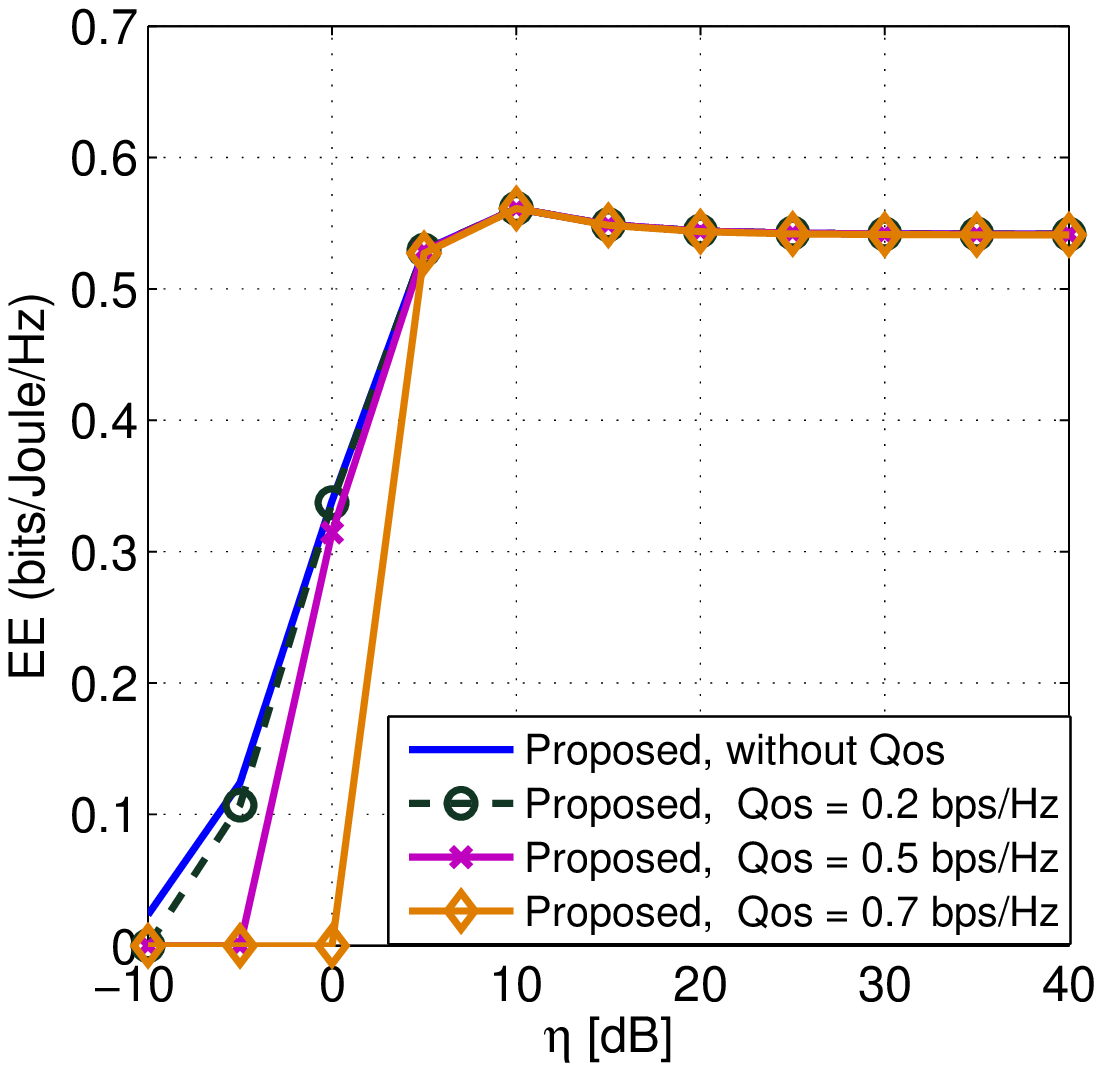}\iftoggle{SINGLE_COL}{\vspace*{-10pt}}{}
    \caption{\small }
    \label{eevsppQoSzf}
  \end{subfigure}
  \iftoggle{SINGLE_COL}{\begin{subfigure}[b]{.40\linewidth}}{\begin{subfigure}[b]{1\linewidth}}
    \includegraphics[width=\linewidth]{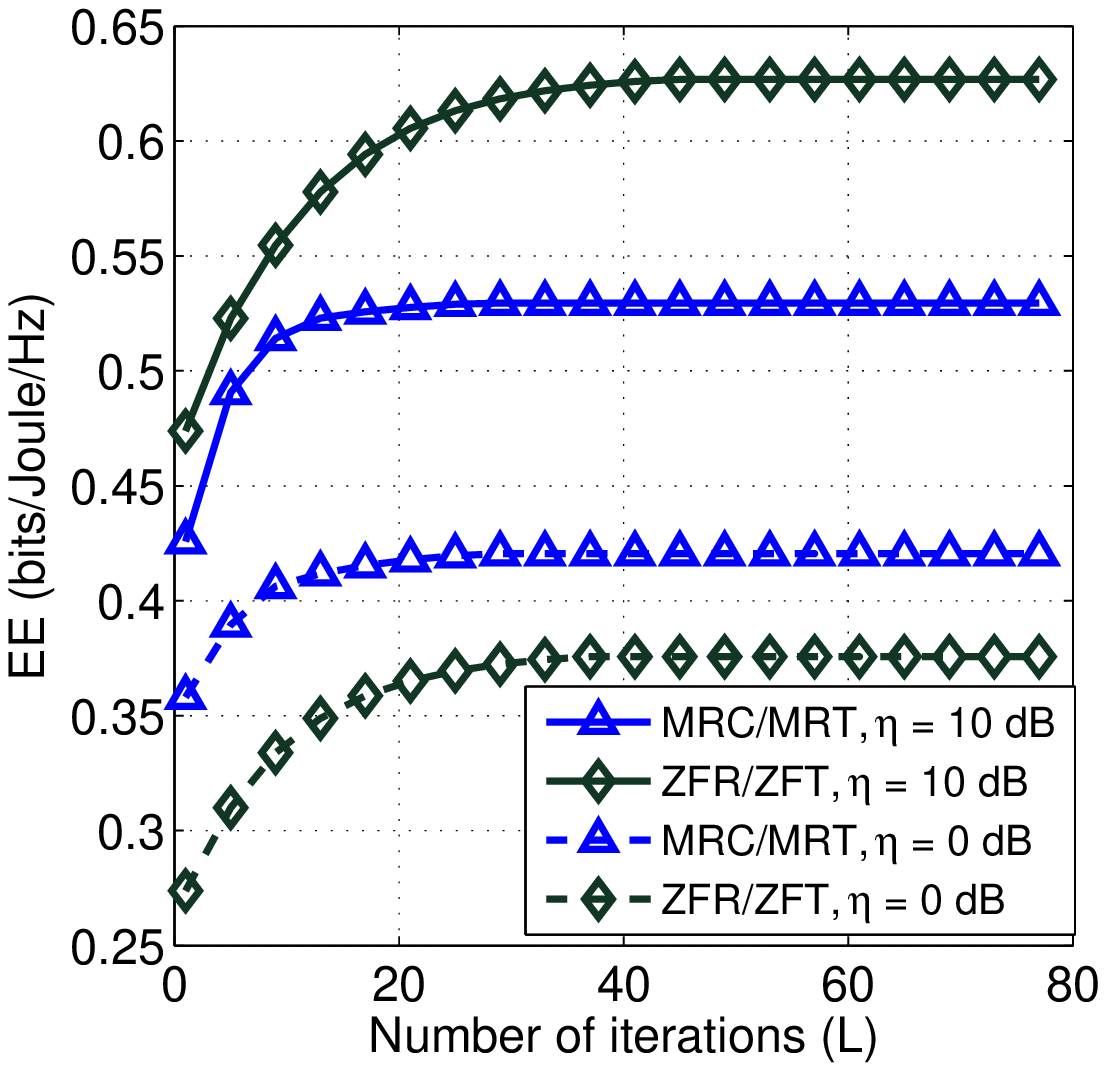}\iftoggle{SINGLE_COL}{\vspace*{-10pt}}{}
    \caption{\small }
    \label{algo_eevsitr}
  \end{subfigure}\iftoggle{SINGLE_COL}{\vspace*{-15pt}}{}
  \caption{\small a) EE versus $\eta\hspace{-0.05in}=\hspace{-0.05in}P_t^{\max}/\sigma^2$ for the proposed optimal power allocation subject to the QoS constraint, considering ZFR/ZFT processing associated with QoS requirements of $0.2$~bps/Hz, $0.5$~bps/Hz and $0.7$~bps/Hz; b) EE versus the number of iterations. For both these figures we have $K=5$, $N=500$, $P_{\rho} = 20$~dBm}
\label{}
  \end{figure}\iftoggle{SINGLE_COL}{\vspace*{-5pt}}{}
%------------------------------------------------------------ 
%---------------------------------------------------------------
\iftoggle{SINGLE_COL}{\vspace{-0.30in}}{}
\subsection{{Convergence of the proposed algorithm}}\iftoggle{SINGLE_COL}{\vspace*{-0.10in}}{}
%---------------------------------------------------------------
We now investigate the number of iterations required for the algorithm to converge for two different $\eta$ values, namely for $\eta=0$~dB, and $\eta=10$~dB. We observe from Fig.~\ref{algo_eevsitr} that the algorithm converges within $20$  iterations for MRC/MRT and within $40$ iterations for ZFR/ZFT. The algorithm has a similar convergence behavior for other $\eta$ values as well.
%%------------------------------------------------------------ 
%\begin{figure}[htp]
%  \begin{center}
%    \iftoggle{SINGLE_COL}{\iftoggle{SINGLE_COL}{\vspace*{-0.1in}}{}\includegraphics[width=0.45\linewidth]{Algo1_MaxEE_EEvsITR_MR_ZF_LIR_0_UI_0_Pow_0_10.eps}}{\includegraphics[width=1\linewidth]{Algo1_MaxEE_EEvsITR_MR_ZF_LIR_0_UI_0_Pow_0_10.eps}}\iftoggle{SINGLE_COL}{\vspace*{-20pt}}{}
%  \caption{\small {EE versus the number of iterations for  $K=5$, $N=500$, and $P_{\rho}=20\mbox{ dBm}$}.}
%  \label{algo_eevsitr}  
%  \end{center}
%    \end{figure}
%%------------------------------------------------------------ 
\iftoggle{SINGLE_COL}{\vspace*{-0.25in}}{}
\subsection{EE optimization under the max-min approach}\iftoggle{SINGLE_COL}{\vspace*{-0.1in}}{}
We now evaluate the performance of the max-min algorithm for ZFR/ZFT processing. For this study, we set both the self-loop interference $\sigma_{LIR}^{2}$ and the inter-user interference $\sigma_{UI}^{2}$ to be $5$~dB below $\sigma^2$. Again, we have to decide the pilot transmission power $P_\rho$, which affects the channel estimation. To decide its value, we first solve our max-min optimization problem by varying $P_\rho$ from $-10$ dBm to $40$ dBm for a fixed $\eta=10$~dB. We observe from Fig.~\ref{eevspp_ZF_algo2} that $P_\rho=20$~dBm attains the maximum EE for the worst user. We therefore set $P_{\rho}=20$~dBm for the max-min analysis. The max EE provided by the proposed optimization is higher than \cite{DBLP:journals/jsac/ZhangCSX16} for all values of pilot power. We see from Fig.~\ref{eevsNPA_ZF_algo2} that the proposed algorithm  significantly improves the EE of the worst user, when compared to equal power allocation. We also see that the system is capable of satisfying the QoS of $0.5$ bps/Hz above $\eta=10$~dB without compromising the EE of the worst~user. %-------------------------------------------------------------------------
%\iftoggle{SINGLE_COL}{\begin{figure}[htbp]}{\begin{figure}[tbp]}
%    \centering\iftoggle{SINGLE_COL}{\vspace*{-5pt}}{}
%  \iftoggle{SINGLE_COL}{\begin{subfigure}[b]{.45\linewidth}}{\begin{subfigure}[b]{1\linewidth}}
%    \includegraphics[width=\linewidth]{Algo2_MaxEE_EEvsPP_ZF_LIR_ne5_UI_ne5_Pow_5.eps}\iftoggle{SINGLE_COL}{\vspace*{-10pt}}{}
%    \caption{\small }
%    \label{eevspp_ZF_algo2}
%  \end{subfigure}
%  \iftoggle{SINGLE_COL}{\begin{subfigure}[b]{.45\linewidth}}{\begin{subfigure}[b]{1\linewidth}}
%    \includegraphics[width=\linewidth]{Algo2_MaxEE_EEvsPow_ZF_LIR_ne5_UI_ne5_EPA_QoS}\iftoggle{SINGLE_COL}{\vspace*{-10pt}}{}
%    \caption{\small }
%    \label{eevsNPA_ZF_algo2}
%  \end{subfigure}\iftoggle{SINGLE_COL}{\vspace*{-15pt}}{}
%  \caption{\small Minimum EE for both equal and optimal power allocation for ZFR/ZFT processing a) versus $P_{\rho}$, where $\eta = 10$~dB; and b) versus $\eta$, where $P_{\rho} = 20$~dBm. For both these figures we have $K=5$ and $N=500$.}
%\label{}
%  \end{figure}
%%------------------------------------------------------------ 
%-------------------------------------------------------------------------
\iftoggle{SINGLE_COL}{\begin{figure}[htbp]}{\begin{figure}[tbp]}
    \centering\iftoggle{SINGLE_COL}{\vspace*{-8pt}}{}
  \iftoggle{SINGLE_COL}{\begin{subfigure}[b]{.32\linewidth}}{\begin{subfigure}[b]{0.8\linewidth}}
    \includegraphics[width=\linewidth]{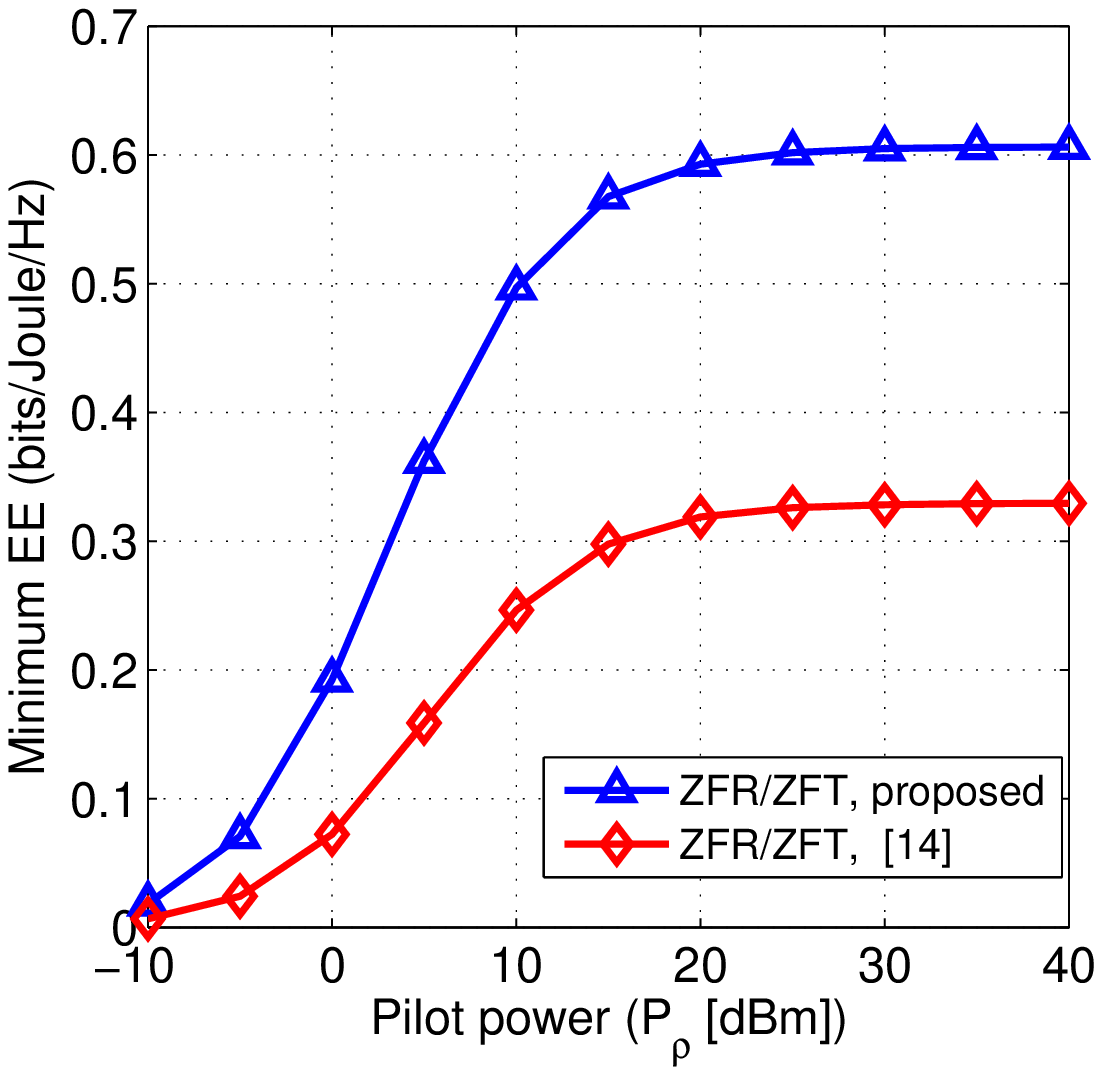}\iftoggle{SINGLE_COL}{\vspace*{-10pt}}{}
    \caption{\small }
    \label{eevspp_ZF_algo2}
  \end{subfigure}
  \iftoggle{SINGLE_COL}{\begin{subfigure}[b]{.33\linewidth}}{\begin{subfigure}[b]{0.8\linewidth}}
    \includegraphics[width=\linewidth]{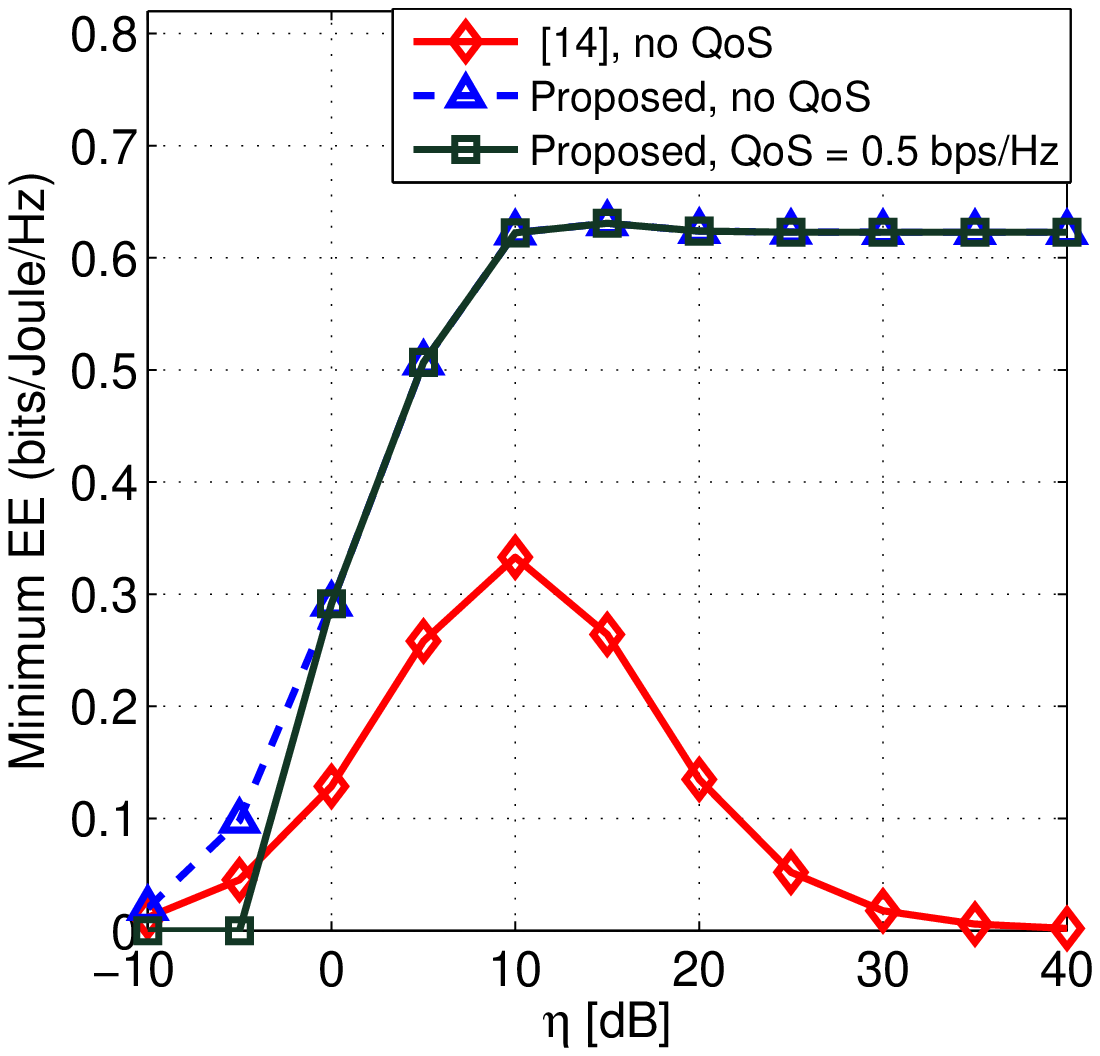}\iftoggle{SINGLE_COL}{\vspace*{-10pt}}{}
    \caption{\small }
    \label{eevsNPA_ZF_algo2}
  \end{subfigure}
\iftoggle{SINGLE_COL}{\begin{subfigure}[b]{.33\linewidth}}{\begin{subfigure}[b]{0.8\linewidth}}
    \includegraphics[width=\linewidth]{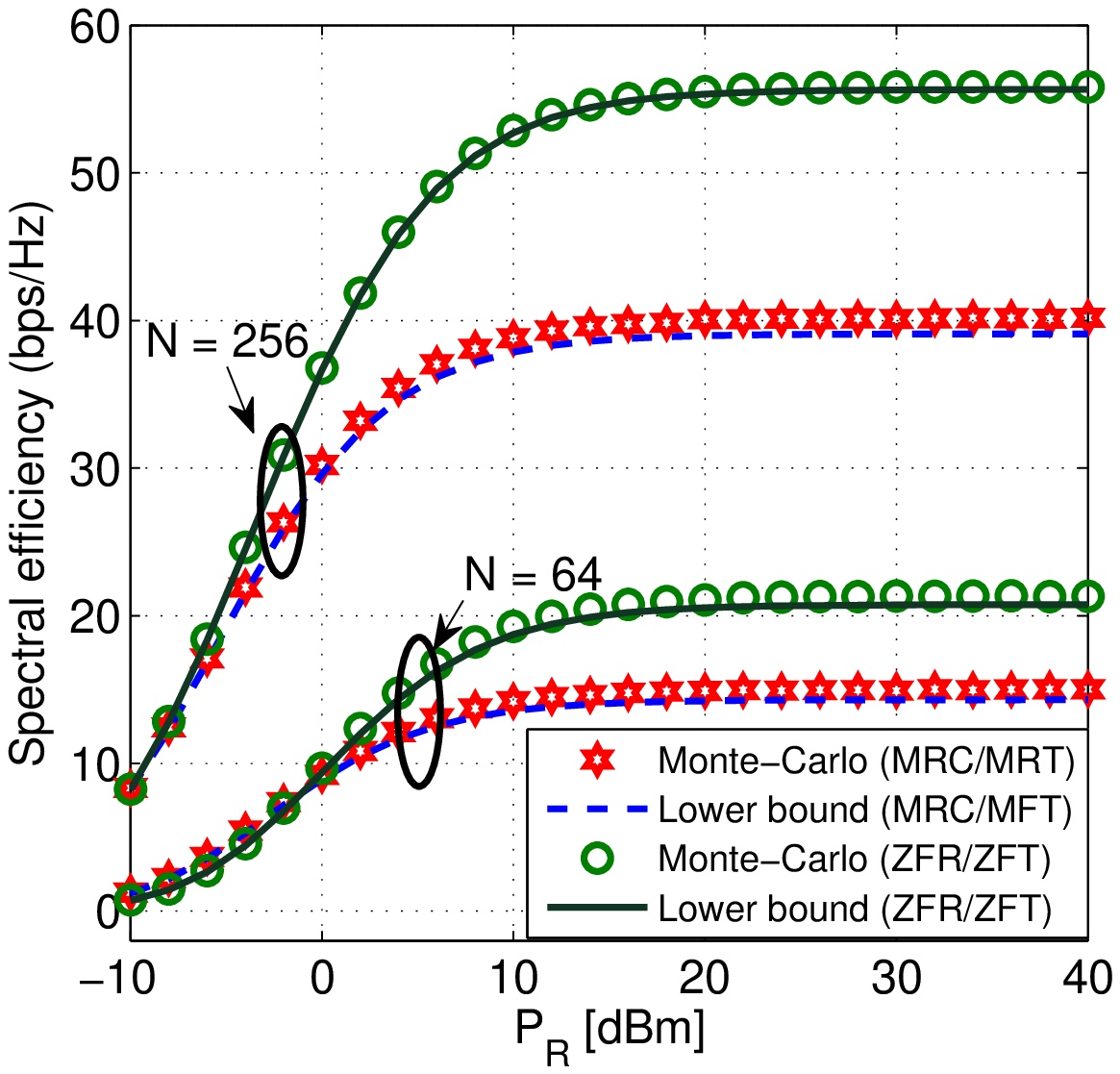}\iftoggle{SINGLE_COL}{\vspace*{-10pt}}{}
    \caption{\small }
    \label{sevssnrmrczf}
  \end{subfigure}  
  \iftoggle{SINGLE_COL}{\vspace*{-35pt}}{}
  \caption{\small Minimum EE for both equal and optimal power allocation for ZFR/ZFT processing a) versus $P_{\rho}$, where $\eta = 10$~dB; and b) versus $\eta$, where $P_{\rho} = 20$~dBm. For both these figures we have $K=5$ and $N=500$; c) SE efficiency of MRC/MRT and ZFR/ZFT versus $P_{R}$, where we have $K=10$, $P_{\rho}=10$~dBm.}
\label{}
  \end{figure}\iftoggle{SINGLE_COL}{\vspace*{-12pt}}{}

 In this paper, we also derived the analytical lower bounds of the achievable rate for both MRC/MRT and ZFR/ZFT processing with MMSE channel estimation which were formulated in Theorem~\ref{theorem1} and Theorem~\ref{theorem2}. In this section, we compare the SE\footnote{Recall from \eqref{R} that SE is obtained by scaling the achievable rate with pilot overhead.} lower bounds with their exact expressions in \eqref{sumexact} for $N=64$ and $N=256$ relay antennas and for $K=10$ user pairs.  We set $P_{\rho}=10$~dBm and allocate equal power to all users i.e. we have $p_{i}=P_{R}/2K,\forall i=1,2,3,\cdots,2K$. {The large scale fading coefficients $\sigma^{2}_{g,i}=\sigma^{2}_{f,i}, \forall i,j=1,2,3,\cdots,2K$, and the self-loop interference $\sigma_{LIR}^{2}$ as well as the  inter-user interference $\sigma_{UI}^{2}=\sigma^2$ are set to $0$~dB with respect to $\sigma^{2}$}.  We see from Fig.~\ref{sevssnrmrczf} that the derived lower bound and the exact expression overlap for ZFR/ZFT processing for $N=256$ relay antennas.  For MRC/MRT, the lower bound marginally differs from the exact expression. We also observe that the SE, of high $P_R$ values saturates for both MRC/MRT and ZFR/ZFT. This is because the relay self-loop interference also increases proportionally upon increasing $P_R$.
\iftoggle{SINGLE_COL}{\vspace*{-0.25in}}{}
\subsection{SE comparison with existing full-duplex designs}\iftoggle{SINGLE_COL}{\vspace*{-0.10in}}{}
We note that the EE maximization framework considered in this work can also maximize the SE by maximizing the numerator of the Problem $\mathbf{P3}$ -- the modified problem is a GP.  We can consequently also maximize the SE both for MRC/MRT and for ZFR/ZFT processing. In contrast to~\cite{Zhang2016Chen}, which derives the lower bound and maximizes the SE (and not the EE and max-min EE), for the MRC/MRT  processing alone, whilst relying on LS channel estimation. The current work is therefore more general than~\cite{Zhang2016Chen}. The objective of this study is to compare the SE of the proposed system to that of the existing full-duplex relaying system in~\cite{Zhang2016Chen}, and  to that of~\cite{DBLP:journals/jsac/ZhangCSX16} where the latter has developed equal-power MRC/MRT and ZFR/ZFT for full-duplex~relaying~designs.
%-------------------------------------------------------------------------
\iftoggle{SINGLE_COL}{\begin{figure}[htbp]}{\begin{figure}[tbp]}
    \centering\iftoggle{SINGLE_COL}{\vspace*{-5pt}}{}
  \iftoggle{SINGLE_COL}{\begin{subfigure}[b]{.40\linewidth}}{\begin{subfigure}[b]{1\linewidth}}
    \includegraphics[width=\linewidth]{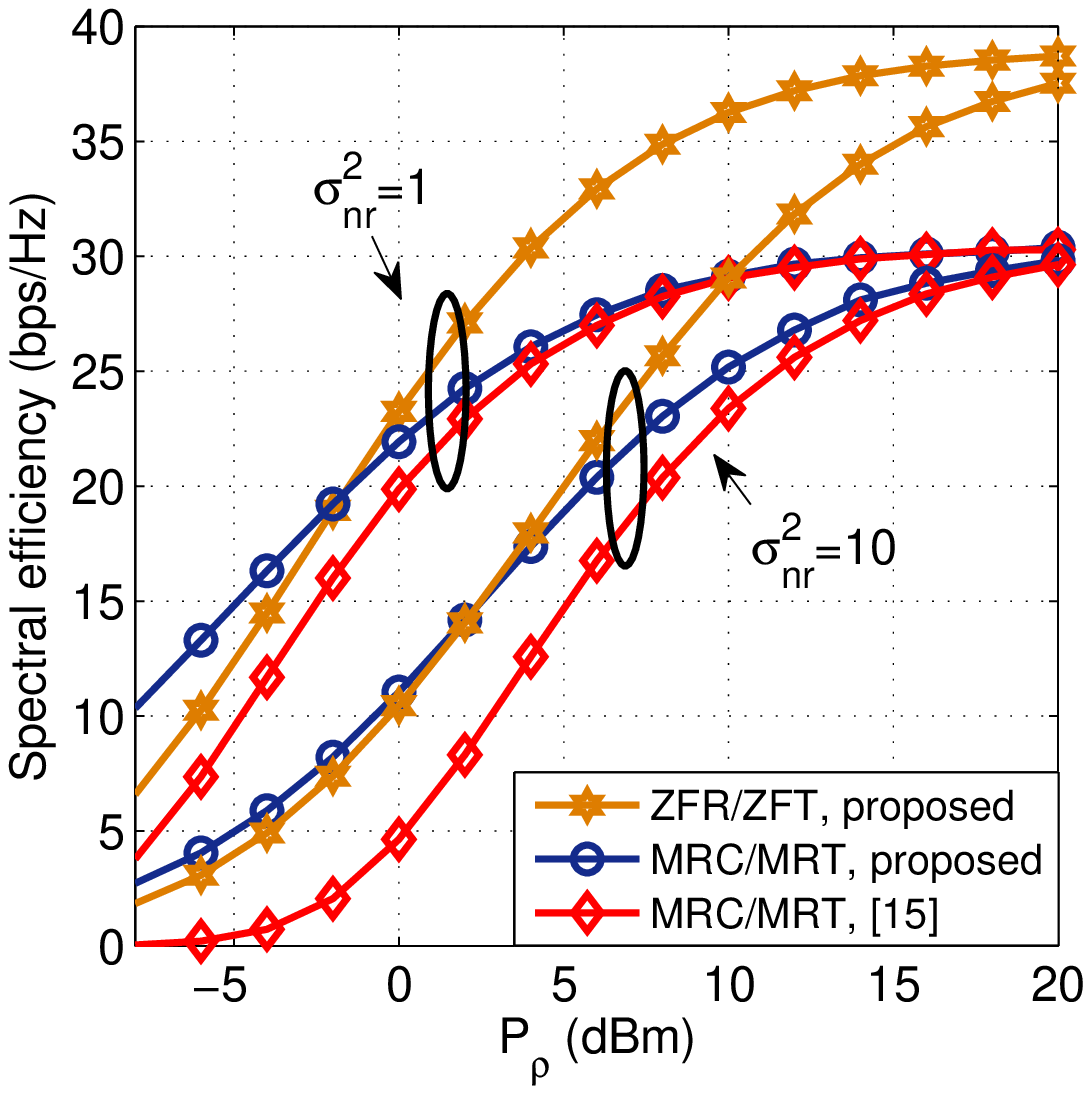}\iftoggle{SINGLE_COL}{\vspace*{-10pt}}{}
    \caption{\small }
    \label{mmse_ls}
  \end{subfigure}
  \iftoggle{SINGLE_COL}{\begin{subfigure}[b]{.40\linewidth}}{\begin{subfigure}[b]{1\linewidth}}
    \includegraphics[width=\linewidth]{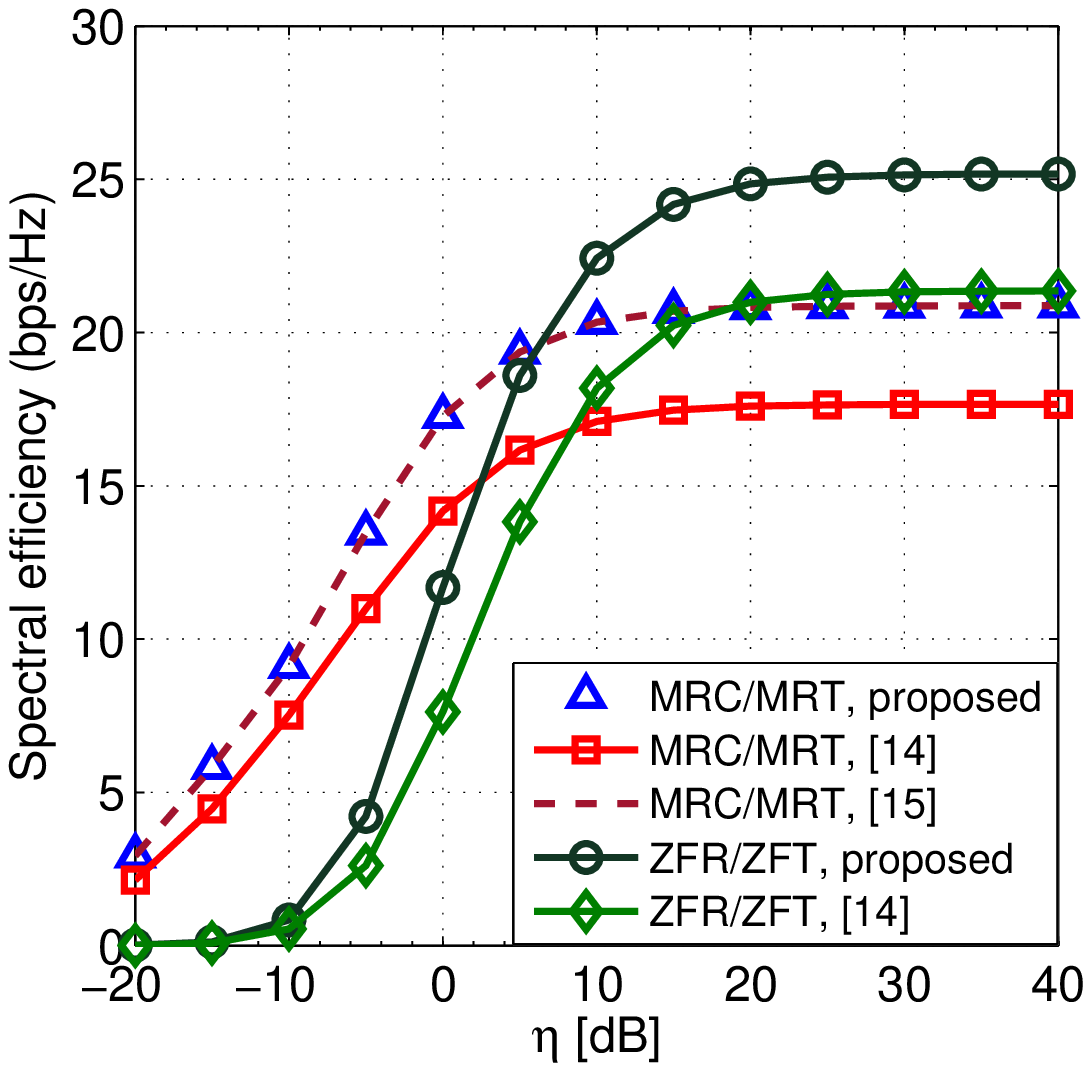}\iftoggle{SINGLE_COL}{\vspace*{-10pt}}{}
    \caption{\small }
    \label{fig:a}
  \end{subfigure}\iftoggle{SINGLE_COL}{\vspace*{-15pt}}{}
  \caption{\small SE of MRC/MRT and ZFR/ZFT versus: a) $P_{\rho}$, comparing the performance of MMSE and LS estimators, where $K=5$, $N=256$ and $P_{R}=10$~dBm; b) $\eta$, where $K=5$, $P_{\rho}=20$~dBm, $N=500$.}
\label{}
  \end{figure}\iftoggle{SINGLE_COL}{\vspace*{-12pt}}{}
%------------------------------------------------------------ 

We commence by comparing the SE to that of the design in \cite{Zhang2016Chen}. We plot in Fig.~\ref{mmse_ls} the SE versus $P_{\rho}$, where we observe that for MRC/MRT processing the MMSE estimator outperforms the LS estimator \cite{Zhang2016Chen} {both} for a lower pilot power $P_{\rho}$, and a higher noise power~$\sigma_{nr}^{2}$. We also see that the ZFR/ZFT {associated} with $P_{\rho}>0$~dB performs much better than the MRC/MRT~processing.

We now evaluate the SE by varying $\eta$ for the proposed algorithm considering both MRC/MRT and ZFR/ZFT processing and compare it to the designs in references \cite{DBLP:journals/jsac/ZhangCSX16,Zhang2016Chen}. We see from Fig.~\ref{fig:a} that the algorithm proposed for both the MRC/MRT and ZFR/ZFT designs considerably improves the SE, when compared to their counterparts from \cite{DBLP:journals/jsac/ZhangCSX16}, which considers an equal power allocation. We also see that the MRC/MRT associated with the proposed algorithm outperforms even ZFR/ZFT for $\eta<5$~dB. Furthermore, the performance of the MRC/MRT processing associated with the proposed algorithm overlaps with that of~\cite{Zhang2016Chen}. This fact can also be justified by observing Fig.~\ref{mmse_ls}, where for $P_{\rho}=20$~dBm, the performance of the proposed system and \cite{Zhang2016Chen} is the same for the MRC/MRT processing.
%Similarly, the SE versus $P_{\rho}$ considering optimal and equal power allocation is shown in Fig.~\ref{fig:b} with $N=500$ antennas.
% observe from Fig.~\ref{fig:a} that the proposed ZFR/ZFT processing outperforms the proposed MRC/MRT for $N\ge 200$.
\iftoggle{SINGLE_COL}{\vspace{-0.25in}}{}
\subsection{Full-duplex versus half-duplex comparison}\iftoggle{SINGLE_COL}{\vspace{-0.05in}}{}
In this section, we compare both the SE and the EE of the proposed full-duplex system to those of the existing half-duplex systems \cite{DBLP:journals/twc/DaiD16,mm_relay_hong} for different values of self-loop interference $\sigma_{LIR}^{2}$ and inter-user interference $\sigma_{UI}^{2}$. 
The objective is to characterize the values of $\sigma_{LIR}^{2}$ and $\sigma_{UI}^{2}$ for which  the proposed full-duplex relaying system {using MMSE channel estimation as well as MRC/MRT and ZFR/ZFT processing} has better SE and EE than their half-duplex counterparts. Reference \cite{mm_relay_hong} derives the asymptotic spectral and energy efficiencies of the half-duplex system by allocating equal power to all users, whereas reference \cite{DBLP:journals/twc/DaiD16}
allocates power to maximize the SE of the half-duplex system. We observe from Fig.~\ref{sevseta_fdhd_mrc} that for the full-duplex system using MRC/MRT processing, the $\sigma_{LIR}^{2}$~dB and $\sigma_{UI}^{2}$~dB should be around $0$~dB for it to achieve a better SE than the half-duplex system with optimal power allocation~\cite{DBLP:journals/twc/DaiD16}. As expected for ZFR/ZFT processing, the full-duplex system requires much higher self-loop interference and inter-user interference suppression. We see from Fig. \ref{sevseta_fdhd_zf} that both $\sigma_{LIR}^{2}$ and $\sigma_{UI}^{2}$ should have values~around~$-10$~dB.

\iftoggle{SINGLE_COL}{\begin{figure}[htbp]}{\begin{figure}[tbp]}
    \centering\iftoggle{SINGLE_COL}{\vspace*{-5pt}}{}
  \iftoggle{SINGLE_COL}{\begin{subfigure}[b]{.33\linewidth}}{\begin{subfigure}[b]{0.8\linewidth}}
    \includegraphics[width=\linewidth]{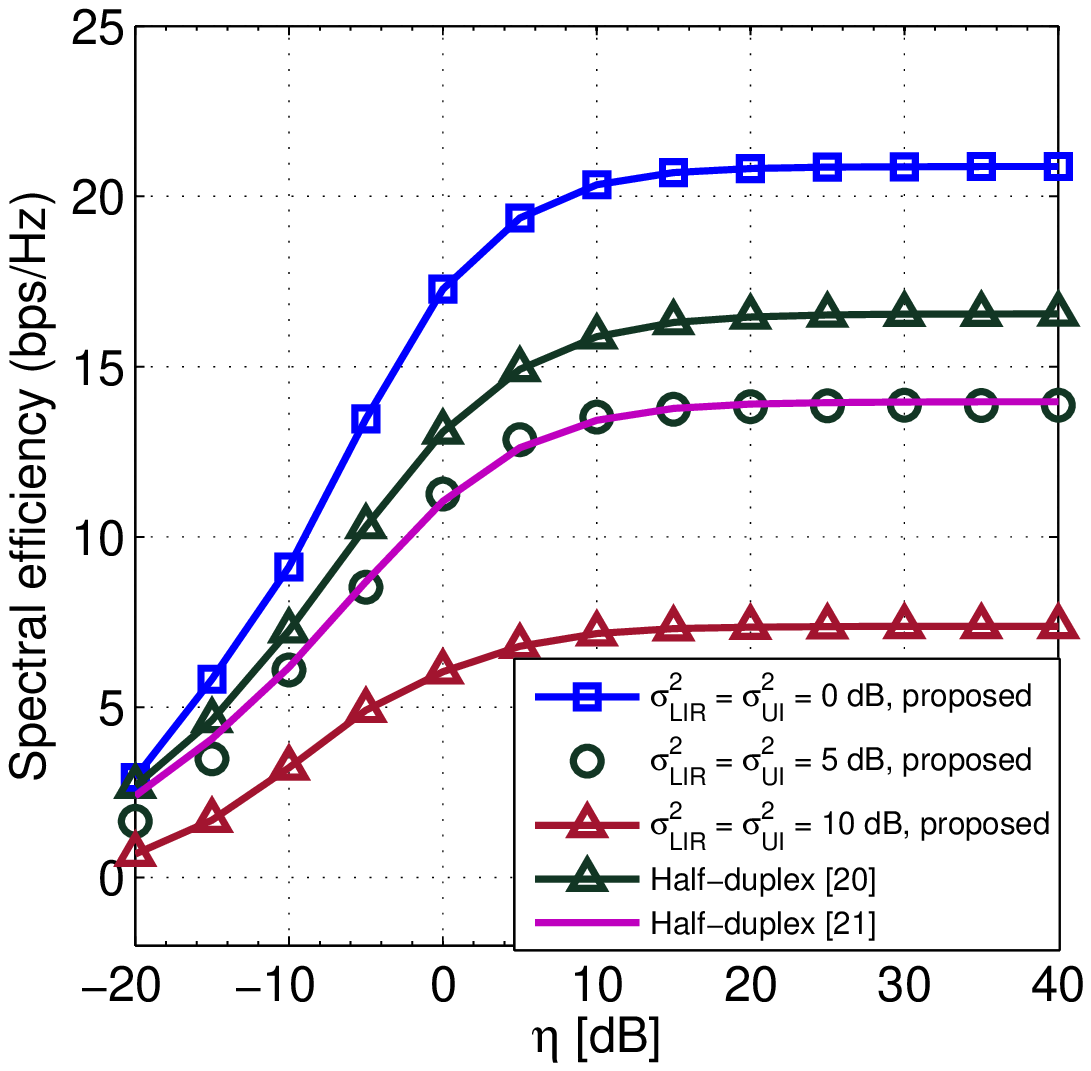}\iftoggle{SINGLE_COL}{\vspace*{-10pt}}{}
    \caption{\small }
    \label{sevseta_fdhd_mrc}
  \end{subfigure}
  \iftoggle{SINGLE_COL}{\begin{subfigure}[b]{.33\linewidth}}{\begin{subfigure}[b]{0.8\linewidth}}
    \includegraphics[width=\linewidth]{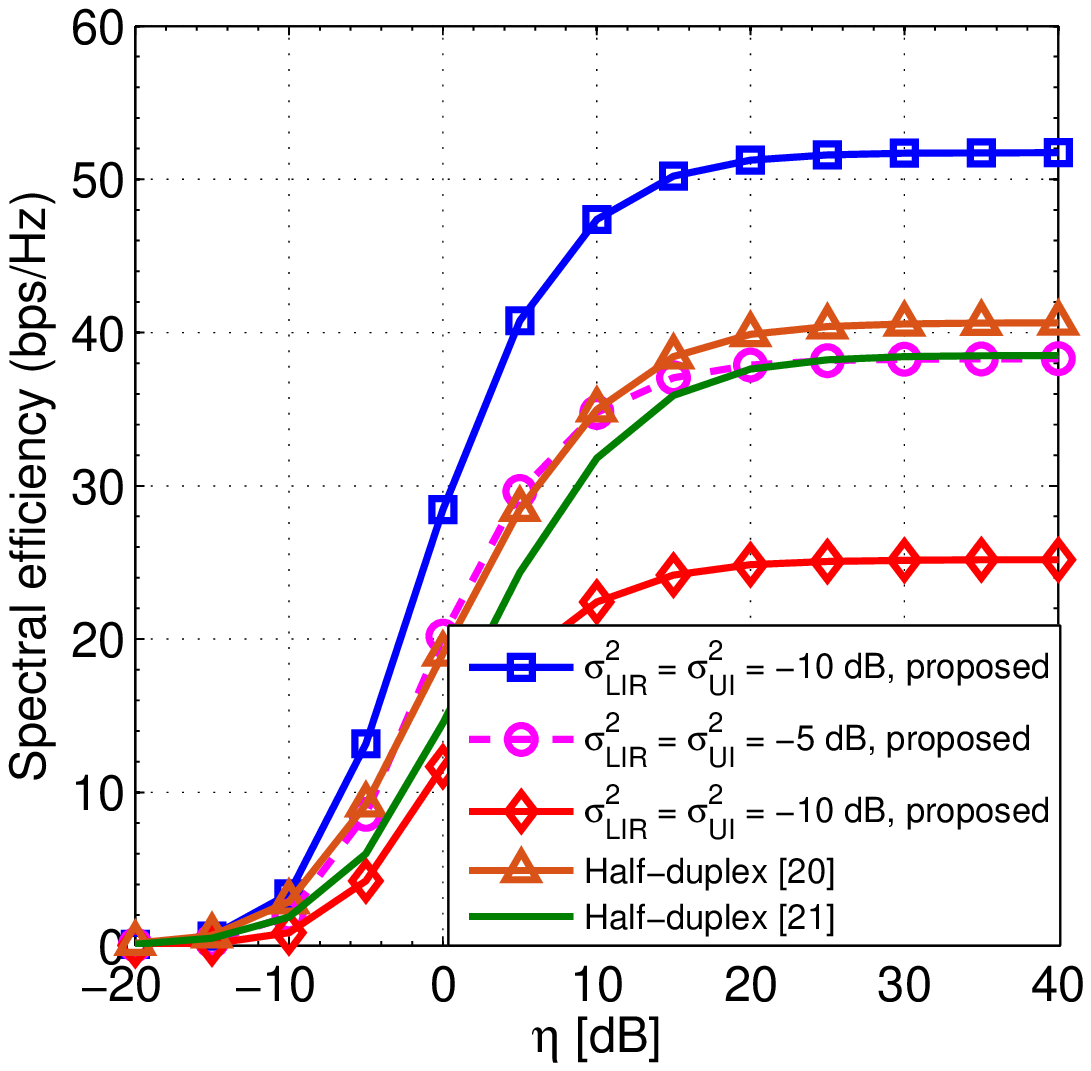}\iftoggle{SINGLE_COL}{\vspace*{-10pt}}{}
    \caption{\small }
    \label{sevseta_fdhd_zf}
  \end{subfigure}
  \iftoggle{SINGLE_COL}{\begin{subfigure}[b]{.32\linewidth}}{\begin{subfigure}[b]{0.8\linewidth}}
    \includegraphics[width=\linewidth]{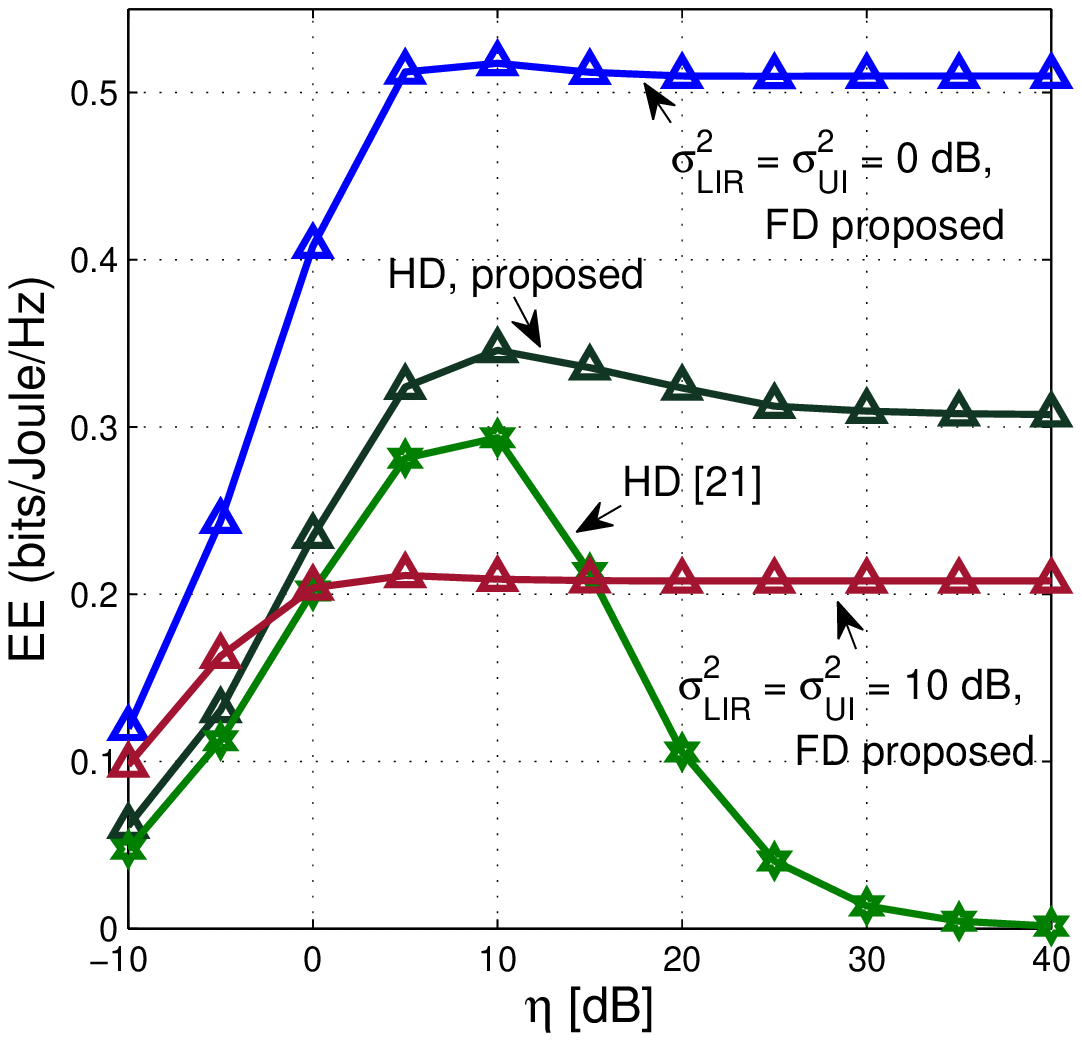}\iftoggle{SINGLE_COL}{\vspace*{-10pt}}{}
    \caption{\small }
    \label{EEvseta_mrc_fdhd}
  \end{subfigure}\iftoggle{SINGLE_COL}{\vspace*{-15pt}}{}
  \caption{\small {Performance comparison of full- and half-duplex systems: a) SE versus $\eta=P_t^{\max}/\sigma^2$ for MRC/MRT processing b) SE versus $\eta=P_t^{\max}/\sigma^2$ for ZFR/ZFT processing; c) EE versus $\eta$ for MRC/MRT processing. For all of these figures we have $K=5$, $N=500$ and $P_{\rho} = 20$~dB.}}
\label{}
  \end{figure}\iftoggle{SINGLE_COL}{\vspace*{-12pt}}{}
%------------------------------------------------------------

%%------------------------------------------------------------ 
%\begin{figure}[htp]
%  \begin{center}
%    \iftoggle{SINGLE_COL}{\iftoggle{SINGLE_COL}{\vspace*{-0.1in}}{}\includegraphics[width=0.45\linewidth]{Algo1_MaxEE_EEvsPow_MR_HDvsFD_OPA_EPA.eps}}{\includegraphics[width=1\linewidth]{Algo1_MaxEE_EEvsPow_MR_HDvsFD_OPA_EPA.eps}}\iftoggle{SINGLE_COL}{\vspace*{-20pt}}{}
%  \caption{\small {EE versus $\eta$ considering MRC/MRT processing, where we have $K=5$, $N=500$, and $P_{\rho}=20\mbox{ dBm}$}.}
%  \label{EEvseta_mrc_fdhd}  
%  \end{center}
%    \end{figure}
%%------------------------------------------------------------ 

We now investigate the EE of full- and half-duplex relaying for different values of $\sigma_{LIR}^{2}$~and $\sigma_{UI}^{2}$ in conjunction with MRC/MRT processing.  We note that the EE of an AF half-duplex massive MIMO relaying has not been investigated in the literature. In the proposed framework, the EE of half-duplex massive MIMO relaying can be evaluated by setting the self-loop interference  and inter-user interference to zero. We observe from Fig.~\ref{EEvseta_mrc_fdhd} that for $\sigma_{LIR}^{2}\hspace{-0.06in}=\hspace{-0.06in}\sigma_{UI}^{2}\hspace{-0.06in}=\hspace{-0.06in}0$~dB the full duplex system has higher EE than a half-duplex system. However, the half-duplex performs better for $\sigma_{LIR}^{2}=\sigma_{LIR}^{2}=10$~dB. The proposed EE power allocation algorithm, which works for a half-duplex system, has better EE than the equal-power half-duplex MRC/MRT scheme~of~\cite{mm_relay_hong}.
%The performance of \cite{DBLP:journals/jsac/ZhangCSX16,mm_relay_hong} is always inferior to that of the proposed system, as they do not allocate power to maximize the EE.
%Reference \cite{DBLP:journals/jsac/ZhangCSX16} investigated the EE with equal power allocation for all users.
\iftoggle{SINGLE_COL}{\vspace*{-0.5in}}{}
\section{Conclusion}
\label{conclude_ref}\iftoggle{SINGLE_COL}{\vspace*{-0.1in}}{}
We considered a multi-pair AF full-duplex massive MIMO two-way relay associated with full-duplex users. We derived closed-form lower bounds for the achievable rate of both MRC/MRT and ZFR/ZFT relay processing in conjunction with MMSE channel estimation and an arbitrary number of relay antennas.  We numerically showed that these lower bounds match with their exact expressions. Furthermore, we used these bounds to design algorithms for jointly allocating power for the users and the relay based on the EE maximization and max-min fairness criterion.
%maximize the EE, and the EE of the worst user, subject to peak power constraints. 
We investigated the performance of both algorithms, with and without QoS constraints, and demonstrated their improved performance over existing full-duplex systems. We exhaustively compared the spectral and energy efficiencies of the proposed relaying system to the existing state-of-the-art half-duplex relaying systems, and characterized the values of self-loop and inter-user interference, for which the proposed system becomes capable of outperforming the~existing~ones.

%SE and EE of the proposed processing with the existing full-duplex and half-duplex relaying systems and showed its improved performance.

%MRC/MRT and ZFR/ZFT processing, and provided the number of relay antenna and pilot transmission power for the scenario when MRC/MRT outperforms~ZFR/ZFT~and~vice-versa.
\appendices
\iftoggle{SINGLE_COL}{}{
\begin{figure*}
\normalsize
% Store the current equation number.
\setcounter{mytempeqncnt}{\value{equation}}
\setcounter{equation}{41}
\begin{align}
&\mbox{Tr}\left\{\mathbb{E}\left[\hat{\mathbf{G}}\mathbf{P}\mathbf{P}^{H}\hat{\mathbf{G}}^{H}\hat{\mathbf{G}}\mathbf{T}^{H}\hat{\mathbf{F}}^{T}\hat{\mathbf{F}}^{*}\mathbf{T}\hat{\mathbf{G}}^{H}\right]\right\} \hspace{-0.1in}\,\,\,\,\stackrel{(a)}{=} \mbox{Tr}\left\{\sum\limits_{k=1}^{2K}{p}_{k}\mathbb{E}\left[\hat{\mathbf{g}}_{k}\hat{\mathbf{g}}_{k}^{H}\sum\limits_{i=1}^{K}\left(\hat{\mathbf{g}}_{2i-1}\hat{\mathbf{f}}_{2i}^{T}+\hat{\mathbf{g}}_{2i}\hat{\mathbf{f}}_{2i-1}^{T}\right)\left(\hat{\mathbf{f}}_{2i-1}^{*}\hat{\mathbf{g}}_{2i}^{H}+\hat{\mathbf{f}}_{2i}^{*}\hat{\mathbf{g}}_{2i-1}^{H}\right)\right]\right\}\label{derp1}\\
&\stackrel{(b)}{=}N^{2}\sum\limits_{k=1,k\neq 2i,2i-1}^{2K}p_{k}\sum\limits_{i=1}^{K}\hat\sigma_{g,k}^{2}\left(\hat\sigma_{g,2i}^{2}\hat\sigma_{f,2i-1}^{2}+\hat\sigma_{g,2i-1}^{2}\hat\sigma_{f,2i}^{2}\right)+N^{2}\sum\limits_{i=1}^{K}p_{2i}\hat\sigma_{g,2i}^{2}\left(\hat\sigma_{g,2i-1}^{2}\hat\sigma_{f,2i}^{2}+(N+1)\hat\sigma_{g,2i}^{2}\hat\sigma_{f,2i-1}^{2}\right)\nonumber\\
&\qquad+N^{2}\sum\limits_{i=1}^{K}p_{2i-1}\hat\sigma_{g,2i-1}^{2}\left(\hat\sigma_{g,2i}^{2}\hat\sigma_{f,2i-1}^{2}+(N+1)\hat\sigma_{g,2i-1}^{2}\hat\sigma_{f,2i}^{2}\right)\nonumber\\
&\stackrel{(c)}{=} N^{2}\sum\limits_{k=1,k\neq 2i,2i-1}^{2K}\sum\limits_{i=1}^{K}p_{k}\hat\sigma_{g,k}^{2}\hat{\Phi}_{i}+N^{2}\sum\limits_{i=1}^{K}\left(p_{2i-1}\hat\sigma_{g,2i-1}^{2}+p_{2i}\hat\sigma_{g,2i}^{2}\right)\hat{\Phi}_{i}+ N^{3}\sum\limits_{i=1}^{K}\left(p_{2i-1}\hat\sigma_{g,2i-1}^{4}\hat\sigma_{f,2i}^{2}+p_{2i}\hat\sigma_{g,2i}^{4}\hat\sigma_{f,2i-1}^{2}\right)\nonumber\\
&\stackrel{(d)}{=}N^{2}\sum\limits_{j=1,j\neq i}^{K}\sum\limits_{i=1}^{K}\left(p_{2j-1}\hat\sigma_{g,2j-1}^{2}+p_{2j}\hat\sigma_{g,2j}^{2}\right)\hat{\Phi}_{i}+N^{2}\sum\limits_{i=1}^{K}\left(p_{2i-1}\hat\sigma_{g,2i-1}^{2}+p_{2i}\hat\sigma_{g,2i}^{2}\right)\hat{\Phi}_{i}+ N^{3}\hat{\Upsilon}\nonumber\\
&\stackrel{(e)}{=} N^{2}\sum\limits_{j=1}^{K}\sum\limits_{i=1}^{K}\left(p_{2j-1}\hat\sigma_{g,2j-1}^{2}+p_{2j}\hat\sigma_{g,2j}^{2}\right)\hat{\Phi}_{i}+ N^{3}\hat{\Upsilon}= N^{2}\sum\limits_{j=1}^{K}\sum\limits_{i=1}^{K}\hat{\Psi}_{j}\hat{\Phi}_{i}+ N^{3}\hat{\Upsilon}= N^{2}\hat{\Psi}\hat{\Phi}+ N^{3}\hat{\Upsilon}\label{derp1simp}\setcounter{equation}{45}
\end{align}
\setcounter{equation}{39}
\hrulefill
\end{figure*}}
\iftoggle{SINGLE_COL}{\vspace{-0.25in}}{}
\section{}\label{alpha_mrc_append}\iftoggle{SINGLE_COL}{\vspace{-0.15in}}{}
The term $\mathbb{E}\left[\|\mathbf{W}\tilde{\mathbf{G}}\mathbf{x}\|^{2}\right]$ in the denominator of \eqref{alpha} can be expressed using (\ref{wmr}) as
\begin{align}
\iftoggle{SINGLE_COL}{}{&}\mathbb{E}\left[\left\|\hat{\mathbf{F}}^{*}\mathbf{T}\hat{\mathbf{G}}^{H}\tilde{\mathbf{G}}\mathbf{x}\right\|^{2}\right]\nonumber\iftoggle{SINGLE_COL}{}{\\
&}\iftoggle{BIG_EQUATION}{}{&}=\mbox{Tr}\left\{\mathbb{E}\left[(\hat{\mathbf{F}}^{*}\mathbf{T}\hat{\mathbf{G}}^{H}{\mathbf{G}}\mathbf{P}\mathbf{P}^{H}{\mathbf{G}}^{H}\hat{\mathbf{G}}\mathbf{T}^{H}\hat{\mathbf{F}}^{T})\right]\right\}\nonumber\iftoggle{SINGLE_COL}{\\[-8pt]}{\\}
&= \mbox{Tr}\left\{\mathbb{E}\left[\hat{\mathbf{F}}^{*}\mathbf{T}\hat{\mathbf{G}}^{H}(\hat{\mathbf{G}}+\mathbf{E}_{g})\mathbf{P}\mathbf{P}^{H}(\hat{\mathbf{G}}^{H}+\mathbf{E}_{g}^{H})\hat{\mathbf{G}}\mathbf{T}^{H}\hat{\mathbf{F}}^{T}\right]\right\}\nonumber\iftoggle{SINGLE_COL}{\\[-8pt]}{\\}
&\stackrel{(a)}{=} \mbox{Tr}\left\{\mathbb{E}\left[\hat{\mathbf{G}}\mathbf{P}\mathbf{P}^{H}\hat{\mathbf{G}}^{H}\hat{\mathbf{G}}\mathbf{T}^{H}\hat{\mathbf{F}}^{T}\hat{\mathbf{F}}^{*}\mathbf{T}\hat{\mathbf{G}}^{H}\right]\right\}\label{eq01}\iftoggle{SINGLE_COL}{\\[-8pt]}{\\}
&+ \mbox{Tr}\left\{\mathbb{E}\left[{\mathbf{E}_{g}}\mathbf{P}\mathbf{P}^{H}{\mathbf{E}_{g}}^{H}\hat{\mathbf{G}}\mathbf{T}^{H}\hat{\mathbf{F}}^{T}\hat{\mathbf{F}}^{*}\mathbf{T}\hat{\mathbf{G}}^{H}\right]\right\}.\label{eq02}
\end{align}
The {equality} in $(a)$ holds, because $\hat{\mathbf{G}}$ and ${\mathbf{E}}_{g}$ are independent and $\mbox{Tr}\left(\mathbf{A}\mathbf{B}\right)=\mbox{Tr}\left(\mathbf{B}\mathbf{A}\right)$. We now want to simplify (\ref{eq01}) and (\ref{eq02}).
We first expand (\ref{eq01}) as in \iftoggle{SINGLE_COL}{}{(\ref{derp1}) (shown at the top of next page).}
\iftoggle{BIG_EQUATION}{}{
\begin{align}
&\mbox{Tr}\left\{\mathbb{E}\left[\hat{\mathbf{G}}\mathbf{P}\mathbf{P}^{H}\hat{\mathbf{G}}^{H}\hat{\mathbf{G}}\mathbf{T}^{H}\hat{\mathbf{F}}^{T}\hat{\mathbf{F}}^{*}\mathbf{T}\hat{\mathbf{G}}^{H}\right]\right\} \hspace{-0.1in}\,\,\,\,\nonumber\iftoggle{SINGLE_COL}{\\[-8pt]}{\\}
&= \mbox{Tr}\left\{\sum\limits_{k=1}^{2K}{p}_{k}\mathbb{E}\left[\hat{\mathbf{g}}_{k}\hat{\mathbf{g}}_{k}^{H}\sum\limits_{i=1}^{K}\left(\hat{\mathbf{g}}_{2i-1}\hat{\mathbf{f}}_{2i}^{T}+\hat{\mathbf{g}}_{2i}\hat{\mathbf{f}}_{2i-1}^{T}\right)\left(\hat{\mathbf{f}}_{2i-1}^{*}\hat{\mathbf{g}}_{2i}^{H}+\hat{\mathbf{f}}_{2i}^{*}\hat{\mathbf{g}}_{2i-1}^{H}\right)\right]\right\}.\label{derp1}
\end{align}}
To simplify (\ref{derp1}), we decompose the summation therein for  $k \neq 2i \text{  or } 2i-1$, $k = 2i-1$ and $k = 2i$. For these $k$ values, Eq. \eqref{derp1} can respectively be simplified as 
\begin{align}\setcounter{equation}{42}
&\mbox{Tr}\hspace{-0.03in}\left\{\hspace{-0.03in}\mathbb{E}\hspace{-0.03in}\left[\hat{\mathbf{g}}_{k}\hat{\mathbf{g}}_{k}^{H}\hspace{-0.03in}\left(\hspace{-0.03in}\hat{\mathbf{g}}_{2i-1}\hat{\mathbf{f}}_{2i}^{T}+\hat{\mathbf{g}}_{2i}\hat{\mathbf{f}}_{2i-1}^{T}\right)\hspace{-0.06in}\left(\hat{\mathbf{f}}_{2i-1}^{*}\hat{\mathbf{g}}_{2i}^{H}\hspace{-0.03in}+\hspace{-0.03in}\hat{\mathbf{f}}_{2i}^{*}\hat{\mathbf{g}}_{2i-1}^{H}\hspace{-0.03in}\right)\hspace{-0.03in}\right]\right\}\iftoggle{SINGLE_COL}{}{\nonumber\\
&}\hspace{-0.03in}=\hspace{-0.03in}N^{2}\hat\sigma_{g,k}^{2}\hspace{-0.03in}\left(\hspace{-0.03in}\hat\sigma_{g,2i}^{2}\hat\sigma_{f,2i-1}^{2}\hspace{-0.03in}+\hspace{-0.03in}\hat\sigma_{g,2i-1}^{2}\hat\sigma_{f,2i}^{2}\right)\hspace{-0.03in},\label{c1}%\iftoggle{SINGLE_COL}{\\[-5pt]}{\\}
\end{align}
\begin{align}
&\mbox{Tr}{\left\{\mathbb{E}\left[{\hat{\mathbf{g}}_{2i-1}}{\hat{\mathbf{g}}_{2i-1}^{H}}({\hat{\mathbf{g}}}_{2i-1}{\hat{\mathbf{f}}}_{2i}^{T}+{\hat{\mathbf{g}}}_{2i}{\hat{\mathbf{f}}}_{2i-1}^{T})({\hat{\mathbf{f}}}_{2i-1}^{*}{\hat{\mathbf{g}}}_{2i}^{H}+{\hat{\mathbf{f}}}_{2i}^{*}{\hat{\mathbf{g}}}_{2i-1}^{H})\right]\right\}}\nonumber\iftoggle{SINGLE_COL}{\\[-8pt]}{\\}
&\iftoggle{BIG_EQUATION}{}{\hspace{2.5in}}=N^{2}\hat\sigma_{g,2i-1}^{2}\left(\hat\sigma_{g,2i}^{2}\hat\sigma_{f,2i-1}^{2}+(N+1)\hat\sigma_{g,2i-1}^{2}\hat\sigma_{f,2i}^{2}\right),\label{c2}\iftoggle{SINGLE_COL}{\\[-6pt]}{\\}
&\mbox{Tr}{\left\{\mathbb{E}\left[\hat{\mathbf{g}}_{2i}\hat{\mathbf{g}}_{2i}^{H}(\hat{\mathbf{g}}_{2i-1}\hat{\mathbf{f}}_{2i}^{T}+\hat{\mathbf{g}}_{2i}\hat{\mathbf{f}}_{2i-1}^{T})(\hat{\mathbf{f}}_{2i-1}^{*}\hat{\mathbf{g}}_{2i}^{H}+\hat{\mathbf{f}}_{2i}^{*}\hat{\mathbf{g}}_{2i-1}^{H})\right]\right\}}\nonumber\iftoggle{SINGLE_COL}{\\[-6pt]}{\\}
&\iftoggle{BIG_EQUATION}{}{\hspace{2.5in}}=N^{2}\hat\sigma_{g,2i}^{2}\left(\hat\sigma_{g,2i-1}^{2}\hat\sigma_{f,2i}^{2}+(N+1)\hat\sigma_{g,2i}^{2}\hat\sigma_{f,2i-1}^{2}\right).\label{c3}
\end{align}
In deriving the {equalities} in \eqref{c1}, \eqref{c2}, and \eqref{c3}, we exploit the following facts: i) for any arbitrary vectors $\mathbf{x}$ and $\mathbf{y}$, $\mathbf{x}^{*}\mathbf{y}^{T}=\mathbf{y}\mathbf{x}^{H}$; and ii) $\mathbb{E}\left[\hat{\mathbf{g}}_{i}^{H}\hat{\mathbf{g}}_{i}\right] = N\hat\sigma_{g,i}^{2}$, $\mathbb{E}\left[\hat{\mathbf{g}}_{i}^{H}\hat{\mathbf{g}}_{j}\right] = 0$, $\mathbb{E}\left[\left|\hat{\mathbf{g}}_{i}^{H}\hat{\mathbf{g}}_{j}\right|^{2}\right] = N\hat\sigma_{g,i}^{2}\hat{\sigma}_{g,j}^{2}$, $\mathbb{E}\left[\hat{\mathbf{g}}_{i}\hat{\mathbf{g}}_{i}^{H}\hat{\mathbf{g}}_{i}\hat{\mathbf{g}}_{i}^{H}\right] = (N+1)\hat\sigma_{g,i}^{4}\mathbf{I}_{N}$\cite{CIT-001}.
%Using (\ref{c1}), (\ref{c2}) and (\ref{c3}), Eq. (\ref{derp1}) can be simplified to its final expression which is given in \iftoggle{SINGLE_COL}{}{(\ref{derp1simp})}.
\iftoggle{BIG_EQUATION}{}{Furthermore, we can simplify \eqref{derp1} as\setlength{\belowdisplayskip}{0pt}
\begin{align}
&\mbox{Tr}\left\{\mathbb{E}\left[\hat{\mathbf{G}}\mathbf{P}\mathbf{P}^{H}\hat{\mathbf{G}}^{H}\hat{\mathbf{G}}\mathbf{T}^{H}\hat{\mathbf{F}}^{T}\hat{\mathbf{F}}^{*}\mathbf{T}\hat{\mathbf{G}}^{H}\right]\right\} \hspace{-0.1in}\,\,\,\,\nonumber\iftoggle{SINGLE_COL}{\\[-4pt]}{\\}
&\stackrel{(a)}{=} \mbox{Tr}\left\{\sum\limits_{k=1}^{2K}{p}_{k}\mathbb{E}\left[\hat{\mathbf{g}}_{k}\hat{\mathbf{g}}_{k}^{H}\sum\limits_{i=1}^{K}\left(\hat{\mathbf{g}}_{2i-1}\hat{\mathbf{f}}_{2i}^{T}+\hat{\mathbf{g}}_{2i}\hat{\mathbf{f}}_{2i-1}^{T}\right)\left(\hat{\mathbf{f}}_{2i-1}^{*}\hat{\mathbf{g}}_{2i}^{H}+\hat{\mathbf{f}}_{2i}^{*}\hat{\mathbf{g}}_{2i-1}^{H}\right)\right]\right\}\nonumber\iftoggle{SINGLE_COL}{\\[-4pt]}{\\}
&\stackrel{(b)}{=}N^{2}\sum\limits_{k\neq 2i,2i-1}^{2K}p_{k}\sum\limits_{i=1}^{K}\hat\sigma_{g,k}^{2}\left(\hat\sigma_{g,2i}^{2}\hat\sigma_{f,2i-1}^{2}+\hat\sigma_{g,2i-1}^{2}\hat\sigma_{f,2i}^{2}\right)+N^{2}\sum\limits_{i=1}^{K}\left\{p_{2i}\hat\sigma_{g,2i}^{2}\left(\hat\sigma_{g,2i-1}^{2}\hat\sigma_{f,2i}^{2}\right.\right.\nonumber\iftoggle{SINGLE_COL}{\\[-12pt]}{\\}
&\qquad\left.\left. +(N+1)\hat\sigma_{g,2i}^{2}\hat\sigma_{f,2i-1}^{2}\right)\right\}+N^{2}\sum\limits_{i=1}^{K}p_{2i-1}\hat\sigma_{g,2i-1}^{2}\left(\hat\sigma_{g,2i}^{2}\hat\sigma_{f,2i-1}^{2}+(N+1)\hat\sigma_{g,2i-1}^{2}\hat\sigma_{f,2i}^{2}\right)\nonumber\iftoggle{SINGLE_COL}{\\[-8pt]}{\\}
&\stackrel{(c)}{=}N^{2}\hspace{-0.1in}\sum\limits_{k\neq 2i,2i-1}^{2K}\hspace{-0.1in}\sum\limits_{i=1}^{K}p_{k}\hat\sigma_{g,k}^{2}\hat{\Phi}_{i}+N^{2}\sum\limits_{i=1}^{K}\left(p_{2i-1}\hat\sigma_{g,2i-1}^{2}+p_{2i}\hat\sigma_{g,2i}^{2}\right)\hat{\Phi}_{i}\nonumber\iftoggle{SINGLE_COL}{\\[-4pt]}{\\}
&\qquad+ N^{3}\sum\limits_{i=1}^{K}\left(p_{2i-1}\hat\sigma_{g,2i-1}^{4}\hat\sigma_{f,2i}^{2}+p_{2i}\hat\sigma_{g,2i}^{4}\hat\sigma_{f,2i-1}^{2}\right)\nonumber\iftoggle{SINGLE_COL}{\\[-6pt]}{\\}
&\stackrel{(d)}{=} N^{2}\sum\limits_{j=1,j\neq i}^{K}\sum\limits_{i=1}^{K}\left(p_{2j-1}\hat\sigma_{g,2j-1}^{2}+p_{2j}\hat\sigma_{g,2j}^{2}\right)\hat{\Phi}_{i}+N^{2}\sum\limits_{i=1}^{K}\left(p_{2i-1}\hat\sigma_{g,2i-1}^{2}+p_{2i}\hat\sigma_{g,2i}^{2}\right)\hat{\Phi}_{i}+ N^{3}\hat{\Upsilon}\nonumber\iftoggle{SINGLE_COL}{\\[-4pt]}{\\}
&\stackrel{(e)}{=} \iftoggle{SINGLE_COL}{\hspace{-0.05in}}{}N^{2}\sum\limits_{j=1}^{K}\iftoggle{SINGLE_COL}{\hspace{-0.05in}}{}\sum\limits_{i=1}^{K}\iftoggle{SINGLE_COL}{\hspace{-0.05in}}{}\left(p_{2j-1}\hat\sigma_{g,2j-1}^{2}\iftoggle{SINGLE_COL}{\hspace{-0.05in}}{}+\iftoggle{SINGLE_COL}{\hspace{-0.05in}}{}p_{2j}\hat\sigma_{g,2j}^{2}\right)\hat{\Phi}_{i}\iftoggle{SINGLE_COL}{\hspace{-0.05in}}{}+\iftoggle{SINGLE_COL}{\hspace{-0.05in}}{} N^{3}\hat{\Upsilon}= N^{2}\sum\limits_{j=1}^{K}\iftoggle{SINGLE_COL}{\hspace{-0.05in}}{}\sum\limits_{i=1}^{K}\hat{\Psi}_{j}\hat{\Phi}_{i}\iftoggle{SINGLE_COL}{\hspace{-0.05in}}{}+\iftoggle{SINGLE_COL}{\hspace{-0.05in}}{} N^{3}\hat{\Upsilon}\iftoggle{SINGLE_COL}{\hspace{-0.05in}}{}= N^{2}\hat{\Psi}\hat{\Phi}+ N^{3}\hat{\Upsilon}\label{derp1simp}.
\end{align}
}
We now discuss the steps used for simplifying \eqref{derp1}. Equality in $(b)$ is obtained by using the results in \eqref{c1}, \eqref{c2} and \eqref{c3}. Equality in  $(c)$ is obtained by defining $\hat{\Phi}_{i}=\left(\hat\sigma_{g,2i-1}^{2}\hat\sigma_{f,2i}^{2}+\hat\sigma_{g,2i}^{2}\hat\sigma_{f,2i-1}^{2}\right)$, and by rearranging the terms. Equality in $(d)$ is obtained by exploiting the fact that \iftoggle{SINGLE_COL}{\\}{} $\sum_{k=1}^{2K}x_{k}=\sum_{j=1}^{K}\left(x_{2j-1}+x_{2j}\right)$, and by defining $\hat{\Upsilon} = \sum_{i=1}^{K}\left(p_{2i-1}\hat\sigma_{g,2i-1}^{4}\hat\sigma_{f,2i}^{2}+p_{2i}\hat\sigma_{g,2i}^{4}\hat\sigma_{f,2i-1}^{2}\right)$. Equality in $(e)$ is obtained by defining $\hat{\Phi}=\sum_{i=1}^{K}\hat{\Phi}_{i}=\sum_{i=1}^{2K}\hat\sigma_{g,i}^{2}\hat\sigma_{f,i^{'}}^{2}$ and $\hat{\Psi} = \sum_{i=1}^{K}\hat{\Psi}_{i}=\sum_{i=1}^{K}\hspace{-0.05in}\left(\hspace{-0.03in}p_{2i-1}\hat\sigma_{g,2i-1}^{2}\hspace{-0.05in}+\hspace{-0.03in}p_{2i}\hat\sigma_{g,2i}^{2}\right)\hspace{-0.05in}=\hspace{-0.05in}\sum_{i=1}^{2K}p_{i}\hat\sigma_{g,i}^{2}$.
By using an approach similar to (\ref{eq01}), we simplify~(\ref{eq02})~as
\begin{align}\setcounter{equation}{46}
&\mbox{Tr}\left\{\mathbb{E}\left[{\mathbf{E}_{g}}\mathbf{P}\mathbf{P}^{H}{\mathbf{E}_{g}}^{H}\hat{\mathbf{G}}\mathbf{T}^{H}\hat{\mathbf{F}}^{T}\hat{\mathbf{F}}^{*}\mathbf{T}\hat{\mathbf{G}}^{H}\right]\right\}\iftoggle{SINGLE_COL}{}{\nonumber\\
&}=\mbox{Tr}\left\{\mathbb{E}\left[{\mathbf{E}_{g}}\mathbf{P}\mathbf{P}^{H}\mathbf{E}_{g}^{H}\right]\mathbb{E}\left[\hat{\mathbf{G}}\mathbf{T}^{H}\hat{\mathbf{F}}^{T}\hat{\mathbf{F}}^{*}\mathbf{T}\hat{\mathbf{G}}^{H}\right]\right\} \nonumber\iftoggle{SINGLE_COL}{\\[-4pt]}{\\}
&= \mbox{Tr}\left\{\sum\limits_{k=1}^{2K}{p}_{k}\hat\sigma_{\xi g,k}^{2}\mathbb{E}\left[\sum\limits_{i=1}^{K}\left(\hat{\mathbf{g}}_{2i-1}\hat{\mathbf{f}}_{2i}^{T}+\hat{\mathbf{g}}_{2i}\hat{\mathbf{f}}_{2i-1}^{T}\right)\iftoggle{SINGLE_COL}{}{\right.\right.\nonumber\\
&\hspace{1.0in}\left.\left.}\left(\hat{\mathbf{f}}_{2i-1}^{*}\hat{\mathbf{g}}_{2i}^{H}+\hat{\mathbf{f}}_{2i}^{*}\hat{\mathbf{g}}_{2i-1}^{H}\right)\right]\right\}\nonumber\iftoggle{SINGLE_COL}{\\[-6pt]}{\\}
&= \sum\limits_{k=1}^{2K}{p}_{k}\hat\sigma_{\xi g,k}^{2}\sum\limits_{i=1}^{K}N^{2}\left(\hat\sigma_{g,2i-1}^{2}\hat\sigma_{f,2i}^{2}+\hat\sigma_{g,2i}^{2}\hat\sigma_{f,2i-1}^{2}\right)\iftoggle{SINGLE_COL}{}{\nonumber\\
&}= N^{2}\hat{\Phi}\sum\limits_{k=1}^{2K}{p}_{k}\hat\sigma_{\xi g,k}^{2}.\label{eq02simp}
\end{align}
We now simplify the second term in the denominator of (\ref{alpha}).
\begin{align}
&\mathbb{E}\left[\|\mathbf{W}\mathbf{G}_{RR}\hat{\mathbf{x}}_{R}\|^{2}\right]\iftoggle{SINGLE_COL}{}{\nonumber\\
&}=\frac{P_{R}}{N}\mbox{Tr}\left\{ \mathbb{E}\left[\hat{\mathbf{F}}^{*}\mathbf{T}\hat{\mathbf{G}}^{H}\mathbf{G}_{RR}\mathbf{G}_{RR}^{H}\hat{\mathbf{G}}\mathbf{T}^{H}\hat{\mathbf{F}}^{T}\right]\right\}\nonumber\iftoggle{SINGLE_COL}{\\[-8pt]}{\\}
&=P_{R}\sigma_{LIR}^{2}\sum\limits_{i=1}^{K}N^{2}\left(\hat\sigma_{g,2i-1}^{2}\hat\sigma_{f,2i}^{2}+\hat\sigma_{g,2i}^{2}\hat\sigma_{f,2i-1}^{2}\right)\iftoggle{SINGLE_COL}{}{\nonumber\\
&}=N^{2}P_{R}\sigma_{LIR}^{2}\hat{\Phi}.\label{wgzrsimp}
\end{align}
The last term $\mathbb{E}\left[\|\mathbf{W}\mathbf{z}_{R}\|^{2}\right]$ in the denominator of (\ref{alpha}) can be simplified as 
\begin{align}
&\hspace{-0.15in}\mathbb{E}\left[\|\mathbf{W}\mathbf{z}_{R}\|^{2}\right]= \mbox{Tr}\left\{\sigma_{nr}^{2} \mathbb{E}\left[\hat{\mathbf{F}}^{*}\mathbf{T}\hat{\mathbf{G}}^{H}\hat{\mathbf{G}}\mathbf{T}^{H}\hat{\mathbf{F}}^{T}\right]\right\}\nonumber\iftoggle{SINGLE_COL}{\\[-0pt]}{\\}
&\hspace{-0.15in}=\sigma_{nr}^{2}\sum\limits_{i=1}^{K}N^{2}\left(\hat\sigma_{g,2i-1}^{2}\hat\sigma_{f,2i}^{2}+\hat\sigma_{g,2i}^{2}\hat\sigma_{f,2i-1}^{2}\right)=N^{2}\sigma_{nr}^{2}\hat{\Phi}.\label{wzrsimp}
\end{align}
By substituting the  expression of $\mathbb{E}\left[\|\mathbf{W}\tilde{\mathbf{G}}\mathbf{x}\|^{2}\right]$ from \eqref{derp1simp} as well as \eqref{eq02simp}, and the  expressions of $\mathbb{E}\left[\|\mathbf{W}\mathbf{G}_{RR}\hat{\mathbf{x}}_{R}\|^{2}\right]$ and $\mathbb{E}\left[\|\mathbf{W}\mathbf{z}_{R}\|^{2}\right]$ from (\ref{wgzrsimp}) and (\ref{wzrsimp}) respectively in (\ref{alpha}), we get (\ref{alphamrproof}). 
\iftoggle{SINGLE_COL}{}{
\begin{figure*}
\normalsize
% Store the current equation number.
\setcounter{mytempeqncnt}{\value{equation}}
\setcounter{equation}{51}
\begin{eqnarray}\label{omegasimp}
\mbox{Tr}\left\{\mathbb{E}\left[\hat{\mathbf{\Lambda}}_{F}^{*}\mathbf{T}\hat{\mathbf{\Lambda}}_{G}\mathbf{T}\right]\right\}=\sum_{j=1}^{2K}\left(\mathbb{E}\left[w^{*}_{f,j,j^{'}}\right]\mathbb{E}\left[w_{g,j,j^{'}}\right]+\mathbb{E}\left[w^{*}_{f,j,j}\right]\mathbb{E}\left[w_{g,j^{'},j^{'}}\right]\right)=\sum_{j=1}^{2K}\frac{1}{\left(N-2K-1\right)^{2}\hat\sigma_{f,j}^{2}\hat\sigma_{g,	j^{'}}^{2}}\triangleq\hat{\eta}
\end{eqnarray}
\setcounter{equation}{49}
\hrule
\end{figure*}
}
\iftoggle{SINGLE_COL}{\vspace{-0.25in}}{}
\section{}\label{alpha_zf_append}\iftoggle{SINGLE_COL}{\vspace{-0.05in}}{}
To derive this result, we will first simplify $\mathbb{E}\left[\|\mathbf{W}\tilde{\mathbf{G}}\mathbf{x}\|^{2}\right]$ in the denominator of \eqref{alpha} using~(\ref{wzf}).
\begin{align}
&\mathbb{E}\left[\|\mathbf{W}\tilde{\mathbf{G}}\mathbf{x}\|^{2}\right]=
\mathbb{E}\left[\|\hat{\bar{\mathbf{F}}}^{*}\mathbf{T}\hat{\bar{\mathbf{G}}}^{H}\tilde{\mathbf{G}}\mathbf{x}\|^{2}\right]\iftoggle{SINGLE_COL}{}{\nonumber\\
&}=\mbox{Tr}\left\{\mathbb{E}\left[(\hat{\bar{\mathbf{F}}}^{*}\mathbf{T}\hat{\bar{\mathbf{G}}}^{H}{\mathbf{G}}\mathbf{P}\mathbf{P}^{H}{\mathbf{G}}^{H}\hat{\bar{\mathbf{G}}}\mathbf{T}^{H}\hat{\bar{\mathbf{F}}}^{T})\right]\right\}\nonumber\\
&=\mbox{Tr}\left\{\mathbb{E}\left[(\hat{\bar{\mathbf{F}}}^{*}\mathbf{T}\hat{\bar{\mathbf{G}}}^{H}(\hat{\mathbf{G}}+\mathbf{E}_{g})\mathbf{P}\mathbf{P}^{H}(\hat{\mathbf{G}}^{H}+\mathbf{E}_{g}^{H})\hat{\bar{\mathbf{G}}}\mathbf{T}^{H}\hat{\bar{\mathbf{F}}}^{T})\right]\right\}\nonumber\\
&=\mbox{Tr}\left\{\mathbb{E}\left[\hat{\bar{\mathbf{F}}}^{*}\mathbf{T}\hat{\bar{\mathbf{G}}}^{H}\hat{\mathbf{G}}\mathbf{P}\mathbf{P}^{H}\hat{\mathbf{G}}^{H}\hat{\bar{\mathbf{G}}}\mathbf{T}^{H}\hat{\bar{\mathbf{F}}}^{T}\right]\right\}\iftoggle{SINGLE_COL}{}{\nonumber\\
&} + \mbox{Tr}\left\{\mathbb{E}\left[{\mathbf{E}_{g}}\mathbf{P}\mathbf{P}^{H}{\mathbf{E}_{g}}^{H}\hat{\bar{\mathbf{G}}}\mathbf{T}^{H}\hat{\bar{\mathbf{F}}}^{T}\hat{\bar{\mathbf{F}}}^{*}\mathbf{T}\hat{\bar{\mathbf{G}}}^{H}\right]\right\}\label{wgxp2}\iftoggle{SINGLE_COL}{\\[-0pt]}{\\}
&\stackrel{(a)}{=}\mbox{Tr}\left\{\mathbb{E}\left[\hat{\bar{\mathbf{F}}}^{*}\mathbf{T}\mathbf{P}\mathbf{P}^{H}\mathbf{T}\hat{\bar{\mathbf{F}}}^{T}\right]\right\}\iftoggle{SINGLE_COL}{}{\nonumber\\
&} + \sum_{i=1}^{2K}p_{i}\sigma_{\xi,g,i}^{2}\mbox{Tr}\left\{\mathbb{E}\left[\hat{\bar{\mathbf{G}}}\mathbf{T}^{H}\hat{\bar{\mathbf{F}}}^{T}\hat{\bar{\mathbf{F}}}^{*}\mathbf{T}\hat{\bar{\mathbf{G}}}^{H}\right]\right\}\nonumber\iftoggle{SINGLE_COL}{\\[-0pt]}{\\}
&\stackrel{(b)}{=} \sum_{i=1}^{2K}p_{i^{'}}\mathbb{E}\left[\hat{\bar{\mathbf{f}}}_{i}^{H}\hat{\bar{\mathbf{f}}}_{i}\right]+ \sum_{i=1}^{2K}p_{i}\sigma_{\xi,g,i}^{2}\mbox{Tr}\left\{\mathbb{E}\left[\hat{\mathbf{\Lambda}}_{F}^{*}\mathbf{T}\hat{\mathbf{\Lambda}}_{G}\mathbf{T}\right]\right\}\iftoggle{SINGLE_COL}{}{\nonumber\\
&}\stackrel{(c)}{=} \hat{\lambda} + \sum_{i=1}^{2K} p_{i}\sigma_{\xi,g,i}^{2}\hat{\eta}.\label{firstzfalpha}
\end{align}
The equality  in $(a)$ is obtained by exploiting the fact that $\hat{\bar{\mathbf{G}}}^{H}\hat{\mathbf{G}} = \hat{\mathbf{G}}^{H}\hat{\bar{\mathbf{G}}}=\mathbf{I}_{2K}$, and  \iftoggle{SINGLE_COL}{\\}{} $\mathbb{E}\left[{\mathbf{E}_{g}}\mathbf{P}\mathbf{P}^{H}{\mathbf{E}_{g}}^{H}\right]=\sum_{i=1}^{2K}p_{i}\sigma_{\xi,g,i}^{2}\mathbf{I}_{N}$. In the equality $(b)$, we define $\hat{\mathbf{\Lambda}}_{F}\triangleq(\hat{\bar{\mathbf{F}}}^{H}\hat{\bar{\mathbf{F}}})=\left({\bar{\mathbf{F}}}^{H}{\bar{\mathbf{F}}}\right)^{-1}$
as well as $\hat{\mathbf{\Lambda}}_{G}\triangleq\left(\hat{\bar{\mathbf{G}}}^{H}\hat{\bar{\mathbf{G}}}\right)=\left({\bar{\mathbf{G}}}^{H}{\bar{\mathbf{G}}}\right)^{-1}$ and {exploits} the fact that $\mathbf{T}\mathbf{P}\mathbf{P}^{H}\mathbf{T}=\mbox{diag}\{p_{2},p_{1},\cdots,p_{2K},p_{2K-1}\}$.
To derive the equality in $(c)$, we first note that the random matrices $\hat{\mathbf{\Lambda}}_{F}$ and $\hat{\mathbf{\Lambda}}_{G}$ have inverse Wishart distribution, i.e. we have $\hat{\mathbf{\Lambda}}_{F}\sim\mathcal{W}^{-1}(\hat{\mathbf{D}}_{d}^{-1},2K)$, $\hat{\mathbf{\Lambda}}_{G}\sim\mathcal{W}^{-1}(\hat{\mathbf{D}}_{u}^{-1},2K)$, where $\hat{\mathbf{D}}_{d}$ and $\hat{\mathbf{D}}_{u}$ are covariance matrices of the rows of the estimated channels $\hat{\mathbf{G}}$ and $\hat{\mathbf{F}}$, which are given after \eqref{est_ch_ref}.  
We denote $\hat{w}_{f,i,j} = \left(\hat{\mathbf{\Lambda}}_{F}\right)_{i,j},\,\hat{w}_{g,i,j} = \left(\hat{\mathbf{\Lambda}}_{G}\right)_{i,j},\,\forall i,j = 1,2,3,...,2K$ and $\mathbb{E}\left[\hat{\mathbf{\Lambda}}_{F}\right] = \frac{\hat{\mathbf{D}}_{d}^{-1}}{N-2K-1}$, $\mathbb{E}\left[\hat{\mathbf{\Lambda}}_{G}\right] = \frac{\hat{\mathbf{D}}_{u}^{-1}}{N-2K-1}$ \cite{graczyk2003complex}. We also have $\mathbb{E}\left[\hat{\bar{\mathbf{f}}}_{i}^{H}\hat{\bar{\mathbf{f}}}_{j}\right]=\mathbb{E}\left[\hat{w}_{f,i,j}\right] =\frac{1}{(N-2K-1)\hat\sigma_{f,i}^{2}}, \forall i=j, \,\mbox{and}\, 0 \, \mbox{otherwise}$. Similarly, $\mathbb{E}\left[\hat{\bar{\mathbf{g}}}_{i}^{H}\hat{\bar{\mathbf{g}}}_{j}\right]=\mathbb{E}\left[\hat{w}_{g,i,j}\right] =\frac{1}{(N-2K-1)\hat\sigma_{g,i}^{2}},\forall i=j, \,\mbox{and}\, 0\, \mbox{otherwise}$. Given the above equalities, we obtain the equality in $(c)$, where $\hat\lambda = \sum_{i=1}^{2K}\frac{p_{i^{'}}}{\left(N-2K-1\right)\hat\sigma_{f,i}^{2}}$. The expression  $\mbox{Tr}\left\{\mathbb{E}\left[\hat{\mathbf{\Lambda}}_{F}^{*}\mathbf{T}\hat{\mathbf{\Lambda}}_{G}\mathbf{T}\right]\right\}$ is simplified~as \iftoggle{SINGLE_COL}{}{in (\ref{omegasimp}).}
\iftoggle{BIG_EQUATION}{}{\iftoggle{SINGLE_COL}{\vspace*{-0.00in}}{}
\begin{eqnarray}\label{omegasimp}
\mbox{Tr}\left\{\mathbb{E}\left[\hat{\mathbf{\Lambda}}_{F}^{*}\mathbf{T}\hat{\mathbf{\Lambda}}_{G}\mathbf{T}\right]\right\}&=&\sum_{j=1}^{2K}\left(\mathbb{E}\left[w^{*}_{f,j,j^{'}}\right]\mathbb{E}\left[w_{g,j,j^{'}}\right]+\mathbb{E}\left[w^{*}_{f,j,j}\right]\mathbb{E}\left[w_{g,j^{'},j^{'}}\right]\right)\nonumber\iftoggle{SINGLE_COL}{\\[-8pt]}{\\}
&=&\sum_{j=1}^{2K}\frac{1}{\left(N-2K-1\right)^{2}\hat\sigma_{f,j}^{2}\hat\sigma_{g,	j^{'}}^{2}}\triangleq\hat{\eta}.
\end{eqnarray}
}

Following similar lines, the second term in the denominator of \eqref{alpha} can be simplified as
%can solve $\mathbb{E}\left[\|\hat{\bar{\mathbf{F}}}^{*}\mathbf{T}\hat{\bar{\mathbf{G}}}^{H}\mathbf{G}_{RR}\tilde{\mathbf{x}}_{R}\|^{2}\right]$ as
\begin{align}\setcounter{equation}{52}
&\mathbb{E}\left[\|\mathbf{W}\mathbf{G}_{RR}\tilde{\mathbf{x}}_{R}\|^{2}\right]=
\mathbb{E}\left[\|\hat{\bar{\mathbf{F}}}^{*}\mathbf{T}\hat{\bar{\mathbf{G}}}^{H}\mathbf{G}_{RR}\tilde{\mathbf{x}}_{R}\|^{2}\right]\iftoggle{SINGLE_COL}{}{\nonumber\\
&}= \frac{P_{R}}{N}\mbox{Tr}\left\{\mathbb{E}\left[\hat{\bar{\mathbf{F}}}^{*}\mathbf{T}\hat{\bar{\mathbf{G}}}^{H}\mathbf{G}_{RR}\mathbf{G}_{RR}^{H}\hat{\bar{\mathbf{G}}}\mathbf{T}^{H}\hat{\bar{\mathbf{F}}}^{T}\right]\right\} \nonumber\iftoggle{SINGLE_COL}{\\[-0pt]}{\\}
&= P_{R}\sigma_{LIR}^{2}\mbox{Tr}\left\{\mathbb{E}\left[\hat{\bar{\mathbf{F}}}^{T}\hat{\bar{\mathbf{F}}}^{*}\mathbf{T}\hat{\bar{\mathbf{G}}}^{H}\hat{\bar{\mathbf{G}}}\mathbf{T}^{H}\right]\right\}\iftoggle{SINGLE_COL}{}{\nonumber\\
&}=P_{R}\sigma_{LIR}^{2}\mbox{Tr}\left\{\mathbb{E}\left[\hat{\mathbf{\Lambda}}_{F}^{*}\mathbf{T}\hat{\mathbf{\Lambda}}_{G}\mathbf{T}\right]\right\}= P_{R}\sigma_{LIR}^{2}\hat{\eta}.\label{Thirdzfalpha}
\end{align}
The last term in the denominator of (\ref{alpha}) can be simplified to
\begin{align}
&\hspace{-0.3in}\mathbb{E}\left[\|\hat{\bar{\mathbf{F}}}^{*}\mathbf{T}\hat{\bar{\mathbf{G}}}^{H}\mathbf{z}_{R}\|^{2}\right]= \sigma_{nr}^{2}\mbox{Tr}\left\{\mathbb{E}\left[\hat{\bar{\mathbf{F}}}^{*}\mathbf{T}\hat{\bar{\mathbf{G}}}^{H}\hat{\bar{\mathbf{G}}}\mathbf{T}^{H}\hat{\bar{\mathbf{F}}}^{T}\right]\right\} \iftoggle{SINGLE_COL}{}{\nonumber\\
&}=\sigma_{nr}^{2}\mbox{Tr}\left\{\mathbb{E}\left[\hat{\mathbf{\Lambda}}_{F}^{*}\mathbf{T}\hat{\mathbf{\Lambda}}_{G}\mathbf{T}\right]\right\}= \sigma_{nr}^{2}\hat{\eta}.\label{seczfalpha}
\end{align}
By using (\ref{firstzfalpha}), (\ref{Thirdzfalpha}) and (\ref{seczfalpha}), we get the simplified expression of $\alpha$ in (\ref{alphazfproof}).

\iftoggle{SINGLE_COL}{\vspace*{-0.25in}}{}
\section{}
\label{gammrc}\iftoggle{SINGLE_COL}{\vspace*{-0.15in}}{}
We will start by simplifying the numerator of (\ref{gammalower}), with $\mathbf{f}_{k} = \hat{\mathbf{f}}_{k}+\mathbf{e}_{f,k}$, $\mathbf{g}_{k}= \hat{\mathbf{g}}_{k}+\mathbf{e}_{g,k}$, and using the fact that $\hat{\mathbf{f}}_{k}$, $\hat{\mathbf{g}}_{k}$, $\mathbf{e}_{f,k}$, $\mathbf{e}_{g,k}$ are independent, as follows
\begin{align}\label{absfwk}
&\mathbb{E}\left[|\mathbf{f}_{k}^{T}\mathbf{W}\mathbf{g}_{k^{'}}|\right]= \mathbb{E}\left[\left(\hat{\mathbf{f}}_{k}+\mathbf{e}_{f,k}\right)^{T}\mathbf{W}\left(\hat{\mathbf{g}}_{k^{'}}+\mathbf{e}_{g,k^{'}}\right)\right]\iftoggle{SINGLE_COL}{}{\nonumber\\
&} =\mathbb{E}\left[\hat{\mathbf{f}}_{k}^{T}\hat{\mathbf{F}}_{k}^{*}\mathbf{T}\hat{\mathbf{G}}^{H}\hat{\mathbf{g}}_{k^{'}}\right]\nonumber\iftoggle{SINGLE_COL}{\\[-2pt]}{\\}
& = \mathbb{E}\left[\hat{\mathbf{f}}_{k}^{T}\sum\limits_{i=1}^{K}\left(\hat{\mathbf{f}}_{2i-1}^{*}\hat{\mathbf{g}}_{2i}^{H}+\hat{\mathbf{f}}_{2i}^{*}\hat{\mathbf{g}}_{2i-1}^{H}\right)\hat{\mathbf{g}}_{k^{'}}\right]\iftoggle{SINGLE_COL}{}{\nonumber\\
&}= \mathbb{E}\left[\hat{\mathbf{f}}_{k}^{T}\hat{\mathbf{f}}_{k}^{*}\hat{\mathbf{g}}_{k^{'}}^{H}\hat{\mathbf{g}}_{k^{'}}+\hat{\mathbf{f}}_{k}^{T}\hat{\mathbf{f}}_{k{'}}^{*}\hat{\mathbf{g}}_{k}^{H}
\hat{\mathbf{g}}_{k^{'}}\right]\nonumber\iftoggle{SINGLE_COL}{\\[-2pt]}{\\}
&=\mathbb{E}\left[\hat{\mathbf{f}}_{k}^{H}\hat{\mathbf{f}}_{k}\right]\mathbb{E}\left[\hat{\mathbf{g}}_{k^{'}}^{H}\hat{\mathbf{g}}_{k^{'}}\right]+\mathbb{E}\left[\hat{\mathbf{f}}_{k{'}}^{H}\hat{\mathbf{f}}_{k}\right]\mathbb{E}\left[\hat{\mathbf{g}}_{k}^{H}\hat{\mathbf{g}}_{k^{'}}\right]\iftoggle{SINGLE_COL}{}{\nonumber\\
&} \stackrel{(a)}{=} N^{2}\hat\sigma_{f,k}^{2}\hat\sigma_{g,{k^{'}}}^{2}.
\end{align}
The equality in $(a)$ is because $\mathbb{E}\left[\hat{\mathbf{g}}_{i}^{H}\hat{\mathbf{g}}_{i}\right] = N\hat\sigma_{g,i}^{2}$, $\mathbb{E}\left[\hat{\mathbf{g}}_{i}^{H}\hat{\mathbf{g}}_{j}\right] = 0$ \cite{CIT-001}.

We now simplify the first term in the denominator of \eqref{gammalower},  the variance of  ${\mathbf{f}}_{k}^{T}\mathbf{W}{\mathbf{g}}_{k^{'}}=\mbox{var}\left[{\mathbf{f}}_{k}^{T}\mathbf{W}{\mathbf{g}}_{k^{'}}\right] = \mathbb{E}\left[|{\mathbf{f}}_{k}^{T}\mathbf{W}{\mathbf{g}}_{k^{'}}|^{2}\right]-\left|\mathbb{E}\left[|{{\mathbf{f}}}_{k}^{T}\mathbf{W}{\mathbf{g}}_{k^{'}}|\right]\right|^{2}$, where we have: %\iftoggle{SINGLE_COL}{\vspace*{-0.3in}}{}
\begin{align}
&\iftoggle{SINGLE_COL}{\hspace{-0.7in}}{}\mathbb{E}\left[|{\mathbf{f}}_{k}^{T}\mathbf{W}{\mathbf{g}}_{k^{'}}|^{2}\right] = \mathbb{E}\left[{\mathbf{f}}_{k}^{T}\mathbf{W}{\mathbf{g}}_{k^{'}}{\mathbf{g}}_{k^{'}}^{H}\mathbf{W}^{H}{\mathbf{f}}_{k}^{*}\right]\nonumber\\
&= \underbrace{\mathbb{E}\left[\hat{\mathbf{f}}_{k}^{T}\mathbf{W}\hat{\mathbf{g}}_{k^{'}}\hat{\mathbf{g}}_{k^{'}}^{H}\mathbf{W}^{H}\hat{\mathbf{f}}_{k}^{*}\right]}_{\text{(I)}}+\underbrace{\mathbb{E}\left[\hat{\mathbf{f}}_{k}^{T}\mathbf{W}\mathbf{e}_{g,k^{'}}\mathbf{e}_{g,k^{'}}^{H}\mathbf{W}^{H}\hat{\mathbf{f}}_{k}^{*}\right]}_{\text{(II)}}\nonumber\iftoggle{SINGLE_COL}{\\[-2pt]}{\\}
&\,+\underbrace{\mathbb{E}\left[\mathbf{e}_{f,k}^{T}\mathbf{W}\hat{\mathbf{g}}_{k^{'}}\hat{\mathbf{g}}_{k^{'}}^{H}\mathbf{W}^{H}\mathbf{e}_{f,k}^{*}\right]}_{\text{(III)}}+\underbrace{\mathbb{E}\left[\mathbf{e}_{f,k}^{T}\mathbf{W}\mathbf{e}_{g,k^{'}}\mathbf{e}_{g,k^{'}}^{H}\mathbf{W}^{H}\mathbf{e}_{f,k}^{*}\right]}_{\text{(IV)}}.\label{fwg}\iftoggle{SINGLE_COL}{}{\nonumber\\
}
\end{align}
\iftoggle{SINGLE_COL}{}{
\begin{figure*}
\normalsize
% Store the current equation number.
\setcounter{mytempeqncnt}{\value{equation}}
\setcounter{equation}{68}
\begin{eqnarray}\label{denapp3}
p_{k^{'}}\mbox{var}\left[{\mathbf{f}}_{k}^{T}\mathbf{W}{{\mathbf{g}_{k^{'}}}}\right]+p_{k}\mbox{SI}_{k}+\mbox{IP}_{k}+\mbox{NR}_{k}+\mbox{LIR}_{k}+\frac{1}{\alpha^{2}}\mbox{UI}_{k}+\frac{1}{\alpha^{2}}\mbox{NU}_{k}\nonumber\\
&&\hspace{-4.0in}= N^{2}\left\{\sum_{i=1}^{2K}p_{i}\left[\eta_{k,i}+\left(\hat{\Phi}\sigma_{g,i}^{2}+N\hat\sigma_{g,i}^{4}\hat\sigma_{f,i^{'}}^{2}\right)P_{R}^{-1}\left(\sigma_{n}^{2}+\sum_{i,k\in U_{k}}p_{i}\sigma_{k,i}^{2}\right)\right]+c_{k}p_{k}+P_{R}\sigma_{LIR}^{2}\left(\sigma_{f,k}^{2}\hat{\Phi}+N\hat\sigma_{f,k}^{4}\hat\sigma_{g,k^{'}}^{2}\right)\right.\nonumber\\
&&\hspace{-4.0in}\left.\qquad +P_{R}^{-1}\sigma_{nr}^{2}\hat{\Phi}\left(\sigma_{n}^{2}+\sum_{i,k\in U_{k}}p_{i}\sigma_{k,i}^{2}\right)+ \sum_{i,k\in U_{k}}p_{i}\sigma_{k,i}^{2}\sigma_{LIR}^{2}\hat{\Phi}+\left(\sigma_{LIR}^{2}\sigma_{n}^{2}+\sigma_{nr}^{2}\sigma_{f,k}^{2}\right)\hat{\Phi}+N\sigma_{nr}^{2}\hat\sigma^{4}_{f,k}\hat\sigma_{g,k^{'}}^{2}\right\}
\end{eqnarray}
\setcounter{equation}{56}
\hrule
\end{figure*}
}
The term-(I) in (\ref{fwg}) is simplified as following.
\begin{align}
&{}\mathbb{E}\left[\hat{\mathbf{f}}_{k}^{T}\mathbf{W}\hat{\mathbf{g}}_{k^{'}}\hat{\mathbf{g}}_{k^{'}}^{H}\mathbf{W}^{H}\hat{\mathbf{f}}_{k}^{*}\right]\iftoggle{SINGLE_COL}{}{\nonumber\\
&}\stackrel{(a)}{=}\mathbb{E}\left[\hat{\mathbf{f}}_{k}^{T}\sum\limits_{i=1}^{K}\left(\hat{\mathbf{f}}_{2i-1}^{*}\hat{\mathbf{g}}_{2i}^{H}+\hat{\mathbf{f}}_{2i}^{*}\hat{\mathbf{g}}_{2i-1}^{H}\right)\hat{\mathbf{g}}_{k^{'}}\hat{\mathbf{g}}_{k^{'}}^{H}\iftoggle{SINGLE_COL}{}{\right.\nonumber\iftoggle{SINGLE_COL}{\\[-2pt]}{\\}
&\left.\hspace{1in}}\left(\hat{\mathbf{g}}_{2i}\hat{\mathbf{f}}_{2i-1}^{T}+\hat{\mathbf{g}}_{2i-1}\hat{\mathbf{f}}_{2i}^{T}\right)\hat{\mathbf{f}}_{k}^{*}\right]\nonumber\iftoggle{SINGLE_COL}{\\[-4pt]}{\\}
&\stackrel{(b)}{=} N^{2}\sum\limits_{i=1,i\neq \lceil \frac{k}{2} \rceil}^{K}\hat\sigma_{f,k}^{2}\hat\sigma_{g,k^{'}}^{2}\left(\hat\sigma_{f,2i-1}^{2}\hat\sigma_{g,2i}^{2} + \hat\sigma_{f,2i}^{2}\hat\sigma_{g,2i-1}^{2}\right)\iftoggle{SINGLE_COL}{}{\nonumber\\
&\quad}+N^{2}\hat\sigma_{f,k}^{2}\hat\sigma_{f,k^{'}}^{2}\hat\sigma_{g,k}^{2}\hat\sigma_{g,k^{'}}^{2}+N^{2}(N+1)^
{2}\hat\sigma_{f,k}^{4}\hat\sigma_{g,k^{'}}^{4}\nonumber\iftoggle{SINGLE_COL}{\\[-4pt]}{\\}
&\stackrel{(c)}{=}  N^{2}\hat\sigma_{f,k}^{2}\hat\sigma_{g,k^{'}}^{2}\sum\limits_{i=1}^{K}\left(\hat\sigma_{f,2i-1}^{2}\hat\sigma_{g,2i}^{2} + \hat\sigma_{f,2i}^{2}\hat\sigma_{g,2i-1}^{2}\right)\iftoggle{SINGLE_COL}{}{\nonumber\\
&\quad}+N^{3}(N+2)\hat\sigma_{f,k}^{4}\hat\sigma_{g,k^{'}}^{4}\nonumber\iftoggle{SINGLE_COL}{\\[-2pt]}{\\}
&\stackrel{(d)}{=} N^{2}\hat\sigma_{f,k}^{2}\hat\sigma_{g,k^{'}}^{2}\left(N(N+2)\hat\sigma_{f,k}^{2}\hat\sigma_{g,k^{'}}^{2} + \hat{\Phi}\right)\label{I}.
\end{align}
The equality in $(b)$ is obtained by expanding $(a)$ for two different cases: i) $\{k,k^{'}\}=\{2i,2i-1\}$; ii) $\{k,k^{'}\}\neq\{2i,2i-1\}$, and by using the fact that $\mathbb{E}\left[\left|\hat{\mathbf{g}}_{i}^{H}\hat{\mathbf{g}}_{j}\right|^{2}\right] = N\hat\sigma_{g,i}^{2}\hat{\sigma}_{g,j}^{2}$ and $\mathbb{E}\left[\hat{\mathbf{g}}_{i}\hat{\mathbf{g}}_{i}^{H}\hat{\mathbf{g}}_{i}\hat{\mathbf{g}}_{i}^{H}\right] = (N+1)\hat\sigma_{g,i}^{4}\mathbf{I}_{N}$\cite{CIT-001}. The equality in $(c)$ is obtained by simple arithmetic manipulations and by re-arranging terms. The equality in~$(d)$ is obtained by defining $\hat{\Phi}=\sum_{i=1}^{K}\left(\hat\sigma_{g,2i-1}^{2}\hat\sigma_{f,2i}^{2}+\hat\sigma_{g,2i}^{2}\hat\sigma_{f,2i-1}^{2}\right)$.
The term-(II) in (\ref{fwg}) is simplified using similar ideas as follows:
\begin{align}
&\mathbb{E}\left[\hat{\mathbf{f}}_{k}^{T}\mathbf{W}\mathbf{e}_{g,k^{'}}\mathbf{e}_{g,k^{'}}^{H}\mathbf{W}^{H}\hat{\mathbf{f}}_{k}^{*}\right]\nonumber\iftoggle{SINGLE_COL}{\\[-4pt]}{\\}
&= N^{2}\iftoggle{SINGLE_COL}{\hspace{-0.03in}\left(\hspace{-0.03in}}{}\hat\sigma_{f,k}^{2}\hat\sigma_{f,k^{'}}^{2}\hat\sigma_{g,k}^{2}\hat\sigma_{\xi g,k^{'}}^{2}+\iftoggle{SINGLE_COL}{}{N^{2}}(N+1)\hat\sigma_{f,k}^{4}\hat\sigma_{g,k^{'}}^{2}\hat\sigma_{\xi g,k^{'}}^{2}\iftoggle{SINGLE_COL}{}{\nonumber\\
&\quad}+ \iftoggle{SINGLE_COL}{}{N^{2}}\sum\limits_{i\neq \lceil \frac{k}{2} \rceil}^{K}\hat\sigma_{\xi g,k^{'}}^{2}\hat\sigma_{f,k}^{2}\left(\hat\sigma_{f,2i-1}^{2}\hat\sigma_{g,2i}^{2} + \hat\sigma_{f,2i}^{2}\hat\sigma_{g,2i-1}^{2}\right)\iftoggle{SINGLE_COL}{\right)}{}\nonumber\iftoggle{SINGLE_COL}{\\[-14pt]}{\\}
&= N^{2}\hat\sigma_{f,k}^{2}\hat\sigma_{\xi g,k^{'}}^{2}\sum\limits_{i=1}^{K}\left(\hat\sigma_{f,2i-1}^{2}\hat\sigma_{g,2i}^{2} +  \hat\sigma_{f,2i}^{2}\hat\sigma_{g,2i-1}^{2}\right)\iftoggle{SINGLE_COL}{}{\nonumber\\
&\quad}+N^{3}\hat\sigma_{f,k}^{4}\hat\sigma_{\xi g,k^{'}}^{2}\hat\sigma_{g,k^{'}}^{2}\nonumber\iftoggle{SINGLE_COL}{\\[-8pt]}{\\}
&=N^{2}\hat\sigma_{f,k}^{2}\hat\sigma_{\xi g,k^{'}}^{2}\left(N\hat\sigma_{f,k}^{2}\hat\sigma_{g,k^{'}}^{2}+\hat{\Phi}\right). \label{II}
\end{align}
The term-(III) in (\ref{fwg}) is simplified as following.
\begin{align}
&\mathbb{E}\left[\mathbf{e}_{f,k}^{T}\mathbf{W}\hat{\mathbf{g}}_{k^{'}}\hat{\mathbf{g}}_{k^{'}}^{H}\mathbf{W}^{H}\mathbf{e}_{f,k}^{*}\right]\nonumber\iftoggle{SINGLE_COL}{\\[-12pt]}{\\}
&= N^{2}\iftoggle{SINGLE_COL}{\hspace{-0.03in}\left(\hspace{-0.03in}}{}\hat\sigma_{f,k^{'}}^{2}\hat\sigma_{\xi f,k}^{2}\hat\sigma_{g,k}^{2}\hat\sigma_{g,k^{'}}^{2}+\iftoggle{SINGLE_COL}{}{N^{2}}(N+1)\hat\sigma_{g,k^{'}}^{4}\hat\sigma_{f,k}^{2}\hat\sigma_{\xi f,k}^{2}\iftoggle{SINGLE_COL}{}{\nonumber\\
&\quad}+\iftoggle{SINGLE_COL}{}{N^{2}}\sum\limits_{i\neq
\lceil \frac{k}{2}\rceil}^{K}\hat\sigma_{\xi f,k}^{2}\hat\sigma_{g,k^{'}}^{2}\left(\hat\sigma_{f,2i-1}^{2}\hat\sigma_{g,2i}^{2}+\hat\sigma_{f,2i}^{2}\hat\sigma_{g,2i-1}^{2}\right)\iftoggle{SINGLE_COL}{\right)}{}\nonumber\iftoggle{SINGLE_COL}{\\[-14pt]}{\\}
&= N^{2}\hat\sigma_{\xi f,k}^{2}\hat\sigma_{g,k^{'}}^{2}\sum\limits_{i=1}^{K}\left(\hat\sigma_{f,2i-1}^{2}\hat\sigma_{g,2i}^{2}+\hat\sigma_{f,2i}^{2}\hat\sigma_{g,2i-1}^{2}\right)\iftoggle{SINGLE_COL}{}{\nonumber\\
&\quad}+N^{3}\hat\sigma_{\xi f,k}^{2}\hat\sigma_{g,k^{'}}^{4}\hat\sigma_{f,k}^{2}\nonumber\iftoggle{SINGLE_COL}{\\[-6pt]}{\\}
&= N^{2}\hat\sigma_{\xi f,k}^{2}\hat\sigma_{g,k^{'}}^{2}\left(N\hat\sigma_{g,k^{'}}^{2}\hat\sigma_{f,k}^{2}+\hat{\Phi}\right).\label{III}
\end{align}
Finally the term-(IV) in (\ref{fwg}) is simplified as follows:
\begin{align}
\iftoggle{SINGLE_COL}{\hspace{-0.2in}}{&\hspace{-0.3in}}\mathbb{E}\left[\mathbf{e}_{f,k}^{T}\mathbf{W}\mathbf{e}_{g,k^{'}}\mathbf{e}_{g,k^{'}}^{H}\mathbf{W}^{H}\mathbf{e}_{f,k}^{*}\right]\iftoggle{SINGLE_COL}{\hspace{-0.05in}}{\nonumber\\
&}= \iftoggle{SINGLE_COL}{\hspace{-0.05in}}{}N^{2}\sum\limits_{i=1}^{K}\hat\sigma_{\xi f,k}^{2}\hat\sigma_{\xi g,k^{'}}^{2}\left(\hat\sigma_{f,2i-1}^{2}\hat\sigma_{g,2i}^{2}+\hat\sigma_{f,2i}^{2}\hat\sigma_{g,2i-1}^{2}\right)\iftoggle{SINGLE_COL}{}{\nonumber\\
&}\iftoggle{SINGLE_COL}{\hspace{-0.05in}}{}=\iftoggle{SINGLE_COL}{\hspace{-0.05in}}{}N^{2}\hat\sigma_{\xi f,k}^{2}\hat\sigma_{\xi g,k^{'}}^{2}\hat{\Phi}.\label{IV}
\end{align}
%------
Further using (\ref{I})-(\ref{IV}), (\ref{fwg}) can be re-written as
\begin{align}
\iftoggle{SINGLE_COL}{}{&}\mathbb{E}\left[|{\mathbf{f}}_{k}^{T}\mathbf{W}{\mathbf{g}}_{k^{'}}|^{2}\right]\iftoggle{SINGLE_COL}{&}{
\nonumber\\&}= N^{2} \hat{\Phi}\left(\hat\sigma_{f,k}^{2}\hat\sigma_{g,k^{'}}^{2}+\hat\sigma_{f,k}^{2}\hat\sigma_{\xi g,k^{'}}^{2}+\hat\sigma_{\xi f,k}^{2}\hat\sigma_{g,k^{'}}^{2}+\hat\sigma_{\xi f,k}^{2}\hat\sigma_{\xi g,k^{'}}^{2}\right)\nonumber\iftoggle{SINGLE_COL}{\\[-4pt]}{\\}
& \iftoggle{SINGLE_COL}{\quad}{}+  N^{3}\hat\sigma_{f,k}^{2}\hat\sigma_{g,k^{'}}^{2}\left((N+2)\hat\sigma_{f,k}^{2}\hat\sigma_{g,k^{'}}^{2} +\hat\sigma_{f,k}^{2}\hat\sigma_{\xi g,k^{'}}^{2}+\hat\sigma_{\xi f,k}^{2}\hat\sigma_{g,k^{'}}^{2}\right).\label{fwgsimp}
\end{align}
By using (\ref{absfwk}), (\ref{fwgsimp}), and considering the fact that $\sigma_{g,k}^{2}=\hat\sigma_{\xi g,k}^{2}+\hat\sigma_{g,k}^{2}$, $\sigma_{f,k}^{2}=\hat\sigma_{\xi f,k}^{2}+\hat\sigma_{f,k}^{2}$, we~have
\begin{eqnarray}\label{var}
\mbox{var}\left[{\mathbf{f}}_{k}^{T}\mathbf{W}{\mathbf{g}}_{k^{'}}\right]\iftoggle{SINGLE_COL}{}{\nonumber\\
&&\hspace{-1.0in}}= N^{2}\left(\hat{\Phi}\sigma_{f,k}^{2}\sigma_{g,k^{'}}^{2} +  N\hat\sigma_{f,k}^{2}\hat\sigma_{g,k^{'}}^{2}\left(\sigma_{f,k}^{2}\hat\sigma_{g,k^{'}}^{2} +\hat\sigma_{f,k}^{2}\sigma_{g,k^{'}}^{2}\right)\right).\iftoggle{SINGLE_COL}{}{\nonumber\\}
\end{eqnarray}
Similar to the above calculations, we can simplify other terms in the denominator of \eqref{gammalower} as \begin{align}
&\mbox{SI}_{k} = N^{2}\sigma_{\xi f,k}^{2}\sigma_{\xi g,k}^{2}\hat{\Phi}+N^{2}\hat\sigma_{f,k}^{2}\sigma_{\xi g,k}^{2}\left(N\hat\sigma_{f,k}^{2}\hat\sigma_{g,k^{'}}^{2} + \hat{\Phi}\right)\iftoggle{SINGLE_COL}{}{\nonumber\\&} +N^{2}\sigma_{\xi f,k}^{2}\hat\sigma_{g,k}^{2}\left(N\hat\sigma_{g,k}^{2}\hat\sigma_{f,k^{'}}^{2}+\hat{\Phi}\right),\iftoggle{SINGLE_COL}{\\[-2pt]}{\\}
&\mbox{IP}_{k}= N^{2}\sum\limits_{j \neq k,k^{'}}^{2K}p_{j}\left[\sigma_{f,k}^{2}\sigma_{g,j}^{2}\hat{\Phi}+N\left(\sigma_{f,k}^{2}\hat\sigma_{g,j}^{4}\hat\sigma_{f,j^{'}}^{2}\iftoggle{SINGLE_COL}{}{\right.\right.\nonumber\iftoggle{SINGLE_COL}{\\[-2pt]}{\\}
&\qquad\qquad\qquad\qquad\left.\left.}+\sigma_{g,j}^{2}\hat\sigma_{f,k}^{4}\hat\sigma_{g,k^{'}}^{2}\right)\right],\iftoggle{SINGLE_COL}{\\[-8pt]}{\\}
&\mbox{NR}_{k} = N^{2}\sigma_{nr}^{2}\left(N\hat\sigma_{f,k}^{4}\hat\sigma_{g,k^{'}}^{2} + \sigma_{f,k}^{2}\hat{\Phi}\right)\iftoggle{SINGLE_COL}{}{,\\
&}\iftoggle{BIG_EQUATION}{}{,\quad}\mbox{LIR}_{k}= N^{2}P_{R},\sigma_{LIR}^{2}\left(N\hat\sigma_{f,k}^{4}\hat\sigma_{g,k^{'}}^{2} + \sigma_{f,k}^{2}\hat{\Phi}\right),\iftoggle{SINGLE_COL}{\\[-2pt]}{\\}
&\mbox{UI}_{k}= \sum\limits_{i,k\in U{k}}p_{i}\mathbb{E}\left[|\Omega_{k,i}{x_{i}}|^{2}\right]  = \sum\limits_{i,k\in U{k}}p_{i}\sigma_{k,i}^{2}\iftoggle{SINGLE_COL}{}{,\,\mbox{and}\\
&}\iftoggle{BIG_EQUATION}{}{,\quad}\mbox{NU}_{k}= \mathbb{E}\left[|z_{k}|^{2}\right] = \sigma_{n}^{2}.\label{nu}
\end{align}
We can further simplify the denominator of (\ref{gammalower}) using (\ref{var})-(\ref{nu}), as \iftoggle{SINGLE_COL}{}{(\ref{denapp3}) (shown at the top of the previous page), where $\eta_{k,i} = \hat{\Phi}\sigma_{f,k}^{2}\sigma_{g,i}^{2} +  N\left(\sigma_{f,k}^{2}\hat\sigma_{g,i}^{4}\hat\sigma_{f,i^{'}}^{2}+\sigma_{g,i}^{2}\hat\sigma_{f,k}^{4}\hat\sigma_{g,k^{'}}^{2}\right)$ and $c_{k}= -\left(\hat{\Phi}\hat\sigma_{f,k}^{2}\hat\sigma_{g,k}^{2}+N\left(\hat\sigma_{f,k}^{4}\hat\sigma_{g,k}^{2}\hat\sigma_{g,k^{'}}^{2}+\hat\sigma_{f,k}^{2}\hat\sigma_{g,k}^{4}\hat\sigma_{f,k^{'}}^{2}\right)\right)$. The proof of Theorem (\ref{theorem1}) is complete.}
\iftoggle{BIG_EQUATION}{}{
\begin{align}\label{denapp3}
&p_{k^{'}}\mbox{var}\left[{\mathbf{f}}_{k}^{T}\mathbf{W}{{\mathbf{g}_{k^{'}}}}\right]+p_{k}\mbox{SI}_{k}+\mbox{IP}_{k}+\mbox{NR}_{k}+\mbox{LIR}_{k}+\frac{1}{\alpha^{2}}\mbox{UI}_{k}+\frac{1}{\alpha^{2}}\mbox{NU}_{k}\nonumber\iftoggle{SINGLE_COL}{\\[-8pt]}{\\}
&= N^{2}\left\{\sum_{i=1}^{2K}p_{i}\left[\eta_{k,i}+\left(\hat{\Phi}\sigma_{g,i}^{2}+N\hat\sigma_{g,i}^{4}\hat\sigma_{f,i^{'}}^{2}\right)P_{R}^{-1}\left(\sigma_{n}^{2}+\sum_{i,k\in U_{k}}p_{i}\sigma_{k,i}^{2}\right)\right]+c_{k}p_{k}\right.\nonumber\iftoggle{SINGLE_COL}{\\[-8pt]}{\\}
&\left.\qquad+P_{R}\sigma_{LIR}^{2}\left(\sigma_{f,k}^{2}\hat{\Phi}+N\hat\sigma_{f,k}^{4}\hat\sigma_{g,k^{'}}^{2}\right) +P_{R}^{-1}\sigma_{nr}^{2}\hat{\Phi}\left(\sigma_{n}^{2}+\sum_{i,k\in U_{k}}p_{i}\sigma_{k,i}^{2}\right)+ \sum_{i,k\in U_{k}}p_{i}\sigma_{k,i}^{2}\sigma_{LIR}^{2}\hat{\Phi}\right.\nonumber\iftoggle{SINGLE_COL}{\\[-8pt]}{\\}
&\left.\qquad+\left(\sigma_{LIR}^{2}\sigma_{n}^{2}+\sigma_{nr}^{2}\sigma_{f,k}^{2}\right)\hat{\Phi}+N\sigma_{nr}^{2}\hat\sigma^{4}_{f,k}\hat\sigma_{g,k^{'}}^{2}\right\},
\end{align}
where $\eta_{k,i} = \hat{\Phi}\sigma_{f,k}^{2}\sigma_{g,i}^{2} +  N\left(\sigma_{f,k}^{2}\hat\sigma_{g,i}^{4}\hat\sigma_{f,i^{'}}^{2}+\sigma_{g,i}^{2}\hat\sigma_{f,k}^{4}\hat\sigma_{g,k^{'}}^{2}\right)$ and,\\  $c_{k}= -\left(\hat{\Phi}\hat\sigma_{f,k}^{2}\hat\sigma_{g,j^{'}}^{2}+N\left(\hat\sigma_{f,k}^{4}\hat\sigma_{g,j}^{2}\hat\sigma_{g,k^{'}}^{2}+\hat\sigma_{f,k}^{2}\hat\sigma_{g,j}^{4}\hat\sigma_{f,k^{'}}^{2}\right)\right)$. The proof of Theorem (\ref{theorem1}) is complete.
}

%\begin{figure*}
%\begin{eqnarray}
%\mbox{SI}_{k} &=& N^{2}\sigma_{\xi f,k}^{2}\sigma_{\xi g,k}^{2}\hat{\Phi}+N^{2}\hat\sigma_{f,k}^{2}\sigma_{\xi g,k}^{2}\left(N\hat\sigma_{f,k}^{2}\hat\sigma_{g,k^{'}}^{2} + \hat{\Phi}\right) +N^{2}\sigma_{\xi f,k}^{2}\hat\sigma_{g,k}^{2}\left(N\hat\sigma_{g,k}^{2}\hat\sigma_{f,k^{'}}^{2}+\hat{\Phi}\right)\nonumber\\
%\mbox{IP}_{k} &=& \sum\limits_{j \neq k,k^{'}}^{2K}p_{j}\left[N^{2}\sigma_{f,k}^{2}\sigma_{g,j}^{2}\hat{\Phi}+N^{3}\left(\sigma_{f,k}^{2}\hat\sigma_{g,j}^{4}\hat\sigma_{f,j^{'}}^{2}+\sigma_{g,j}^{2}\hat\sigma_{f,k}^{4}\hat\sigma_{g,k^{'}}^{2}\right)\right]=N^{2}\sum\limits_{j \neq k,k^{'}}^{2K}p_{j}\eta_{k,j}\\
%\mbox{NR}_{k} &=& N^{2}\sigma_{nr}^{2}\left(N\hat\sigma_{f,k}^{4}\hat\sigma_{g,k^{'}}^{2} + \sigma_{f,k}^{2}\hat{\Phi}\right),\qquad\,\mbox{LIR}_{k}= N^{2}P_{R}\sigma_{LIR}^{2}\left(N\hat\sigma_{f,k}^{4}\hat\sigma_{g,k^{'}}^{2} + \sigma_{f,k}^{2}\hat{\Phi}\right)\nonumber\\
%\mbox{UI}_{k}&=& \sum\limits_{i,k\in U{k}}p_{i}\mathbb{E}\left[|\Omega_{k,i}{x_{i}}|^{2}\right]  = \sum\limits_{i,k\in U{k}}p_{i}\sigma_{k,i}^{2},\qquad\mbox{NU}_{k}= \mathbb{E}\left[|z_{k}|^{2}\right] = \sigma_{n}^{2}
%\end{eqnarray}
%\hrule
%\end{figure*}

%\section{Proof of Theorem \ref{theorem2}}
\iftoggle{SINGLE_COL}{\vspace*{-0.15in}}{}
\section{}
\label{gamzf}\iftoggle{SINGLE_COL}{\vspace*{-0.15in}}{}
Starting with the numerator of (\ref{gammalower}) considering ZFR/ZFT processing, we have
\begin{align}\iftoggle{SINGLE_COL}{\setcounter{equation}{67}}{\setcounter{equation}{69}}
\mathbb{E}\left[{\mathbf{f}_{k}}^{T}\mathbf{W}\mathbf{g}_{k^{'}}\right] &= \mathbb{E}\left[\left(\hat{\mathbf{f}}_{k}+\mathbf{e}_{f,k}\right)^{T}\hat{\bar{\mathbf{F}}}^{*}\mathbf{T}\hat{\bar{\mathbf{G}}}^{H}\left(\hat{\mathbf{g}}_{k^{'}}+\mathbf{e}_{g,k^{'}}\right)\right]\iftoggle{SINGLE_COL}{}{\nonumber\\
&}= \mathbb{E}\left[\hat{\mathbf{f}}_{k}^{T}\hat{\bar{\mathbf{F}}}^{*}\mathbf{T}\hat{\bar{\mathbf{G}}}^{H}\hat{\mathbf{g}}_{k^{'}}\right]\nonumber\iftoggle{SINGLE_COL}{\\[-8pt]}{\\}
&\stackrel{(a)}{=} \mathbb{E}\left[{\mathbf{1}}_{f,k}^{T}\mathbf{T}{\mathbf{1}}_{g,k^{'}}\right]\iftoggle{SINGLE_COL}{}{\nonumber\\
&}\stackrel{(b)}{=} \mathbb{E}[1] = 1.
\end{align}
\iftoggle{SINGLE_COL}{}{
\begin{figure*}
\normalsize
% Store the current equation number.
\setcounter{mytempeqncnt}{\value{equation}}
\setcounter{equation}{70}
\begin{align}
&\mbox{var}\left[{\mathbf{f}_{k}}^{T}\mathbf{W}\mathbf{g}_{k^{'}}\right]=\mathbb{E}\left[\left|{\mathbf{f}_{k}}^{T}\mathbf{W}\mathbf{g}_{k^{'}}\right|^{2}\right]-\left|\mathbb{E}\left[{\mathbf{f}_{k}}^{T}\mathbf{W}\mathbf{g}_{k^{'}}\right]\right|^{2}\nonumber\\
&\stackrel{(a)}{=} \mathbb{E}\left[\hat{\mathbf{f}}_{k}^{T}\mathbf{W}\hat{\mathbf{g}}_{k^{'}}\hat{\mathbf{g}}_{k^{'}}^{H}\mathbf{W}^{H}\hat{\mathbf{f}}_{k}^{*}\right]-1+\mathbb{E}\left[\mathbf{e}_{f,k}^{T}\mathbf{W}\hat{\mathbf{g}}_{k^{'}}\hat{\mathbf{g}}_{k^{'}}^{H}\mathbf{W}^{H}\mathbf{e}_{f,k}^{*}\right] + \mathbb{E}\left[\hat{\mathbf{f}}_{k}^{T}\mathbf{W}\mathbf{e}_{g,k^{'}}\mathbf{e}_{g,k^{'}}^{H}\mathbf{W}^{H}\hat{\mathbf{f}}_{k}^{*}\right] + \mathbb{E}\left[\mathbf{e}_{f,k}^{T}\mathbf{W}\mathbf{e}_{g,k^{'}}\mathbf{e}_{g,k^{'}}^{H}\mathbf{W}^{H}\mathbf{e}_{f,k}^{*}\right]\nonumber\\
&\stackrel{(b)}{=}\sigma_{\xi,f,k}^{2} \mathbb{E}\left[\hat{\mathbf{g}}_{k^{'}}^{H}\mathbf{W}^{H}\mathbf{W}\hat{\mathbf{g}}_{k^{'}}\right]+ \sigma_{\xi,g,k^{'}}^{2}\mathbb{E}\left[\hat{\mathbf{f}}_{k}^{T}\mathbf{W}\mathbf{W}^{H}\hat{\mathbf{f}}_{k}^{*}\right] + \sigma_{\xi,f,k}^{2}\sigma_{\xi,g,k^{'}}^{2}\mbox{Tr}\left\{\mathbb{E}\left[\mathbf{W}\mathbf{W}^{H}\right]\right\}\nonumber\\
&\stackrel{(c)}{=}\sigma_{\xi,f,k}^{2} \mathbb{E}\left[\hat{\mathbf{g}}_{k^{'}}^{H}\hat{\bar{\mathbf{G}}}\mathbf{T}\hat{\bar{\mathbf{F}}}^{T}\hat{\bar{\mathbf{F}}}^{*}\mathbf{T}\hat{\bar{\mathbf{G}}}^{H}\hat{\mathbf{g}}_{k^{'}}\right] + \sigma_{\xi,g,k^{'}}^{2}\mathbb{E}\left[\hat{\mathbf{f}}_{k}^{T}\hat{\bar{\mathbf{F}}}^{*}\mathbf{T}\hat{\bar{\mathbf{G}}}^{H}\hat{\bar{\mathbf{G}}}\mathbf{T}\hat{\bar{\mathbf{F}}}^{T}\hat{\mathbf{f}}_{k}^{*}\right]
+ \sigma_{\xi,f,k}^{2}\sigma_{\xi,g,k^{'}}^{2}\mbox{Tr}\left\{\mathbb{E}\left[\hat{\bar{\mathbf{F}}}^{*}\mathbf{T}\hat{\bar{\mathbf{G}}}^{H}\hat{\bar{\mathbf{G}}}\mathbf{T}\hat{\bar{\mathbf{F}}}^{T}\right]\right\}\nonumber\\
&\stackrel{(d)}{=}\sigma_{\xi,f,k}^{2} \mathbb{E}\left[\mathbf{1}_{g,k^{'}}^{T}\mathbf{T}\hat{\mathbf{\Lambda}}_{F}^{*}\mathbf{T}\mathbf{1}_{g,k^{'}}\right]+ \sigma_{\xi,g,k^{'}}^{2} \mathbb{E}\left[\mathbf{1}_{f,k}^{T}\mathbf{T}\hat{\mathbf{\Lambda}}_{G}\mathbf{T}\mathbf{1}_{f,k}\right] + \sigma_{\xi,f,k}^{2}\sigma_{\xi,g,k^{'}}^{2}\mbox{Tr}\left\{\mathbb{E}\left[\hat{\mathbf{\Lambda}}_{F}^{*}\mathbf{T}\hat{\mathbf{\Lambda}}_{G}\mathbf{T}\right]\right\}\nonumber\\
%(\because \mathbf{1}_{g,k^{'}}^{T}=\hat{\mathbf{g}}_{k^{'}}^{H}\hat{\bar{\mathbf{G}}},\,\mathbf{1}_{g,k^{'}}=\hat{\bar{\mathbf{G}}}^{H}\hat{\mathbf{g}}_{k^{'}},\,\mathbf{1}_{f,k}^{T}=\hat{\mathbf{f}}_{k}^{T}\hat{\bar{\mathbf{F}}}^{*},\,\mathbf{1}_{f,k}=\hat{\bar{\mathbf{F}}}^{T}\hat{\mathbf{f}}_{k}^{*})\nonumber\\
&\stackrel{(e)}{=}\sigma_{\xi,f,k}^{2} \mathbb{E}\left[\hat{w}^{*}_{f,k,k}\right]+ \sigma_{\xi,g,k^{'}}^{2} \mathbb{E}\left[\hat{w}_{g,k^{'},k^{'}}\right] + \sigma_{\xi,f,k}^{2}\sigma_{\xi,g,k^{'}}^{2}\sum_{j=1}^{2K}\left(\mathbb{E}\left[w^{*}_{f,j,j^{'}}\right]\mathbb{E}\left[w_{g,j,j^{'}}\right]+\mathbb{E}\left[w^{*}_{f,j,j}\right]\mathbb{E}\left[w_{g,j^{'},j^{'}}\right]\right)\nonumber\\
&\stackrel{(f)}{=}\frac{\sigma_{\xi,f,k}^{2}}{(N-2K-1)\hat\sigma_{f,k}^{2}}+ \frac{\sigma_{\xi,g,k^{'}}^{2}}{(N-2K-1)\hat\sigma_{g,k^{'}}^{2}}+ \sigma_{\xi,f,k}^{2}\sigma_{\xi,g,k^{'}}^{2}\hat{\eta}\label{varzf}\\
%\mbox{SI}_{k} &=& \frac{\sigma_{\xi,f,k}^{2}}{(N-2K-1)\hat\sigma_{f,k^{'}}^{2}}+ \frac{\sigma_{\xi,g,k}^{2}}{(N-2K-1)\hat\sigma_{g,k^{'}}^{2}}+ \sigma_{\xi,f,k}^{2}\sigma_{\xi,g,k}^{2}\hat{\eta}\label{sizf}\\
%\mbox{IP}_{k} &=& \sum_{i=1,i\ne k,k^{}}^{2K}p_{i}\left(\frac{\sigma_{\xi,f,k}^{2}}{(N-2K-1)\hat\sigma_{f,i^{'}}^{2}}+ \frac{\sigma_{\xi,g,i}^{2}}{(N-2K-1)\hat\sigma_{g,k^{'}}^{2}}+ \sigma_{\xi,f,k}^{2}\sigma_{\xi,g,i}^{2}\hat{\eta}\right)\label{ipzf}\\
%\mbox{NR}_{k}&=&  \frac{\sigma_{nr}^{2}}{(N-2K-1)\hat\sigma_{g,k^{'}}^{2}}+ \sigma_{\xi,f,k}^{2}\sigma_{nr}^{2}\hat{\eta},\qquad\mbox{LIR}_{k}= \frac{P_{R}\sigma_{LIR}^{2}}{(N-2K-1)\hat\sigma_{g,k^{'}}^{2}}+ P_{R}\sigma_{LIR}^{2}\sigma_{\xi,f,k}^{2}\hat{\eta}\label{lirzf}\\
%\mbox{UI}_{k}&=& \sum\limits_{i,k\in U{k}}p_{i}\mathbb{E}\left[|\Omega_{k,i}{x_{i}}|^{2}\right]  = \sum\limits_{i,k\in U{k}}p_{i}\sigma_{k,i}^{2},\qquad
%\mbox{NU}_{k}= \mathbb{E}\left[|z_{k}|^{2}\right] = \sigma_{n}^{2}\label{nuzf}
\setcounter{equation}{77}
&p_{k^{'}}\mbox{var}\left[{\mathbf{f}}_{k}^{T}\mathbf{W}{{\mathbf{g}_{k^{'}}}}\right]+p_{k}\mbox{SI}_{k}+\mbox{IP}_{k}+\mbox{NR}_{k}+\mbox{LIR}_{k}+\frac{1}{\alpha^{2}}\mbox{UI}_{k}+\frac{1}{\alpha^{2}}\mbox{NU}_{k}\nonumber\\
&= \sum_{i=1}^{2K} \theta_{k,i}p_{i}+ \sigma_{nr}^{2}\left(\frac{1}{(N-2K-1)\hat\sigma_{g,k^{'}}^{2}}+\sigma_{\xi,f,k}^{2}\hat{\eta}\right)+P_{R}\sigma_{LIR}^{2}\left(\frac{1}{(N-2K-1)\hat\sigma_{g,k^{'}}^{2}}+\sigma_{\xi,f,k}^{2}\hat{\eta}\right)\nonumber\\
&\quad+\sum\limits_{i=1}^{2K}p_{i}\left[P_{R}^{-1}\sigma_{n}^{2}\left(\frac{1}{\left(N-2K-1\right)\hat\sigma_{f,i^{'}}^{2}} +\hat{\eta}\sigma_{\xi,g,i}^{2}\right)+\sum_{i,k\in U_{k}}p_{i}P_{R}^{-1}\sigma_{k,i}^{2}\left(\frac{1}{\left(N-2K-1\right)\hat\sigma_{f,i^{'}}^{2}} +\hat{\eta}\sigma_{\xi,g,i}^{2}\right)\right]\nonumber\\
&\quad +\left(P_{R}^{-1}\hat\eta\sigma_{nr}^{2}\sigma_{n}^{2}+\sum_{i,k\in U_{k}}p_{i}P_{R}^{-1}\sigma_{k,i}^{2}\hat\eta\sigma_{nr}^{2}+\sum_{i,k\in U_{k}}p_{i}\sigma_{k,i}^{2}\hat\eta\sigma_{LIR}^{2}+\hat\eta\sigma_{LIR}^{2}\sigma_{n}^{2}\right)\label{denzfsimp}
\end{align}
\setcounter{equation}{71}
\hrule
\end{figure*}
}
The equality in $(a)$ is obtained by using the following results: $\hat{\mathbf{g}}_{k^{'}}^{H}\hat{\bar{\mathbf{G}}}=\mathbf{1}_{g,k^{'}}^{T}$, $\hat{\bar{\mathbf{G}}}^{H}\hat{\mathbf{g}}_{k^{'}}=\mathbf{1}_{g,k^{'}}$, $\hat{\mathbf{f}}_{k}^{T}\hat{\bar{\mathbf{F}}}^{*}=\mathbf{1}_{f,k}^{T}$, $\hat{\bar{\mathbf{F}}}^{T}\hat{\mathbf{f}}_{k}^{*}=\mathbf{1}_{f,k}$. Equality in $(b)$ is attained because ${\mathbf{1}}_{f,k}^{T}\mathbf{T}{\mathbf{1}}_{g,k^{'}}=1$. 

The expression of $\mbox{var}\left[{\mathbf{f}_{k}}^{T}\mathbf{W}\mathbf{g}_{k^{'}}\right]$ in the denominator of \eqref{gammalower} is given by \iftoggle{SINGLE_COL}{}{(\ref{varzf}) (simplified at the top of next page).}
\iftoggle{BIG_EQUATION}{}{
\begin{align}
&\mbox{var}\left[{\mathbf{f}_{k}}^{T}\mathbf{W}\mathbf{g}_{k^{'}}\right]=\mathbb{E}\left[\left|{\mathbf{f}_{k}}^{T}\mathbf{W}\mathbf{g}_{k^{'}}\right|^{2}\right]-\left|\mathbb{E}\left[{\mathbf{f}_{k}}^{T}\mathbf{W}\mathbf{g}_{k^{'}}\right]\right|^{2}\nonumber\iftoggle{SINGLE_COL}{\\[-3pt]}{\\}
&\stackrel{(a)}{=} \mathbb{E}\left[\hat{\mathbf{f}}_{k}^{T}\mathbf{W}\hat{\mathbf{g}}_{k^{'}}\hat{\mathbf{g}}_{k^{'}}^{H}\mathbf{W}^{H}\hat{\mathbf{f}}_{k}^{*}\right]-1+\mathbb{E}\left[\mathbf{e}_{f,k}^{T}\mathbf{W}\hat{\mathbf{g}}_{k^{'}}\hat{\mathbf{g}}_{k^{'}}^{H}\mathbf{W}^{H}\mathbf{e}_{f,k}^{*}\right] + \mathbb{E}\left[\hat{\mathbf{f}}_{k}^{T}\mathbf{W}\mathbf{e}_{g,k^{'}}\mathbf{e}_{g,k^{'}}^{H}\mathbf{W}^{H}\hat{\mathbf{f}}_{k}^{*}\right]\iftoggle{BIG_EQUATION}{}{\nonumber\iftoggle{SINGLE_COL}{\\[-8pt]}{\\}&\qquad} + \mathbb{E}\left[\mathbf{e}_{f,k}^{T}\mathbf{W}\mathbf{e}_{g,k^{'}}\mathbf{e}_{g,k^{'}}^{H}\mathbf{W}^{H}\mathbf{e}_{f,k}^{*}\right]\nonumber\iftoggle{SINGLE_COL}{\\[-0pt]}{\\}
&\stackrel{(b)}{=}\sigma_{\xi,f,k}^{2} \mathbb{E}\left[\hat{\mathbf{g}}_{k^{'}}^{H}\mathbf{W}^{H}\mathbf{W}\hat{\mathbf{g}}_{k^{'}}\right]+ \sigma_{\xi,g,k^{'}}^{2}\mathbb{E}\left[\hat{\mathbf{f}}_{k}^{T}\mathbf{W}\mathbf{W}^{H}\hat{\mathbf{f}}_{k}^{*}\right] + \sigma_{\xi,f,k}^{2}\sigma_{\xi,g,k^{'}}^{2}\mbox{Tr}\left\{\mathbb{E}\left[\mathbf{W}\mathbf{W}^{H}\right]\right\}\nonumber\iftoggle{SINGLE_COL}{\\[-0pt]}{\\}
&\stackrel{(c)}{=}\sigma_{\xi,f,k}^{2} \mathbb{E}\left[\hat{\mathbf{g}}_{k^{'}}^{H}\hat{\bar{\mathbf{G}}}\mathbf{T}\hat{\bar{\mathbf{F}}}^{T}\hat{\bar{\mathbf{F}}}^{*}\mathbf{T}\hat{\bar{\mathbf{G}}}^{H}\hat{\mathbf{g}}_{k^{'}}\right] + \sigma_{\xi,g,k^{'}}^{2}\mathbb{E}\left[\hat{\mathbf{f}}_{k}^{T}\hat{\bar{\mathbf{F}}}^{*}\mathbf{T}\hat{\bar{\mathbf{G}}}^{H}\hat{\bar{\mathbf{G}}}\mathbf{T}\hat{\bar{\mathbf{F}}}^{T}\hat{\mathbf{f}}_{k}^{*}\right]\nonumber\iftoggle{SINGLE_COL}{\\[-0pt]}{\\}
&\qquad+ \sigma_{\xi,f,k}^{2}\sigma_{\xi,g,k^{'}}^{2}\mbox{Tr}\left\{\mathbb{E}\left[\hat{\bar{\mathbf{F}}}^{*}\mathbf{T}\hat{\bar{\mathbf{G}}}^{H}\hat{\bar{\mathbf{G}}}\mathbf{T}\hat{\bar{\mathbf{F}}}^{T}\right]\right\}\nonumber\iftoggle{SINGLE_COL}{\\[-0pt]}{\\}
&\stackrel{(d)}{=}\sigma_{\xi,f,k}^{2} \mathbb{E}\left[\mathbf{1}_{g,k^{'}}^{T}\mathbf{T}\hat{\mathbf{\Lambda}}_{F}^{*}\mathbf{T}\mathbf{1}_{g,k^{'}}\right]+ \sigma_{\xi,g,k^{'}}^{2} \mathbb{E}\left[\mathbf{1}_{f,k}^{T}\mathbf{T}\hat{\mathbf{\Lambda}}_{G}\mathbf{T}\mathbf{1}_{f,k}\right] + \sigma_{\xi,f,k}^{2}\sigma_{\xi,g,k^{'}}^{2}\mbox{Tr}\left\{\mathbb{E}\left[\hat{\mathbf{\Lambda}}_{F}^{*}\mathbf{T}\hat{\mathbf{\Lambda}}_{G}\mathbf{T}\right]\right\}\nonumber\iftoggle{SINGLE_COL}{\\[-0pt]}{\\}
%(\because \mathbf{1}_{g,k^{'}}^{T}=\hat{\mathbf{g}}_{k^{'}}^{H}\hat{\bar{\mathbf{G}}},\,\mathbf{1}_{g,k^{'}}=\hat{\bar{\mathbf{G}}}^{H}\hat{\mathbf{g}}_{k^{'}},\,\mathbf{1}_{f,k}^{T}=\hat{\mathbf{f}}_{k}^{T}\hat{\bar{\mathbf{F}}}^{*},\,\mathbf{1}_{f,k}=\hat{\bar{\mathbf{F}}}^{T}\hat{\mathbf{f}}_{k}^{*})\nonumber\iftoggle{SINGLE_COL}{\\[-4pt]}{\\}
&\stackrel{(e)}{=}\sigma_{\xi,f,k}^{2} \mathbb{E}\left[\hat{w}^{*}_{f,k,k}\right]+ \sigma_{\xi,g,k^{'}}^{2} \hspace{-0.05in}\left(\hspace{-0.05in}\mathbb{E}\left[\hat{w}_{g,k^{'},k^{'}}\right] \hspace{-0.02in}+ \hspace{-0.02in}\sigma_{\xi,f,k}^{2}\hspace{-0.02in}\sum_{j=1}^{2K}\hspace{-0.03in}\left(\mathbb{E}\left[w^{*}_{f,j,j^{'}}\right]\mathbb{E}\left[w_{g,j,j^{'}}\right]\hspace{-0.03in}+\hspace{-0.03in}\mathbb{E}\left[w^{*}_{f,j,j}\right]\mathbb{E}\left[w_{g,j^{'},j^{'}}\right]\hspace{-0.03in}\right)\hspace{-0.05in}\right)\nonumber\iftoggle{SINGLE_COL}{\\[-6pt]}{\\}
&\stackrel{(f)}{=}\frac{\sigma_{\xi,f,k}^{2}}{(N-2K-1)\hat\sigma_{f,k}^{2}}+ \frac{\sigma_{\xi,g,k^{'}}^{2}}{(N-2K-1)\hat\sigma_{g,k^{'}}^{2}}+ \sigma_{\xi,f,k}^{2}\sigma_{\xi,g,k^{'}}^{2}\hat{\eta}.\label{varzf}
\end{align}
}
{The equality in $(b)$ therein is because $\hat{\mathbf{F}}^{T}\mathbf{W}\hat{\mathbf{G}}= \mathbf{T},\,\mbox{i.e.},\, \hat{\mathbf{f}}_{k}^{T}\mathbf{W}\hat{\mathbf{g}}_{j}= 1,\forall j = k^{'}\, \mbox{and}\, 0, \, \mbox{otherwise}$. The equality in $(c)$ is obtained by substituting the value of $\mathbf{W}$ from \eqref{wzf}. Equalities in $(d)$ and $(e)$ are obtained by simple manipulations, and equality in $(f)$ is because $\mathbb{E}\left[\hat{w}_{f,k,k}\right] =\frac{1}{(N-2K-1)\hat\sigma_{f,k}^{2}}$, $\mathbb{E}\left[\hat{w}_{g,k^{'},k^{'}}\right] =\frac{1}{(N-2K-1)\hat\sigma_{g,k^{'}}^{2}}$, and $\hat{\eta}\triangleq\sum_{j=1}^{2K}\frac{1}{\left(N-2K-1\right)^{2}\hat\sigma_{f,j}^{2}\hat\sigma_{g,	j^{'}}^{2}}
$.

Remember that there is no need to perform SIC in the case ZFR/ZFT processing, and hence the self-interference term in the denominator of \eqref{gammalower} can be expressed as 
\begin{align}
&\mbox{SI}_{k} = \frac{1}{(N-2K-1)}\left(\frac{\sigma_{\xi,f,k}^{2}}{\hat\sigma_{f,k^{'}}^{2}}+ \frac{\sigma_{\xi,g,k}^{2}}{\hat\sigma_{g,k^{'}}^{2}}\right)+ \sigma_{\xi,f,k}^{2}\sigma_{\xi,g,k}^{2}\hat{\eta}.\label{sizf}
\end{align}
Similarly, the other terms in the denominator of (\ref{gammalower}) can be written as follows
\begin{align}
&\mbox{IP}_{k}\iftoggle{SINGLE_COL}{}{\nonumber\\
&\hspace{-0.06in}}= \sum_{i\ne k,k^{'}}^{2K}p_{i}\left[\frac{1}{(N-2K-1)}\left(\frac{\sigma_{\xi,f,k}^{2}}{\hat\sigma_{f,i^{'}}^{2}}+ \frac{\sigma_{\xi,g,i}^{2}}{\hat\sigma_{g,k^{'}}^{2}}\right)+ \sigma_{\xi,f,k}^{2}\sigma_{\xi,g,i}^{2}\hat{\eta}\right],\label{ipzf}\iftoggle{SINGLE_COL}{\\[-4pt]}{\\}
&\mbox{NR}_{k}\iftoggle{SINGLE_COL}{\hspace{-0.05in}}{}= \iftoggle{SINGLE_COL}{\hspace{-0.05in}}{} \sigma_{nr}^{2}\left(\iftoggle{SINGLE_COL}{\hspace{-0.05in}}{}\frac{1}{(N-2K-1)\hat\sigma_{g,k^{'}}^{2}}\iftoggle{SINGLE_COL}{\hspace{-0.05in}}{}+\iftoggle{SINGLE_COL}{\hspace{-0.05in}}{} \sigma_{\xi,f,k}^{2}\hat{\eta}\right)\iftoggle{SINGLE_COL}{\hspace{-0.05in}}{}\iftoggle{SINGLE_COL}{}{,\label{NR}\\
&}\iftoggle{BIG_EQUATION}{}{,}\mbox{LIR}_{k}\iftoggle{SINGLE_COL}{\hspace{-0.05in}}{}= \iftoggle{SINGLE_COL}{\hspace{-0.05in}}{}P_{R}\sigma_{LIR}^{2}\left(\iftoggle{SINGLE_COL}{\hspace{-0.05in}}{}\frac{1}{(N-2K-1)\hat\sigma_{g,k^{'}}^{2}}+ \sigma_{\xi,f,k}^{2}\hat{\eta}\right)\iftoggle{SINGLE_COL}{\hspace{-0.05in}}{},\label{lirzf}\iftoggle{SINGLE_COL}{\\[-4pt]}{\\}
&\mbox{UI}_{k}= \sum\limits_{i,k\in U{k}}p_{i}\mathbb{E}\left[|\Omega_{k,i}{x_{i}}|^{2}\right]  = \sum\limits_{i,k\in U{k}}p_{i}\sigma_{k,i}^{2}\iftoggle{SINGLE_COL}{}{,\, \mbox{and}\label{UI}\iftoggle{SINGLE_COL}{\\[-6pt]}{\\}
&}\iftoggle{BIG_EQUATION}{}{,\quad}\mbox{NU}_{k}= \mathbb{E}\left[|z_{k}|^{2}\right] = \sigma_{n}^{2}.\label{nuzf}
\end{align}
Substituting the values obtained from (\ref{varzf}-\ref{nuzf}) in the denominator of (\ref{gammalower}), we obtain \iftoggle{SINGLE_COL}{}{(\ref{denzfsimp}) (shown at the top of next page) where $\theta_{k,i}=\left(\frac{\sigma_{\xi,f,k}^{2}}{(N-2K-1)\hat\sigma_{f,i^{'}}^{2}}+ \frac{\sigma_{\xi,g,i}^{2}}{(N-2K-1)\hat\sigma_{g,k^{'}}^{2}}+ \sigma_{\xi,f,k}^{2}\sigma_{\xi,g,i}^{2}\hat{\eta}\right)$.}
\iftoggle{BIG_EQUATION}{}{
\begin{align}
&p_{k^{'}}\mbox{var}\left[{\mathbf{f}}_{k}^{T}\mathbf{W}{{\mathbf{g}_{k^{'}}}}\right]+p_{k}\mbox{SI}_{k}+\mbox{IP}_{k}+\mbox{NR}_{k}+\mbox{LIR}_{k}+\frac{1}{\alpha^{2}}\mbox{UI}_{k}+\frac{1}{\alpha^{2}}\mbox{NU}_{k}\nonumber\iftoggle{SINGLE_COL}{\\[-8pt]}{\\}
&\hspace{-0.05in}= \sum_{i=1}^{2K} \theta_{k,i}p_{i}+ \sigma_{nr}^{2}\left(\frac{1}{(N-2K-1)\hat\sigma_{g,k^{'}}^{2}}+\sigma_{\xi,f,k}^{2}\hat{\eta}\right)+P_{R}\sigma_{LIR}^{2}\left(\frac{1}{(N-2K-1)\hat\sigma_{g,k^{'}}^{2}}+\sigma_{\xi,f,k}^{2}\hat{\eta}\right)\nonumber\iftoggle{SINGLE_COL}{\\[-6pt]}{\\}
&\hspace{-0.05in}+\hspace{-0.05in}\sum\limits_{i=1}^{2K}\hspace{-0.01in}p_{i}\hspace{-0.02in}\left[\hspace{-0.02in}P_{R}^{-1}\sigma_{n}^{2}\hspace{-0.05in}\left(\hspace{-0.05in}\frac{1}{\left(N-2K-1\right)\hat\sigma_{f,i^{'}}^{2}} +\hat{\eta}\sigma_{\xi,g,i}^{2}\right)\hspace{-0.05in}+\hspace{-0.1in}\sum_{i,k\in U_{k}}p_{i}P_{R}^{-1}\sigma_{k,i}^{2}\left(\frac{1}{\left(N-2K-1\right)\hat\sigma_{f,i^{'}}^{2}} +\hat{\eta}\sigma_{\xi,g,i}^{2}\right)\hspace{-0.05in}\right]\nonumber\iftoggle{SINGLE_COL}{\\[-8pt]}{\\}
&+\left(P_{R}^{-1}\hat\eta\sigma_{nr}^{2}\sigma_{n}^{2}+\sum_{i,k\in U_{k}}p_{i}P_{R}^{-1}\sigma_{k,i}^{2}\hat\eta\sigma_{nr}^{2}+\sum_{i,k\in U_{k}}p_{i}\sigma_{k,i}^{2}\hat\eta\sigma_{LIR}^{2}+\hat\eta\sigma_{LIR}^{2}\sigma_{n}^{2}\right),\label{denzfsimp}
\end{align}
where $\theta_{k,i}=\left(\frac{\sigma_{\xi,f,k}^{2}}{(N-2K-1)\hat\sigma_{f,i^{'}}^{2}}+ \frac{\sigma_{\xi,g,i}^{2}}{(N-2K-1)\hat\sigma_{g,k^{'}}^{2}}+ \sigma_{\xi,f,k}^{2}\sigma_{\xi,g,i}^{2}\hat{\eta}\right)$.
}
\iftoggle{SINGLE_COL}{\vspace*{-1pt}}{}
\bibliographystyle{IEEEtran}
\bibliography{IEEEabrv,Relay_ref}
%\bibliography{Relay_ref}{}

\end{document}